\documentclass[10pt]{iopart}

\expandafter\let\csname equation*\endcsname\relax
\expandafter\let\csname endequation*\endcsname\relax

\usepackage{mhchem}
\usepackage{hyperref}
\usepackage{bm}
\usepackage{physics}
\usepackage{xcolor}
\usepackage{bbold}
\usepackage{nicefrac}
\usepackage{amssymb}
\usepackage{dsfont}
\usepackage{xifthen}
\usepackage{graphicx}
\usepackage{array}
\usepackage{booktabs}
\usepackage{etoolbox}
\usepackage[nocompress]{cite}
\usepackage[protrusion=true,expansion=true,final]{microtype}
\usepackage{siunitx}
\usepackage{iopams}
\usepackage{cleveref}
\crefname{appendix}{}{}
\usepackage{algorithm}
\usepackage{algorithmic}
\makeatletter
    \long\def\@makefntext#1{\parindent 1em\noindent 
    \makebox[1em][l]{\footnotesize\rm$\m@th{\arabic{footnote}}$}%
    \footnotesize\rm \nobreakspace #1}
    \def\@makefnmark{\textsuperscript{\hbox{${\arabic{footnote}}\m@th$}}}
    \def\@thefnmark{\arabic{footnote}}
\makeatother

\DeclareMathOperator{\SU}{SU}
\DeclareMathOperator{\U}{U}
\DeclareMathOperator{\SO}{SO}

\newcommand{\nodag}{{\vphantom\dagger}}
\newcommand{\noprime}{{\vphantom\prime}}
\newcommand{\cc}{c^\nodag}
\newcommand{\cd}{c^\dagger}
\newcommand{\up}{\uparrow}
\newcommand{\down}{\downarrow}

\newcommand{\psibar}{\bar{\psi}}
\newcommand{\etabar}{\bar{\eta}}
\newcommand{\phibar}{\bar{\phi}}
\newcommand{\W}{\mathcal{W}[\etabar, \eta]}
\newcommand{\Wc}{\mathcal{W}_c[\etabar, \eta]}
\newcommand{\eff}{\Gamma[\phibar, \phi]}
\newcommand{\diff}{\partial_{\Lambda}}
\newcommand{\diffsig}{\partial^{\Sigma}_{\Lambda}}
\newcommand{\Wclam}{\mathcal{W}^{\Lambda}_c[\etabar, \eta]}
\newcommand{\efflam}{\Gamma^{\Lambda}[\phibar, \phi]}
\newcommand{\G}[4]{\Gamma^{\Lambda}(#1, #2| #3, #4)}
\newcommand{\Gt}[4]{\tilde{\Gamma}^{\Lambda}(#1, #2| #3, #4)}

\newcommand{\R}{\bm{r}}
\newcommand{\bmR}{\bm{R}}
\newcommand{\Rin}{\bm{R}^\textrm{inv}}

\newcommand{\NPairs}{N_\text{Pairs}}
\newcommand{\NU}{N_\text{Unique}}
\newcommand{\Nsum}{N_\text{sum}}

\newcommand{\Gam}[3]{\Gamma^\Lambda_{#1}(\R_#2,\R_#3)}

\newcommand{\eq}[1]{\cref{#1}}
\newcommand{\eqs}[3][]{eqs.~\ifthenelse{\isempty{#1}}{}{\eqref{#1}, }\eqref{#2}\ifthenelse{\isempty{#1}}{}{,} and \eqref{#3}}
\newcommand{\Sec}[1]{\cref{#1}}
\newcommand{\mRef}[1]{Ref.~\cite{#1}}
\newcommand{\tab}[1]{\cref{#1}}
\newcommand{\Refs}[3][]{Refs.~\ifthenelse{\isempty{#1}}{}{\cite{#1}, }\cite{#2}\ifthenelse{\isempty{#1}}{}{,} and \cite{#3}}
\newcommand{\fig}[1]{\cref{#1}}

\newcommand{\PFSpace}[1]{\widetilde{#1}}

\let\oldi\i
\renewcommand{\i}{\ensuremath{\mathrm{i}}}

\newcolumntype{A}{ >{$} r <{$} @{} >{${}} l <{$} } 
	
\DeclareDocumentCommand{\fdvn}{ s o m g g d() }
{ 
	\IfBooleanTF{#1}
	{\let\fractype\flatfrac}
	{\let\fractype\frac}
	\IfNoValueTF{#4}
	{
		\IfNoValueTF{#6}
		{\fractype{\delta \IfNoValueTF{#2}{}{^{#2}}}{\delta #3\IfNoValueTF{#2}{}{^{#2}}}}
		{\fractype{\partial \IfNoValueTF{#2}{}{^{#2}}}{\delta #3\IfNoValueTF{#2}{}{^{#2}}} \argopen(#6\argclose)}
	}
	{
		\IfNoValueTF{#5}
		{\fractype{\delta \IfNoValueTF{#2}{}{^{#2}} #3}{\partial #4\IfNoValueTF{#2}{}{^{#2}}}}
		{\fractype{\delta^2 #3}{\delta #4 \delta #5}}
	}
}

\newcommand{\vertex}[5]{\Gamma_{{#1}}^{{#2}}\left(\begin{matrix}#3\\#4\\#5\end{matrix}\right)}

\def\dd{\mathrm{d}}

\begin{document}

\review[Pseudo-fermion functional renormalization group for spin models]{Pseudo-fermion functional renormalization group for spin models}

\author{Tobias M\"uller$^{1}$, Dominik Kiese$^{2}$, Nils Niggemann$^{3,4,5}$, Bj\"orn Sbierski$^{6,7}$, Johannes Reuther$^{3,4,5}$, Simon Trebst$^{8}$, Ronny Thomale$^{1,5}$ and Yasir Iqbal$^{5}$}
\address{$^{1}$ Institut f\"ur Theoretische Physik und Astrophysik, Julius-Maximilian-Universit\"at W\"urzburg, Am Hubland, D-97074 W\"urzburg, Germany}
\address{$^2$ Center for Computational Quantum Physics, Flatiron Institute, 162 5th Avenue, New York, NY 10010, USA}
\address{$^{3}$ Dahlem Center for Complex Quantum Systems and Fachbereich Physik, Freie Universit\"at Berlin, D-14195 Berlin, Germany}
\address{$^{4}$ Helmholtz-Zentrum Berlin f\"ur Materialien und Energie, Hahn-Meitner-Platz 1, D-14109 Berlin, Germany}
\address{$^{5}$ Department of Physics and Quantum Centre of Excellence for Diamond and Emergent Materials (QuCenDiEM), Indian Institute of Technology Madras, Chennai 600036, India}
\address{$^6$ Department of Physics and Arnold Sommerfeld Center for Theoretical Physics (ASC), Ludwig-Maximilians-Universit\"at M\"unchen, Theresienstra{\ss}e 37, M\"unchen D-80333, Germany}
\address{$^7$ Munich Center for Quantum Science and Technology (MCQST), Schellingstr. 4, D-80799 M\"unchen, Germany}
\address{$^{8}$ Institut f\"ur Theoretische Physik, Z\"ulpicher Stra{\ss}e 77a, Universit\"at zu K\"oln, 50937 K\"oln, Germany}

\eads{\mailto{tobias.mueller1@uni-wuerzburg.de}, \mailto{yiqbal@physics.iitm.ac.in}}
\vspace{-5pt}

\begin{abstract}
For decades, frustrated quantum magnets have been a seed for scientific progress and innovation in condensed matter. As much as the numerical tools for low-dimensional quantum magnetism have thrived and improved in recent years due to breakthroughs inspired by quantum information and quantum computation, higher-dimensional quantum magnetism can be considered as the final frontier, where strong quantum entanglement, multiple ordering channels, and manifold ways of paramagnetism culminate. At the same time, efforts in crystal synthesis have induced a significant increase in the number of tangible frustrated magnets which are generically three-dimensional in nature, creating an urgent need for quantitative theoretical modeling. We review the pseudo-fermion (PF) and pseudo-Majorana (PM) functional renormalization group (FRG) and their specific ability to address higher-dimensional frustrated quantum magnetism. First developed more than a decade ago, the PFFRG interprets a Heisenberg model Hamiltonian in terms of Abrikosov pseudofermions, which is then treated in a diagrammatic resummation scheme formulated as a renormalization group flow of $m$-particle pseudofermion vertices. The article reviews the state of the art of PFFRG and PMFRG and discusses their application to exemplary domains of frustrated magnetism, but most importantly, it makes the algorithmic and implementation details of these methods accessible to everyone. By thus lowering the entry barrier to their application, we hope that this review will contribute towards establishing PFFRG and PMFRG as the numerical methods for addressing frustrated quantum magnetism in higher spatial dimensions. 
\end{abstract}

\vspace{-5pt}

\maketitle

\ioptwocol

\section{Introduction}

In condensed matter physics, quantum-many body systems can give rise to remarkable collective states of matter that have no classical counterparts,
such as superconductors~\cite{Bardeen-1957}, superfluids~\cite{Khalat-1965} or quantum spin liquids~\cite{Savary_2017}. But connecting such complex emergent behavior to a microscopic picture in terms of short-ranged interactions between the elementary quantum mechanical degrees of freedom has, to this day, remained a fundamental challenge.
Analytical approaches can often provide initial guidance and crude understanding on the level of mean-field theory or effective field theory descriptions, but their validity and underlying abstractions are often a matter of debate. Instead, unbiased numerical simulations are called for 
in verifying these assumptions and providing quantitative guidance, e.g.~by mapping out phase diagrams (in terms of the microscopic interactions) and identifying the respective phase transitions. This has led to a remarkable string of method development including the expansion of Monte Carlo simulation techniques to the quantum realm \cite{gubernatis_kawashima_werner_2016}, the development of dynamical mean-field theory \cite{RevModPhys.68.13}, the formulation of entanglement-based variational approaches such as the density matrix-renormalization group \cite{RevModPhys.77.259} and tensor network approaches \cite{SCHOLLWOCK201196}, and most recently their combination with ideas from machine learning \cite{Carrasquilla2017,CarleoTroyer2017}.
Over the past three decades, this combination of analytical and numerical approaches has led to remarkable progress in understanding the general features of collective states of quantum systems constituted by bosonic degrees of freedom, such as a broad class of quantum magnets or ultracold atomic systems. 

However, there are a number of important outstanding problems that have for decades resisted solution, most prominently the {\it many-fermion problem}. The quantum statistics, which sets fermions apart from bosons, has profound implications not only on the intricate nodal structure of quantum mechanical wavefunctions of many-fermion systems and the resulting, enticingly complex variety of fermionic ground states but also on the ability to simulate many-fermion systems with the most powerful, unbiased numerical approach to quantum many-body systems -- quantum Monte Carlo simulations. As realized early on, the fermionic exchange statistics leads to the infamous sign problem \cite{PhysRevB.41.9301}, i.e.~the occurrence of negative statistical weights in the sampling of fermionic world-line configurations. Overcoming the sign problem by identifying a basis transformation to a sign-free basis (such as the basis of eigenstates) is known to be a {\it NP}-hard problem \cite{PhysRevLett.94.170201}. 
This computational complexity arising from the sign problem also manifests itself in a broad variety of {\it frustrated quantum magnets} -- quantum magnets with competing interactions that cannot be simultaneously satisfied and which thereby give rise to low-temperature physics that is quite distinct from their conventional counterparts \cite{lacroix2011introduction}. This includes the formation of long-range entangled quantum order, emergent gauge theories, 
and fractionalization of the elementary quantum mechanical (spin) degrees of freedom. As such, frustrated quantum magnets have attracted broad interest since they have long served as a fertile ground to develop the basic phenomenology and concepts of quantum many-body systems at large.
However, their numerical exploration has remained challenging as they are often not amenable to path-integral quantum Monte Carlo techniques due to their
intrinsic sign problem (with some exceptions \cite{PhysRevLett.113.197205}), dynamical mean-field theory due to their long-range quantum structure at low temperatures, or tensor-network based approaches as some of the most interesting problems occur in three spatial dimensions such as the formation of quantum spin ice (though DMRG has made some inroads into higher-dimensional problems \cite{Stoudenmire2012}). 

Diagrammatic methods centered around the concept of correlation functions represent an alternative and well established approach to quantum many body physics \cite{abrikosovMethodsQuantum1975}. By construction, diagrammatic methods are both oblivious to frustration and the dimensionality of the system. However, the main obstacle for their direct application to spin systems is the lack of an unconstrained fermionic or bosonic path integral for the latter such that Wick's theorem, the main pillar of diagrammatic perturbation theory, does not hold. Due to the resulting complications, diagrammatic approaches to spin systems - though explored \cite{Vaks-1968,izyumovStatisticalMechanics1988,Maleev-1974,Izyumov-2002} - have not been widely applied.

Once spin operators are represented in terms of (pseudo-)fermionic operators the situation changes and one can apply the well developed diagrammatic toolbox to the resulting - and in general interacting - fermionic problem. However, as it turns out, simple perturbation theory is rarely sufficient or even suffers from infinities so that one has to resort to modern resummation schemes. The resummation scheme of choice for pseudo-fermion Hamiltonians is the fermionic functional renormalization group (FRG) \cite{Metzner2012,salmhofer_book} in vertex expansion. It builds upon Wetterich's generalization \cite{Wetterich1993} of the renormalization group idea by Kadanoff and Wilson \cite{kadanoffScalingLaws1966,Wilson1971,Wilson1974} whereas the latter was originally applied to critical phenomena and the Kondo model.

The pseudo-fermion functional renormalization group (PFFRG), which at its core consists of a hierarchy of flow equations for vertex functions, was formulated in a seminal work by Reuther and W\"olfle \cite{Reuther2010,Reuther_2010}, and subsequently developed over the years~\cite{Reuther-2011a,Reuther-2011b,Reuther-2011c,Reuther-2011d,Goettel2012,Reuther2012,Suttner2014,Reuther2014b,Reuther-2014,Rueck2018,Kiese2020b} to bring much-needed numerical guidance to such frustrated quantum magnets in two~\cite{Singh-2012,Iqbal2015,Rousochatzakis2015,Iqbal2016c,Buessen2016a,Iqbal_tri,Hering2019a,Iida2020,Astrakhantsev2021,Kiese2022,Hering2021,Iqbal-mll2023,niggemann2023_squagome} and three spatial dimensions~\cite{Iqbal2016b,Laubach-2016,Buessen2016a,Iqbal2017,Iqbal2018,Iqbal2019,Ghosh2019,Ghosh2019a,Chillal2020,Zivkovic2021,Hering2021b,Noculak2023,Niggemann2023,Gresista2023,lozanogomez2023,chern2023}. Much technical understanding has been developed over the past fifteen years including ways to reduce the number of flow equations by symmetry-optimization \cite{Buessen2019a}, to reliably distinguish the formation of quantum spin liquids versus the long-range magnetic order, to expand the approach to the limits of large $S$ \cite{Baez2017} and large $N$ \cite{Buessen2018}, along with numerous technical tricks to speed up practical implementations which have been made available as open-source packages \cite{PFFRGSolver,Buessen2021b,PMFRG.jl}. More recent advances include an alternate formulation of the PFFRG approach in terms of auxiliary Majorana fermions along with inroads to quantitatively describe the finite-temperature physics \cite{Niggemann2021,schneiderTamingPseudofermion2022} of frustrated magnets.

We emphasize that the diagrammatic Monte Carlo approach has been applied to PF Hamiltonians \cite{kulaginBoldDiagrammatic2013,kulaginBoldDiagrammatic2013a,Huang2016}. Offering a different type of resummation scheme than the FRG, this method can thus be seen as closely related to PFFRG and we will make appropriate quantitative comparisons below.

It is the purpose of this review to give a pedagogical introduction to the PFFRG approach and to provide extensive technical details on its implementation so that a beginning graduate student might find all the information to set up one's own calculations. We also provide an overview of the many applications of the PFFRG approach over the past fifteen years to a variety of quantum magnets with competing, diagonal and/or off-diagonal spin exchange in two and three dimensional lattice geometries as well as, more recently, to systems with coupled spin-orbital or spin-valley degrees of freedom. 
For those wanting to readily jump to certain parts of this review here is an overview of its structure:
In \cref{sec:model} we introduce the basic microscopic exchange model of a quantum magnet, which we then recast in terms of auxiliary fermions in \cref{sec:auxiliary_fermions} and discuss its fundamental symmetries. We then dive into the technical discussion setting the 
stage with an introduction of the functional renormalization group in \cref{sec:functionalrenormalization}, 
then moving to the PFFRG in \cref{sec:PFFRG}. 
We close our technical discussion with an account of recent extensions to finite-temperature physics in \cref{sec:finiteT}.
Section \ref{sec:applications} is then devoted to a broad overview of applications of the PFFRG approach to fundamental problems
in frustrated quantum magnetism.
We close this review with \cref{sec:challenges} on future directions and some of the challenges that still lay ahead of us
and some conclusions in \cref{sec:conclusions}.

\section{Model}
\label{sec:model}
In the following, we consider models with time-independent spin Hamiltonians in which two spins on lattice sites $i$ and $j$ interact via an exchange interaction $J^{\alpha \beta}_{ij}$. Here, we assume a general interaction with $\alpha,\beta = x,y,z$ that couples the $\alpha$'th component of spin $i$ to the $\beta$'th component of spin $j$
\begin{equation}
\mathcal{H}=\frac{1}{2}\sum_{i,j}\sum_{\alpha,\beta}S^\alpha_{i} J^{\alpha\beta}_{ij} S^\beta_{j}.
\label{eq:mod-ham}
\end{equation}
For spin-$1/2$, the operators can be represented by Pauli matrices $\sigma^\alpha$, i.e. $S^\alpha_{i} = \frac{\hbar}{2} \sigma^\alpha_i$, where $\sigma_i^\alpha$ is defined to only act on site $i$ and thus commute with all other operators that are not acting on $i$. In this general form, \cref{eq:mod-ham} describes a vast multitude of interacting spin models, for instance the isotropic Heisenberg model $J^{\alpha \beta}_{ij} = J_{ij} \delta_{\alpha \beta}$, but can also contain anisotropic interactions, e.g. of Kitaev~\cite{Revelli2019,Buessen2021,Reuther-2014,Reuther-2011d,Fukui2022,Fukui2022a,Fukui2023} or Dzyaloshinsky-Moriya type~\cite{Hering-2017,Buessen2019a,Noculak2023}. Similarly, the real-space extent of the interactions is not only limited to short-range, but also long-range interactions are treatable~\cite{Keles2018,Keles2018a,sbierskiMagnetismTwodimensional2023}

Due to its exponentially large Hilbert space and its strongly interacting nature, an exact solution of \cref{eq:mod-ham} is often impossible. In the following, we discuss the functional renormalization group as a many-body field theoretical method to obtain an approximate solution. Although this approach, in principle, also can handle interactions involving more than two spins, this is numerically not tractable, such that we restrict the discussion in this manuscript of Hamiltonians of the type defined in \cref{eq:mod-ham}.

\section{Auxiliary Fermions}
\label{sec:auxiliary_fermions}
Many standard diagrammatic techniques used to treat quantum many-body systems are not applicable to spin models, due to the peculiar commutator structure of their corresponding operators. The canonical angular momentum commutation relations 
\begin{equation}
  \commutator{S^\alpha}{S^\beta} = \i \hbar \sum\limits_{\gamma=1}^3 \epsilon_{\alpha\beta\gamma} S^\gamma, \label{eq:Scommutator}
\end{equation}
of the spin operator's components $S^\alpha$ ($\alpha=x,y,z$) is neither fermionic nor bosonic. This in turn renders Wick's theorem~\cite{Wick1950}, a fundamental theorem upon which most many-body techniques are based, inapplicable in its standard formulation~\cite{Wang1966,Wang1966a}.

This fact can be remedied by introducing an auxiliary particle representation of the spin operators in terms of pseudo-fermions, first introduced by Abrikosov~\cite{Abrikosov1965}. In the following, we will review this representation and give an overview over the consequences the construction has for the Green's functions of the pseudo-particles.

\subsection{Spin-operator mapping}
\label{sec:operatormapping}

\begin{figure}
     \centering
     \includegraphics[width = \columnwidth]{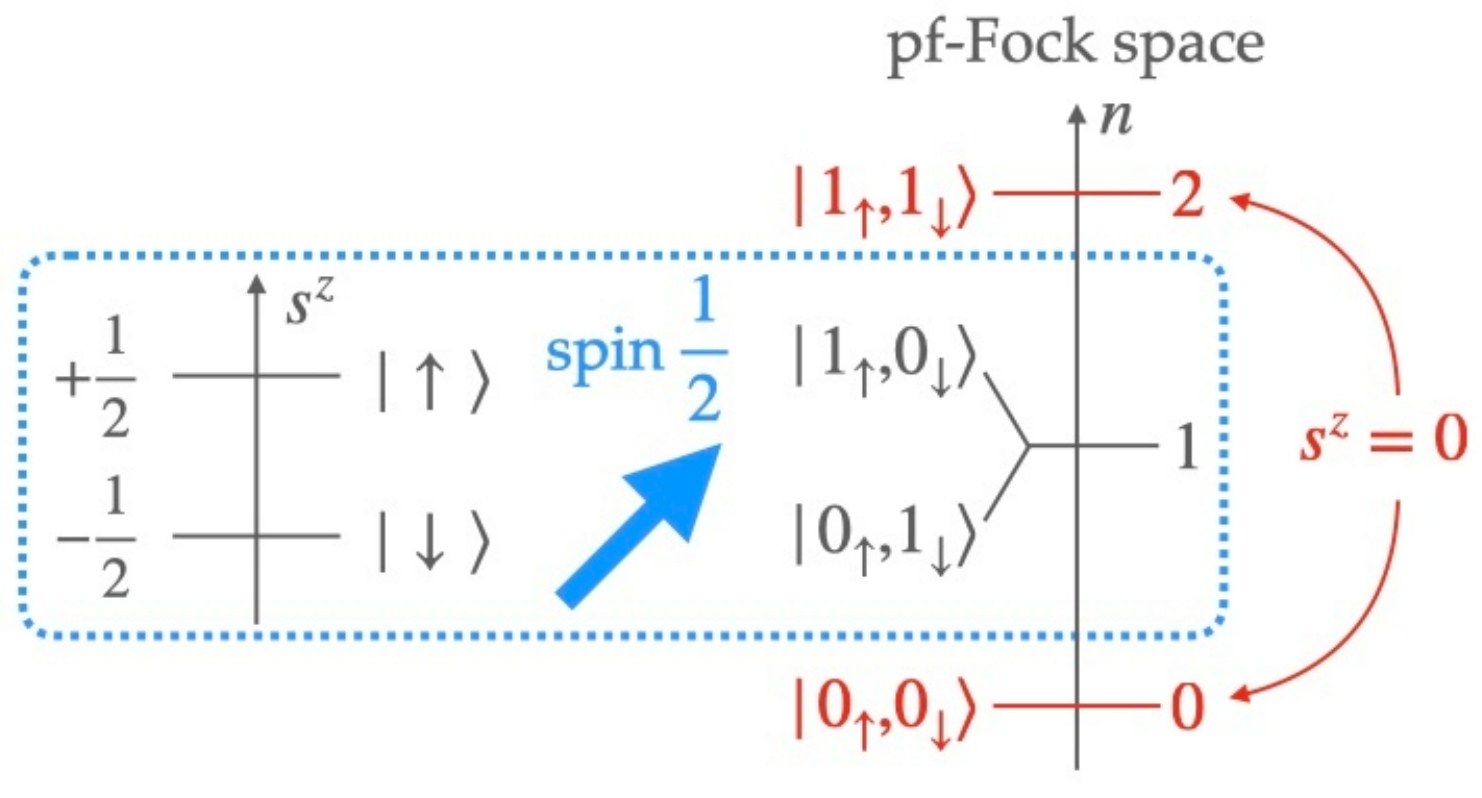}
     \caption{\textbf{Pseudofermion representation of spin-$\tfrac{1}{2}$ operators.} The mapping to a fermionic Fock space doubles the Hilbert space dimension and introduces two unphysical states with $s^z = 0$.}
     \label{fig:spin_to_fermion}
 \end{figure}

We introduce two species of auxiliary (complex) fermions, $\cc_\up$ and $\cc_\down$, to define the operator mapping 
\begin{equation}
  S^\alpha = \frac{1}{2} \sum\limits_{\mu^\noprime,\mu' = \uparrow,\downarrow} \cd_{\mu^\noprime} \sigma^\alpha_{\mu^\noprime \mu'} \cc_{\mu'}, \label{eq:spinfermions}
\end{equation}
where $\sigma^\alpha$ ($\alpha \in \{1,2,3\}$) are the Pauli matrices. As can be readily verified, this representation fulfills the commutation relations \eq{eq:Scommutator}. However, by introducing two different fermionic operators, the Hilbert space now consists of four states
\begin{align}
  \ket{0_\uparrow,0_\downarrow}& 	& \ket{1_\uparrow,1_\downarrow} &= \cd_\uparrow \cd_\downarrow\ket{0_\uparrow,0_\downarrow}\nonumber\\
  \ket{1_\uparrow,0_\downarrow} &= \cd_\uparrow\ket{0_\uparrow,0_\downarrow}
  \quad&\ket{0_\uparrow,1_\downarrow} &= \cd_\downarrow\ket{0_\uparrow,0_\downarrow},
   \label{eq:pfstates}
\end{align}
 where only the singly occupied ones in the second row correspond to the physical up/down spin states $\ket{\uparrow}$/$\ket{\downarrow}$ of the spin model, while the empty and doubly occupied ones do not have a physical counterpart. The structure of the spin-fermion mapping is pictorially shown in Fig.~\ref{fig:spin_to_fermion}.
This overcounting necessitates the introduction of the single occupation constraint in form of the operator identity
\begin{equation}
  \cd_{i\downarrow}\cc_{i\downarrow}+\cd_{i\uparrow}\cc_{i\uparrow}=1. \label{eq:numberconstraint}
\end{equation}
in addition to the mapping in \eq{eq:spinfermions}, to construct a faithful operator mapping. For a many-body system, \eq{eq:numberconstraint} has to hold individually per site $i$.

\subsection{Gauge symmetry}

\label{sec:pseudofermiongaugesymm}

The restriction to half-filling in pseudo-fermion space introduces an ambiguity in the pseudo-fermion description: The physical states can be described as being constructed by filling an unphysical fermionic vacuum, as done in \eq{eq:pfstates}, but equally well as filling holes into an unphysical fully occupied state. This leads to an $\SU(2)$ gauge structure, which allows to rotate freely between creation operators of one and annihilation operators of the other spin species, while simultaneously changing the fermionic vacuum.

To formalize this intuitive picture, we introduce the matrix operator~\cite{Affleck1988}
\begin{equation}
  \Psi = \begin{pmatrix}
                   \cc_{\uparrow} & \cd_{\downarrow}\\
                   \cc_{\downarrow} & -\cd_{\uparrow}\\
                  \end{pmatrix},\label{eq:matrixfermions}
 \end{equation}
which allows us to express the mapping in \eq{eq:spinfermions} as 

 \begin{equation}
  S^\alpha = -\frac{1}{4} \Tr\left[\sigma^\alpha \Psi^\nodag \Psi^\dagger \right]. \label{eq:spinmatrices}
 \end{equation}

 In this form, it becomes clear that the mapping is invariant under a right-multiplication with a $\operatorname{SU}(2)$ matrix according to
 \begin{equation}
  \Psi \to \Psi U^\dagger,\qquad U \in \SU(2),\label{eq:su2right}
 \end{equation}
 i.e., a $\SU(2)$ transformation intermixing creation and annihilation operators of up- and down spins. Consequently, the two possible vacua $\ket{0_\uparrow,0_\downarrow}$ and $\ket{1_\uparrow,1_\downarrow}$~\eq{eq:pfstates} get intermixed. {Given that the physical Hamiltonian \eqref{eq:mod-ham} is a function spin operators, one acquires a SU(2) {\it gauge} symmetry in the fermionic representation~\cite{Affleck1988,Dagotto-1988}. 
 
 As operator expectation values in second quantization are always taken with respect to a specific vacuum, only the $\U(1)$ and $\mathds{Z}_2$ subgroup of the full $\SU(2)$ gauge symmetry can be exploited, as they will lead to a pure, rather than mixed, vacuum state. The former amounts to multiplying the operators with a complex phase, leaving the vacuum invariant, while the latter represents a particle-hole transformation given by 
 \begin{equation}
 U_{\mathds{Z}_2} = \begin{pmatrix}
   0 & 1\\
   -1 & 0
 \end{pmatrix}.\label{eq:phmatrix}
\end{equation}

Under this transformation, $ \ket{0_\uparrow,0_\downarrow}$ and $ \ket{1_\uparrow,1_\downarrow}$ swap their roles. Both these subgroups of the full $\SU(2)$ symmetry, therefore, lead to well-defined expectation values, which can be brought into relation with each other.

For completeness, let us mention that the single occupation constraint \eq{eq:numberconstraint} can also be recast in terms of the matrix operator in \eq {eq:matrixfermions}, by realizing that half filling additionally implies the operator identities
\begin{equation}
  \cd_{i\uparrow}\cd_{i\downarrow} = \cc_{i\uparrow}\cc_{i\downarrow}=0, \label{eq:occupationconstraint}
\end{equation}
as in a half-filled state, we can neither create nor annihilate two fermions at the same time.

Together, \eqs{eq:numberconstraint}{eq:occupationconstraint} can be recast as a vector equation
\begin{equation}
  \mathfrak{G}^\alpha = \frac{1}{4} \Tr\left[\sigma^\alpha \Psi^\dagger_i \Psi^\nodag_i \right] = 0. \label{eq:constraintmatrices}
\end{equation}

Using the cyclic property of the trace, a $\operatorname{SU}(2)$ transformation according to \eq{eq:su2right} corresponds to transforming the Pauli matrix in \eq{eq:constraintmatrices} according to 
\begin{equation}
  \sigma^\alpha \to U^\dagger \sigma^\alpha U = \sum\limits_\beta R^\alpha_\beta \sigma^\beta,\label{eq:sigmatransform}
\end{equation}
where we have used the fact that any $\SU(2)$ transformation on a Pauli matrix will correspond to an $\SO(3)$ rotation $R$ of the Pauli matrix vector. This means, although superficially equivalent to the half-filling constraint, the additional operator identities in \eq{eq:occupationconstraint} will be generated under the action of the $\SU(2)$ gauge symmetry of the pseudo-fermion representation.
The special case of a particle-hole symmetry \eq{eq:phmatrix} does not mix the components of the constraint vector $\mathfrak{G}^\alpha$ defined in \eq{eq:constraintmatrices}, but only flips the sign of its $y$ component.

Lastly, using the relation between $\SO(3)$ and $\SU(2)$ backward, we can straightforwardly show that also any physical rotation in spin space corresponds to another SU(2) transformation of the corresponding fermionic operators, according to
\begin{align}
  \sum\limits_\beta R^\alpha_\beta S^\beta_i &= -\frac{1}{4} \Tr\left[\left(\sum\limits_\beta R^\alpha_\beta \sigma^\beta\right) \Psi^\nodag_i \Psi^\dagger_i \right] \nonumber\\
					&= -\frac{1}{4} \Tr\left[U^\dagger \sigma^\alpha U \Psi^\nodag_i \Psi^\dagger_i \right]\nonumber \\
					&= -\frac{1}{4} \Tr\left[\sigma^\alpha (U \Psi^\nodag_i) (U \Psi_i)^\dagger \right]. \label{eq:so3su2}
\end{align}
In contrast to the symmetry of the operator mapping \cref{eq:spinfermions} given by \cref{eq:su2right}, this left-multiplication with a $\operatorname{SU}(2)$ matrix in \cref{eq:so3su2} does not leave the spin operators invariant, but rather is a representation in matrix operator language of their transformation properties under physical rotations in spin space.

\subsection{Pseudo-fermion Hamiltonian}
Having defined the pseudo-fermion mapping, we are now able to rewrite the spin Hamiltonian, \eq{eq:mod-ham}, in fermionic language, leading to
\begin{equation}
  \PFSpace{\mathcal{H}} = \frac{1}{8}\sum_{i,j}\sum\limits_{\alpha,\beta}\sum\limits_{\mu,\nu,\rho,\sigma} J^{\alpha\beta}_{ij} \cd_{i\mu}\cc_{i\nu}\cd_{j\rho}\cc_{j\sigma} \sigma^\alpha_{\mu\nu}{\sigma}^\beta_{\rho\sigma}. \label{eq:heisenbergfermion}
\end{equation}
The absence of a kinetic term here is generic for pseudo-fermion Hamiltonians. As we will show in \Sec{sec:localU1}, the $\U(1)$ gauge-symmetry of the operator mapping in \eq{eq:spinfermions} does not allow for terms quadratic in operators which are not on-site. This renders spin systems inherently strongly interacting in the pseudo-fermionic picture, preventing any perturbative treatment of the problems around a Gaussian theory.

\subsection{Symmetries of pseudo-fermion Green's functions}
\label{sec:pseudofermionsymmetries}
The spin Hamiltonian in \eq{eq:heisenbergfermion} features, in addition to the $\SU(2)$ gauge symmetry of the pseudo-fermion mapping, a series of physical symmetries, such as hermiticity and time-reversal invariance. In this section, following the presentation in \mRef{Buessen2019b}, we will summarize these symmetries and their implications for the relevant one- and two-particle Green's functions\footnote{In literature, $n$-particle functions are also called $2n$-point functions. We choose the former name, referring to the number of pairs of creation and annihilation operators within an expectation value, signifying the number of particles involved, while the latter refers to the number of operators itself, or equivalently the number of external arguments to the function.}.

To fix notation, we define the one-particle Green's function~\cite{Negele2018}
\begin{equation}
\begin{split}
  G(x'_1|x_1) = -&\int \dd\tau_1' \dd \tau_1 \e^{\i(\tau'_1\omega'_1 - \tau^\noprime_1\omega_1)}\times\\ &\times \ev{\mathrm{T}_\tau \cc_{i^\noprime_1,\mu^\noprime_1}(\tau^\noprime_1) \cd_{i'_1,\mu'_1}(\tau'_1)}\label{eq:oneparticleG}
  \end{split}
\end{equation}
and its two-particle counterpart 
\begin{equation}
  \begin{split}
  G(x'_1,&x'_2|x_1,x_2) =\\ &\int \dd\tau_1' \dd\tau_2' \dd \tau^\noprime_1 \dd \tau^\noprime_2 \e^{\i(\tau'_1\omega'_1+\tau'_2\omega'_2 - \tau^\noprime_1\omega^\noprime_1- \tau^\noprime_2\omega^\noprime_2)}
  \times \\ &\times\ev{\mathrm{T}_\tau \cc_{i^\noprime_1,\mu^\noprime_1}(\tau^\noprime_1)\cc_{i^\noprime_2,\mu^\noprime_2}(\tau^\noprime_2)\cd_{i'_2,\mu'_2}(\tau'_2)\cd_{i'_1,\mu'_1}(\tau'_1)},\label{eq:twoparticleG}
  \end{split}
\end{equation}
where $\mathrm{T}_\tau$ is the imaginary time ordering operator and we use multi-indices $x_j = (i_j, \i\omega_j, \mu_j)$, containing the site index $i_j$, Matsubara frequency\footnote{For clarity we drop the discrete index of the Matsubara frequencies.} $\omega_j$ and spin index $\mu_j$. Primed indices are referred to as outgoing, whereas unprimed ones are called incoming indices.

\subsubsection{Time-translation invariance}
We start our discussion of symmetries with the invariance of the pseudo-fermion Hamiltonian under translations in imaginary time, manifested in the absence of any explicit time or Matsubara frequency dependence in  \eq{eq:heisenbergfermion}. This, in turn, implies Matsubara frequency conservation, leading to a parametrization of one- and two-particle Green's functions as
\begin{equation}
  G(x'_1|x_1) = G(x'_1|x_1) \delta_{\omega'_{1},\omega_{1}}
\end{equation}
and
\begin{equation}
  G(x'_1,x'_2|x_1,x_2) = G(x'_1,x'_2|x_1,x_2) \delta_{\omega'_{1}+\omega'_{2},\omega_{1}+\omega_{2}},
\end{equation}
reducing the frequency dependencies by one frequency each.

\subsubsection{Time-reversal invariance}
The second physical symmetry to be considered is imaginary time-reversal, which can be implemented at the pseudo-femion level by the antiunitary operator $\mathcal{T}$ acting according to
\begin{equation}
  \mathcal{T}\begin{pmatrix} \cd_{i\mu}\\\cc_{i\mu}\end{pmatrix}\mathcal{T}^{-1} = \begin{pmatrix} \e^{\i \pi \mu/2}\cd_{i\bar{\mu}}\\\e^{-\i \pi \mu/2}\cc_{i\bar{\mu}}\end{pmatrix},
\end{equation}
where the notation $\bar{\mu} = -\mu$ is a shorthand to indicate the flip of the spin index with $\mu =\pm 1$ representing spin up/down.

Using this relation for the two-particle Green's function in \eq{eq:oneparticleG}, we find for time-reversal invariant systems the relation
\begin{equation}
  G(x'_1|x_1) = \e^{\i \pi (\mu'_{1}-\mu^\noprime_{1})/2}G(\mathcal{T}x'_1|\mathcal{T}x_1)^*,
\end{equation}
where $\mathcal{T}x_j = (i_j, -\i \omega_j, \bar{\mu}_j)$ and the complex conjugation due to the antiunitarity of time-reversal is only meant to act on the Green's function itself, but not on its arguments.

The phase factor $\e^{\i \pi (\mu'_{1}-\mu^\noprime_{1})/2}$ can be simplified, using the fact that $\mu=\pm1$, to 
\begin{equation}
  \e^{\i \pi (\mu'_{1}-\mu^\noprime_{1})/2} = \mu'_{1} \mu^\noprime_{1},\label{eq:troneparticle}
\end{equation}
as can be easily verified by considering all possible combinations of the spin indices.

Similarly, the two-particle vertex obeys the relation
\begin{equation}
  G(x'_1,x'_2|x_1,x_2) = \mu'_{1}\mu'_{2}\mu^\noprime_{1}\mu^\noprime_{2}G(\mathcal{T}x'_1, \mathcal{T}x'_2|\mathcal{T}x_1,\mathcal{T}x_2)^*.\label{eq:trtwoparticle}
\end{equation}

Note that time-reversal symmetry of a magnetic system is broken when coupling to an external magnetic field $\bm{B}$ via a term $\bm{S} \cdot \bm{B}$ or more generally by interactions involving an odd number of spins. Since, however, time-reversal symmetry will turn out to be crucial for the performance of our calculations, we will refrain from adding such terms, although the formalism is, in principle, capable of handling these.

\subsubsection{Hermiticity}
The Hamiltonian and therefore the thermal density matrix of a pseudo-fermionic system being real directly dictates that the complex conjugate of operator expectation values can be related to the expectation values of their hermitian conjugate. 

From this observation, the relations
\begin{equation}
  G(x'_1|x_1) = G(x_1^*|{x'_1}^*)^*
\end{equation}
for the one-particle Green's function and
\begin{equation}
  G(x'_1,x'_2|x_1,x_2) = G({x_1}^*,{x_2}^*|{x'_1}^*,{x'_2}^*)^*
\end{equation}
for the two-particle Green's function follow directly. Here, we use the shorthand $x_j^* = (i_j, -\i \omega_j, \mu_j)$.
By themselves, these relations already give some insight into the analytical structure of the Green's functions, however, a combination with the time-reversal relations, \eqs{eq:troneparticle}{eq:trtwoparticle}, will lead to new relations for the whole Green's function, rather than their real and imaginary parts separately.

\subsubsection{Lattice symmetries}
\label{sec:latticesymm}
As a last physical symmetry, we want to consider the effect of lattice symmetries associated with the spin Hamiltonian in \eq{eq:heisenbergfermion}. Defining a suitable unitary operator $\mathcal{L}$, the action of the space group on the operators can be expressed as
\begin{equation}
  \mathcal{L}\begin{pmatrix} \cd_{i\mu}\\\cc_{i\mu}\end{pmatrix}\mathcal{L}^{-1} = \begin{pmatrix} \cd_{\operatorname{L}(i)\mu}\\\cc_{\operatorname{L}(i)\mu}\end{pmatrix},
\end{equation}
where the lattice point $i$ is mapped to $\operatorname{L}(i)$ by the transformation. Assuming invariance of the system under all lattice symmetries\footnote{Any symmetry breaking imposed by the specific coupling structure can be treated by defining a new lattice compatible with the coupling symmetries.}, we find for the Green's functions
\begin{equation}
  G(x'_1|x_1) = G(\operatorname{L}x'_1|\operatorname{L}x_1)
\end{equation}
and
\begin{equation}
  G(x'_1,x'_2|x_1,x_2) = G(\operatorname{L}x'_1,\operatorname{L}x'_2|\operatorname{L}x_1,\operatorname{L}x_2).
\end{equation}
Due to the translational subgroup contained in any space group, we can therefore always map at least one of the site indices on which the Green's function depends on, back into a reference unit cell. If, furthermore, all lattice points are symmetry equivalent (an Archimedian lattice), we can use the point group of the lattice to select a single reference point in this unit cell\footnote{In case of $n$ symmetry inequivalent points per unit cell (for a non-Archimedian lattice such as the square-kagome lattice), we have to choose $n$ such reference points.}.

\subsubsection{Crossing symmetries}
Before turning towards the gauge symmetries of the pseudo-fermion mapping, let us briefly mention that, due to the fermionic anticommutation relations, Green's functions have to change sign under pairwise exchange of two incoming or outgoing indices. For the two-particle Green's function, this implies
\begin{equation}
  \begin{split}
  G(x'_1,x'_2|x_1,x_2) &= -G(x'_2,x'_1|x_1,x_2) \\&= -G(x'_1,x'_2|x_2,x_1) \\&= \phantom{-}G(x'_2,x'_1|x_2,x_1)\,.\label{eq:crossing}
  \end{split}
\end{equation}
These relations are commonly referred to as crossing symmetries of the Green's function, as in a diagrammatic language it corresponds to crossing the incoming or outgoing legs of a given diagram.

\subsubsection{\texorpdfstring{Local $\U(1)$ gauge symmetry}{Local U(1) gauge symmetry}}
\label{sec:localU1}
Having exhausted all symmetries of the Green's functions which hold for general fermionic systems, we now want to consider the additional constraints the pseudo-fermion mapping in \eq{eq:spinfermions} imposes on these objects. As already discussed in \Sec{sec:pseudofermiongaugesymm}, the single occupation constraint accompanying the mapping of spin operators to fermions introduces a local $\SU(2)$ gauge symmetry. Since however, the vacuum is not invariant under this group, only two subgroups of the full symmetry can be exploited for expectation values defining Green's functions. The first one, we want to discuss, is the local $\U(1)$ symmetry.

The action of this group amounts to rotating the complex phase of an operator at site $i$ by an arbitrary angle $\phi_i$, i.e., the operators transform as
\begin{equation}
  \mathcal{U}_\phi\begin{pmatrix} \cd_{i\mu}\\\cc_{i\mu}\end{pmatrix}\mathcal{U}_\phi^{-1} = \begin{pmatrix} \e^{\i \phi_i}\cd_{i\mu}\\\e^{-\i \phi_i}\cc_{i\mu}\end{pmatrix}.
\end{equation}

To allow for non-vanishing Green's functions, these phases have to cancel, which implies that pairs of incoming and outgoing parameters have to reside on the same lattice point. This leads to a purely local one-particle Green's function
\begin{equation}
  G(x'_1|x_1) = G(x'_1|x_1) \delta_{i'_{1}i_{1}}, 
\end{equation}
while the two-particle Green's function features two possible combinations of incoming and outgoing sites
\begin{equation}
  \begin{split}
    G(x'_1,x'_2|x_1,x_2) =  &G(x'_1,x'_2|x_1,x_2)\delta_{i'_{1}i_{1}}\delta_{i'_{2}i_{2}}\\ -  &G(x'_2,x'_1|x_1,x_2)\delta_{i'_{1}i_{2}}\delta_{i'_{2}i_{1}}.
  \end{split}
\end{equation}
In the second term, we have already explicitly incorporated the crossing-symmetry, \eq{eq:crossing}, leading to a direct and crossed term in terms of real space indices\footnote{The real-space structure we find here is completely analogous to the one in spin space for $\SU(2)$ symmetric systems, as used in itinerant fermion FRG~\cite{Platt2013}.}.
Similar to a global $\U(1)$ symmetry implying conservation of total particle number, this gauged implementation leads to a particle number conservation per site, in accordance with the local single occupation constraint of the pseudo-fermions.

In analogy, multi-particle Green's functions will become multi-local, which has profound implications for the natural basis we will treat pseudo-fermions in: While itinerant particles tend to delocalize, electronic systems are usually best treated in a momentum-space picture, whereas for pseudo-fermions real space is more appropriate.

\subsubsection{Local particle-hole conjugation}
The second subgroup of the $\SU(2)$ gauge symmetry, we want to discuss, is the $\mathds{Z}_2$ subgroup, which amounts to a local particle-hole conjugation\footnote{In PFFRG literature, this transformation is usually called particle-hole symmetry, which would imply an antiunitary implementation. As the local $\mathds{Z}_2$, however, is unitary, we prefer the term conjugation.}
\begin{equation}
  \mathcal{Z}_i\begin{pmatrix} \cd_{i\mu}\\\cc_{i\mu}\end{pmatrix}\mathcal{Z}_i^{-1} = \begin{pmatrix} \mu\cc_{i\bar{\mu}}\\\mu\cd_{i\bar{\mu}}\end{pmatrix},  
\end{equation}
which also swaps spin sectors. Applying this transformation to the locally parameterized Green's function from the previous section, we find
\begin{equation}
  G(x'_1|x_1) \delta_{i'_{1}i_{1}} = -\mu'_{1}\mu_1 G(\mathcal{T}x_1|\mathcal{T}x'_1)\delta_{i'_{1}i^\noprime_{1}}
\end{equation}
for the one-particle case. The negative sign is due to an anticommutation within the expectation value defining the Green's function, while the inversion of frequency is due to the swapping of creation and annihilation operators, which flips the energy spectrum.

For the two-particle case, we analogously find, by applying the local conjugation to the two independent sites separately
\begin{align}
  \begin{split}
    G(x'_1,x'_2|x_1,x_2)&\delta_{i'_{1}i^\noprime_{1}}\delta_{i'_{2}i_{2}}\\ &= -\mu'_{1}\mu_{1} G(\mathcal{T}x_1,x'_2|\mathcal{T}x'_1,x_2)\delta_{i'_{1}i^\noprime_{1}}\delta_{i'_{2}i^\noprime_{2}}\label{eq:ph1}
  \end{split}\\ &=-\mu'_{2}\mu_{2} G(x'_1,\mathcal{T}x_2|x_1,\mathcal{T}x'_2)\delta_{i'_{1}i^\noprime_{1}}\delta_{i'_{2}i^\noprime_{2}}.\label{eq:ph2}
\end{align}

Note that even in the case of $i_1 = i_2$, the two relations \eqs{eq:ph1}{eq:ph2} hold separately, as the symmetry is connected to the decomposition of the spin operators involved, as labelled by $1$ and $2$, rather than the sites the fermions reside on.

\subsubsection{Summary of the symmetries}

\begin{table}
  \setlength{\tabcolsep}{4pt}
  \setlength\extrarowheight{2pt}
  \centering
  \begin{tabular}{Ar}
  \toprule
    G(x'_1|x_1) &= G(x'_1|x_1) \delta_{i'_{1}i^\noprime_{1}} &  ($\U(1)$)\\
    G(x'_1|x_1)  \delta_{i'_{1}i^\noprime_{1}} &= G(\operatorname{L}x'_1|\operatorname{L}x_1)  \delta_{i'_{1}i^\noprime_{1}} &  (L)\\
    G(x'_1|x_1) \delta_{i'_{1}i^\noprime_{1}} &= G(x'_1|x_1) \delta_{i'_{1}i^\noprime_{1}} \delta_{\omega'_{1},\omega^\noprime_{1}} & (TT)\\
    G(x'_1|x_1) \delta_{i'_{1}i^\noprime_{1}} &= -\mu'_{1}\mu_{1} G(\mathcal{T}x_1|\mathcal{T}x'_1)\delta_{i'_{1}i^\noprime_{1}} &(PH)\\
    G(x'_1|x_1) \delta_{i'_{1}i^\noprime_{1}} &= \mu'_{1} \mu_{1} G(\mathcal{T}x'_1|\mathcal{T}x_1)^* \delta_{i_{1'}i^\noprime_{1}} & (TR)\\
    G(x'_1|x_1) \delta_{i'_{1}i^\noprime_{1}} &= G(x_1^*|{x'_1}^*)^* \delta_{i'_{1}i^\noprime_{1}} & (H)\\
  \bottomrule
  \end{tabular}
  \caption{Symmetry relations of the one-particle correlation function for pseudo-fermion Hamiltonians. The combined index $x_j=(i_j, \i\omega_j,\mu_j)$ is used as shorthand for site index $i_j$, Matsubara frequency $\i\omega_j$ and spin index $\mu_j$ for incoming (unprimed) and outgoing (primed) parameters. The labels are shorthands for the underlying symmetry of the relations: TR denotes time-reversal, TT time translation, L lattice symmetries, H hermitian conjugation, X crossing symmetry in both incoming and outgoing particles and PH is a particle-hole transformation. \label{tab:onepartsymm}}
\end{table}

\begin{table*}
  \setlength{\tabcolsep}{4pt}
  \setlength\extrarowheight{2pt}
  \centering
  \begin{tabular}{Ar}
  \toprule
  G(x'_1,x'_2|x_1,x_2) &=  G(x'_1,x'_2|x_1,x_2)\delta_{i'_{1}i^\noprime_{1}}\delta_{i'_{2}i^\noprime_{2}}\\ -  &G(x'_2,x'_1|x_1,x_2)\delta_{i'_{1}i^\noprime_{2}}\delta_{i'_{2}i^\noprime_{1}} &  ($\U(1)$)\\
  G(x'_1,x'_2|x_1,x_2)\delta_{i'_{1}i^\noprime_{1}}\delta_{i'_{2}i^\noprime_{2}} &= G(\operatorname{L}x'_1,\operatorname{L}x'_2|\operatorname{L}x_1,\operatorname{L}x_2)\delta_{i'_{1}i^\noprime_{1}}\delta_{i'_{2}i^\noprime_{2}} &  (L)\\
  G(x'_1,x'_2|x_1,x_2)\delta_{i'_{1}i^\noprime_{1}}\delta_{i'_{2}i^\noprime_{2}} &= G(x'_1,x'_2|x_1,x_2) \delta_{i'_{1}i^\noprime_{1}}\delta_{i'_{2}i^\noprime_{2}}  \delta_{\omega'_{1}+\omega'_{2},\omega_{1}+\omega_{2}} & (TT)\\

  G(1',2';1,2)\delta_{i'_{1}i^\noprime_{1}}\delta_{i'_{2}i^\noprime_{2}} &= -\mu'_{1}\mu_{1} G(\mathcal{T}x_1,x'_2|\mathcal{T}x'_1,x_2)\delta_{i'_{1}i_{1}}\delta_{i'_{2}i^\noprime_{2}} &(PH1)\\
  G(x'_1,x'_2|x_1,x_2)\delta_{i'_{1}i^\noprime_{1}}\delta_{i'_{2}i_{2}} &=-\mu'_{2}\mu_{2} G(x'_1,\mathcal{T}x_2|x_1,\mathcal{T}x'_2)\delta_{i'_{1}i^\noprime_{1}}\delta_{i'_{2}i^\noprime_{2}} &(PH2)\\
  G(x'_1,x'_2|x_1,x_2)\delta_{i'_{1}i^\noprime_{1}}\delta_{i'_{2}i^\noprime_{2}} &= \mu'_{1}\mu'_{2}\mu_{1}\mu_{2}G({x_1}^*,{x_2}^*|{x'_1}^*,{x'_2}^*)^*\delta_{i'_{1}i^\noprime_{1}}\delta_{i'_{2}i^\noprime_{2}} & (TR)\\
  G(x'_1,x'_2|x_1,x_2)\delta_{i'_{1}i^\noprime_{1}}\delta_{i'_{2}i^\noprime_{2}} &= G({x_1}^*,{x_2}^*|{x'_1}^*,{x'_2}^*)^*\delta_{i'_{1}i^\noprime_{1}}\delta_{i'_{2}i^\noprime_{2}} & (H)\\
  G(x'_1,x'_2|x_1,x_2)\delta_{i'_{1}i^\noprime_{1}}\delta_{i'_{2}i^\noprime_{2}} &= G(x'_2,x'_1|x_2,x_1)\delta_{i'_{1}i^\noprime_{1}}\delta_{i'_{2}i^\noprime_{2}} & (X)\\
  \bottomrule
  \end{tabular}
  \caption{Symmetry relations of the two-particle correlation function for pseudo-fermion Hamiltonians. The combined index $x_j=(i_j, \i\omega_j,\mu_j)$ is used as shorthand for site index $i_j$, Matsubara frequency $\i\omega_j$ and spin index $\mu_j$ for incoming (unprimed) and outgoing (primed) parameters. The labels are shorthands for the underlying symmetry of the relations, for details see main text.\label{tab:twopartsymm}}
\end{table*}

For reference, we summarize all symmetry relations of the one- and two-particle Green's function in \tab{tab:onepartsymm} and \tab{tab:twopartsymm}, respectively.

We can divide the symmetries into two groups. The first one reduces the dependence of the Green's functions on the external degrees of freedom: The local $\U(1)$ symmetry renders the one(two)-particle function (bi-)local, greatly reducing their spatial dependence. 
Additionally, lattice symmetries (L) allow to fix one site as a reference point within the unit cell. In frequency-space, time-translational invariance (TT) has a similar effect, reducing the number of frequency arguments by one.

The second group of symmetries establishes relations within the remaining structure of the Green's functions: This is the case for the remaining part of the lattice symmetries and the local particle-hole conjugation, which induces one symmetry relation (PH) in the one-particle case and two, (PH1) and (PH2), in the two-particle one, one for each site index. Time-reversal (TR) and Hermitian (H) symmetry relate the real and imaginary parts of the Green's function. For the case of the two-particle Green's function, we also have the combined crossing symmetry in both incoming and outgoing arguments (X). As the bilocality constraint following from ($\U(1)$), already decomposes the vertex into two components, which are related by an individual crossing symmetry, as shown in \eq{eq:crossing}, crossing symmetry in incoming or outgoing particles seperately is already accounted for.

\subsection{Gauge invariance of Lagrangian}
\label{sec:gaugefield}

Having discussed the symmetries of both the spin Hamiltonian and the Green's functions, we still have to see, how the gauge symmetry affects the Lagrangian, which we need for a field-theoretic treatment of pseudo-fermion systems.
To this end, we bring the spin Hamiltonian \cref{eq:heisenbergfermion} in a more convenient form for our purpose 
\begin{equation}
  \PFSpace{\mathcal{H}} = -\frac{1}{32} \sum\limits_{i,j,\alpha,\beta} J_{ij}^{\alpha\beta} \Tr\left[\sigma^\alpha \Psi_i^\nodag \Psi_i^\dagger \right] \Tr\left[\sigma^\beta \Psi_j^\nodag \Psi_j^\dagger \right]. \label{eq:heisenbergmatrixfermion}
\end{equation}

Here, it is manifest that the Heisenberg model with $J^{\alpha\beta}\propto\delta_{\alpha\beta}$ is invariant both under a local SU(2) gauge transformation according to \eq{eq:su2right}, 
\begin{equation}
  \Psi_i \to \Psi_i U_i^\dagger,\qquad U_i \in \SU(2),\label{eq:su2rightlocal}
\end{equation}
where the matrix $U_i$ can be site-dependent, as well as a global rotation in spin space given by \eq{eq:so3su2}. For field-theoretical treatments, we will need the Lagrangian of this system~\cite{Affleck1988}
 \begin{equation}
   L = \i \sum\limits_{i,\mu} \cd_{i\mu} \pdv{t} \cc_{i\mu} - H,
 \end{equation} 
which, by means of integration by parts, is up to a constant term, equivalent to the manifestly gauge invariant form
\begin{equation}
  L = \frac{\i}{2} \sum\limits_{i} \Tr\left[\Psi_{i} \pdv{t} \Psi^\dagger_{i}\right] - H.\label{eq:matrixlagrangetimedep}
\end{equation}

To incorporate the single-occupation per site constraint in this formulation, we add three Lagrange multipliers $A^\alpha$ enforcing the three components of \eq{eq:constraintmatrices}, by adding a term $\Tr[ \Psi_{i} (\bm{A}\cdot\bm{\sigma}) \Psi^\dagger_{i}]$ to the Lagrangian. This term, however, is nothing else than a coupling of the matrix valued field $\Psi$ to the temporal component of the $\SU(2)$ gauge field $\bm{A}\cdot\bm{\sigma}$, given we allow for fluctuations of the Lagrange multipliers, promoting them to fields. 

This approach even allows for time-dependent gauge transformations, which would not leave the quadratic term in \eq{eq:matrixlagrangetimedep} invariant, due to the time derivative terms of the transformation not being cancelled. Demanding a suitable transformation of the gauge field
\begin{equation}
  \bm{A}\cdot\bm{\sigma} \to U^\dagger \left(\bm{A}\cdot\bm{\sigma}- \i \pdv{t}\right) U,
\end{equation}
restores this invariance, thereby promoting the local $\SU(2)$ gauge invariance to a time-dependent one. The Lagrangian fully invariant under the local and time-dependent gauge symmetry of the $\SU(2)$ symmetry of the pseudo-fermions therefore reads as
\begin{equation}
  L = \frac{1}{2} \sum\limits_{i} \Tr\left[\Psi_{i} (\i \pdv{t} + \bm{A}\cdot\bm{\sigma})  \Psi^\dagger_{i}\right] - H.\label{eq:matrixlagrangetimedepsu2invar}
\end{equation}

\section{Functional Renormalization Group} 
\label{sec:functionalrenormalization}
The reformulation of the general spin Hamiltonian in terms of auxiliary spinon operators as presented in the previous section opens up the possibility of employing established many-body techniques developed for interacting fermions. In contrast to itinerant systems, however, the fermionized spin model lacks a quadratic term as a result of the aforementioned $\SU(2)$ gauge invariance, such that perturbative approaches based on a small parameter $t/J$, where $t$ characterizes the kinetic energy scale and $J$ the spin interactions, are inapplicable.

The functional renormalization group~\cite{Wegner1973,Polchinski1984,Wetterich1993} first emerged in high-energy physics, where it has been successfully applied to, e.g., electroweak physics~\cite{Reuter1993}, quantum chromodynamics~\cite{Schaefer2008,Li2012,Schaefer2012,Yokota2016} and models of quantum gravity~\cite{Eichhorn2013,Castro2021}. The general idea behind FRG is the successive inclusion of low-energy fluctuations during a renormalization group flow, which evolves the many-body interactions of a microscopic theory in terms of an infrared cutoff. In this sense, it naturally extends concepts of Wilsonian RG~\cite{Wilson1971,Wilson1974,Shankar_scaling}, namely, running couplings and an effective action, to coupling functions (\textit{vertices}) and their generating functionals. Nowadays, FRG calculations are also widely used in condensed matter research, ranging from applications to zero dimensional systems such as quantum dots~\cite{Hedden2004,Karrasch2006,Karrasch2008,Karrasch2011}, over studies of Luttinger-liquid physics in 1D~\cite{Andergassen2004,Andergassen2006,Metzner2005}, to extensive characterizations of Fermi liquid instabilities in variants of the Hubbard model \cite{Halboth2000,Zanchi2000,Honerkamp2001,Honerkamp2001a,Honerkamp2001b,Honerkamp2008,Uebelacker2012,Eberlein2014,Tagliavini2019,Hille_quantitative,Hille2020a,Vilardi2020,Metzner2012}.

In this section, the functional renormalization group approach to correlated fermionic systems is introduced on a general level, closely following the derivations presented in Refs.~\cite{Reuther2011,Buessen2019b}. Given the vast amount of literature that exists on the matter~\cite{Hedden2004,Salmhofer2004,Metzner2005,Kopietz2010,Metzner2012,Platt2013,Dupuis2021,Kopietz2010}, we aim at keeping the discussion concise and, if feasible, encourage the reader to follow the given references for further detail beyond the scope of this review. For a practical implementation of FRG for pseudo-fermions, see \cref{sec:PFFRG}.

\subsection{Generating functionals}
\label{sec:generating_functionals}

We consider a fermionic action of the form 
\begin{align}
    S[\psibar, \psi] = -(\psibar, G^{-1}_{0} \psi) + S_{\text{int}}[\psibar, \psi] \,,
    \label{eq:action}
\end{align}
where $(\psibar, G^{-1}_{0} \psi) \equiv \sum_{x'_1, x_1} \psibar_{x'_1} [G^{-1}_{0}]_{x'_1 x_1} \psi_{x_1}$. Here, $\psibar, \psi$ denote fermionic Grassmann fields and summations over their multi-indices $x_i$, which could comprise e.g. spin projections or Matsubara frequencies, are to be understood as sums (integrals) over their discrete (continuous) components. Furthermore, we assume a quartic interaction
\begin{align}
    S_{\text{int}}[\psibar, \psi] =\sum_{x'_1, x'_2, x_1, x_2} V_{x'_1 x'_2 | x_1 x_2} \psibar_{x'_1} \psibar_{x'_2} \psi_{x_2} \psi_{x_1} \,,
    \label{eq:interaction}
\end{align}
in which the interaction tensor $V$ is antisymmetric with respect to permutations $(x'_1 \leftrightarrow x'_2)$ and $(x_1 \leftrightarrow x_2)$, as indicated by a vertical line separating the respective index sets. 

For a given action, the central goal is to compute the corresponding $n$-particle Green's functions, i.e., expectation values of the form
\begin{align}
    \langle \psibar_{x'_1} ... \psibar_{x'_n} \psi_{x_n} ... \psi_{x_1} \rangle \equiv G_n(x'_1, ..., x'_n| x_1, ..., x_n) \,,
\end{align}
where the (thermal) average $\langle \ . \ \rangle$ is defined with respect to the partition function
\begin{align}
    Z = \int D[\psibar, \psi] \ \e^{-S[\psibar, \psi]} \,.
\end{align}
Defining the functional
\begin{align}
    \W = \frac{1}{Z_0} \int D[\psibar, \psi] \e^{-S[\psibar, \psi] - (\psibar, \eta) - (\etabar, \psi)} \,,
\end{align}
where $Z_0$ is the Gaussian partition function, the \textit{disconnected} Green's functions can be obtained by considering functional derivatives of $\mathcal{W}$ with respect to the fermionic sources $\etabar, \eta$ and setting them to zero afterwards:
\begin{align}
   G_n(x'_1, &..., x'_n| x_1, ..., x_n) = \notag \\ 
   &\frac{\delta^{n}}{\delta \etabar_{x_1} ... \delta \etabar_{x_n}} \frac{\delta^{n}}{\delta \eta_{x'_n} ... \delta \eta_{x'_1}} \W \bigg{|}_{\etabar = \eta = 0} \,.
\end{align}
For practical purposes it is more convenient to work with \textit{fully-connected} correlators\footnote{Statistically speaking, this corresponds to considering the cumulants of the distribution $\tfrac{1}{Z} \e^{-S[\psibar, \psi]}$ instead of its moments.}, as disconnected diagrams contain redundant information from Greens's functions involving less particles, effectively mixing information about different particle number sectors. These are generated by the so-called Schwinger functional 
\begin{align}
    \Wc = \text{ln}(\W) \,.
\end{align}
Although this new functional reduces the superfluous information contained in $\mathcal{W}$ by excluding fully-disconnected contributions, there is still some redundancy left in this description: some terms can be separated into two mutually disconnected parts by removing a single propagator $G \equiv G_1 = G^c_1$\footnote{Here, $G^c_1$ corresponds to the second functional derivative of $\mathcal{W}_c$ with vanishing sources.}. To obtain a complete description of the physical system, it therefore suffices to compute precisely those \textit{one-particle irreducible (1PI)} correlation functions or \textit{vertices} from which external legs have been amputated~\cite{Negele2018}. Their respective generator $\Gamma$ is given by the functional Legendre transform of $\mathcal{W}_c$, i.e.,
\begin{align}
    \eff = -\Wc - (\phibar, \eta) - (\etabar, \phi) + (\phibar, G^{-1}_0 \phi) \,,
    \label{eq:effective_action}
\end{align}
where $\phi = -\frac{\delta \Wc}{\delta \etabar}$ and $\phibar = \frac{\delta \Wc}{\delta \eta}$ are the conjugate sources. The one-particle vertex $\Gamma_1$, for example, corresponds to the fermionic self-energy $\Sigma$ up to a minus sign, i.e., $\Gamma_1 = -\Sigma$. The 1PI vertices thus resemble the \textit{effective} $n$-body interactions of the system, and their generating functional $\Gamma$ is therefore commonly referred to as the \textit{effective action} \cite{Negele2018,Kopietz2010,Metzner2012}. It turns out, that of these three functionals only the 1PI formulation allows for well-defined initial conditions for the renormalization group equations we will derive in the following~\cite{Kopietz2010}.

\subsection{Exact flow equations}
\label{subsec:exact_flow_eq}
In order to set up the functional renormalization group approach, a proper RG transformation needs to be defined. To this end, an infrared cutoff $\Lambda$ is introduced into the bare propagator $G_0$ such that it vanishes in the ultraviolet limit, $G^{\Lambda \to \infty}_0 = 0$, and again coincides with $G_0$ when approaching the infrared limit, $G^{\Lambda \to 0}_0 = G_0$. This is usually achieved by virtue of a multiplicative regulator function $R(\Lambda)$ with $G^{\Lambda}_0 = R(\Lambda) G_0$, such that $R(0) = 1$ and $R(\Lambda\to\infty) = 0$. 
This procedure renders the original action $S$ and likewise the generating functionals $\mathcal{W}$, $\mathcal{W}_c$ and, most importantly, $\Gamma$, cutoff dependent. Considering its derivative with respect to $\Lambda$, the evolution of the effective action from $\Lambda \to \infty$ to $\Lambda \to 0$ can thus be described by an ordinary differential equation (ODE) which reads 
\begin{align}
    \diff \efflam = &-\diff \mathcal{W}^{\Lambda}_c[\etabar^{\Lambda}, \eta^{\Lambda}] - \left(\phibar, \diff \eta^{\Lambda} \right) \notag \\ 
    &- \left(\diff \etabar^{\Lambda}, \phi \right) + \left(\phibar, Q^{\Lambda} \phi \right) \,,
    \label{eq:diff_eff}
\end{align}
with $Q^{\Lambda} \equiv \diff (G^{\Lambda}_0)^{-1}$. Note that a $\Lambda$-dependence needs to be added to the $\etabar, \eta$ source fields to make up for the change of variables in the Legendre transformation. The cutoff derivative of the generator $\mathcal{W}_c$ hereby computes to 
\begin{align}
    \diff \Wclam &= -\mathrm{Tr}\left(Q^{\Lambda} G^{\Lambda}_0 \right) + \mathrm{Tr}\left(Q^{\Lambda} \frac{\delta^{2} \Wclam}{\delta \etabar \delta \eta}\right) \notag \\ 
    &- \left(\frac{\delta \Wclam}{\delta \eta}, Q^{\Lambda} \frac{\delta \Wclam}{\delta \etabar} \right) \,,
    \label{eq:diff_Wc}
\end{align}
which, due to Eq.~\eqref{eq:effective_action}, motivates\footnote{This is because the derivative $\tfrac{\delta^{2} \Wclam}{\delta \etabar \delta \eta}$, which appears in $\diff \Wclam$, can also be expressed in terms of second order field derivatives of $\Gamma$.} the definition of a matrix $M^{\Lambda}$ capturing the second functional derivatives of $\Gamma$ with respect to the conjugate source fields, i.e.
\begin{gather}
    M^{\Lambda} = \notag \\ \left[\mathbb{1} - 
    \begin{pmatrix}
          -G^{\Lambda} & 0  \\
                    0 & \left(G^{\Lambda} \right)^{T}
    \end{pmatrix}
    \begin{pmatrix}
          U^{\Lambda} & \frac{\delta^{2} \efflam}{\delta \phibar \delta \phibar}  \\
          \frac{\delta^{2} \efflam}{\delta \phi \delta \phi} & -(U^{\Lambda})^T
    \end{pmatrix}
    \right]^{-1} \,,
\end{gather}
with $U^{\Lambda} = \frac{\delta^{2} \efflam}{\delta \phibar \delta \phi} - \Gamma^{\Lambda}_1$. Using this matrix, Eq.~\eqref{eq:diff_eff} can be written in a more compact form 
\begin{align}
    \diff \efflam = \mathrm{Tr}\left(Q^{\Lambda} G^{\Lambda}_0 \right) - \mathrm{Tr}\left(G^{\Lambda} Q^{\Lambda} M^{\Lambda}_{11} \right) \,,
    \label{eq:diff_eff_compact}
\end{align}
where $M^{\Lambda}_{11}$ denotes the upper left element of $M^{\Lambda}$. In practice, it is more convenient to rephrase this functional equation as a hierarchy of ODEs for the vertices, which represent ordinary functions. To this end, one Taylor-expands the effective action on both sides of Eq.~\eqref{eq:diff_eff_compact} as 
\begin{align}
    &\efflam = \sum_{n = 0}^{\infty} \frac{(-1)^n}{(n!)^2} \sum_{x'_1, ..., x'_n} \sum_{x_1, ..., x_n} \notag \\ &\Gamma^{\Lambda}_n(x'_1, ..., x'_n|x_1, ..., x_n) \phibar_{x'_1} ... \phibar_{x'_n} \phi_{x_n} ... \phi_{x_1} \,,
\end{align}
and carries out the matrix valued geometric series in $M^{\Lambda}$ explicitly. By comparing the coefficients for a given power of the fields on the left and right hand side of the so-expanded Eq.~\eqref{eq:diff_eff_compact}, we can finally find the flow equations for the $n$-particle vertices. Henceforth, we limit the discussion to the flow of the self-energy $\Sigma$ and two-particle vertex, which we simply refer to as \textit{the vertex} from now on and, for the sake of brevity, denote it by $\Gamma$ instead of $\Gamma_2$. For the flow of $\Sigma$ one finds
\begin{align}
    \diff \Sigma^{\Lambda}(x'_1|x_1) &= -\sum_{x'_2, x_2} \G{x'_1}{x'_2}{x_1}{x_2} S^{\Lambda}(x_2 | x'_2)  \notag \\ 
    &\equiv - [\Gamma^{\Lambda} \bullet S^{\Lambda}](x'_1|x_1) \,,
    \label{eq:self_flow_general}
\end{align}
where we defined the \textit{single-scale} propagator\footnote{$\diffsig G^{\Lambda}$ is a shorthand notation for $\diff G^{\Lambda}|_{\Sigma^{\Lambda} = \text{const.}}.$}

\begin{align}
    S^{\Lambda} \equiv G^{\Lambda} Q^{\Lambda} G^{\Lambda} = -\diffsig G^{\Lambda} \,,\label{eq:singlescaledef}
\end{align}
as well as the \textit{tadpole contraction} $\bullet$, which connects an incoming and an outgoing line at an $n$-particle vertex with a fermionic propagator. For a compact representation of the vertex flow, we resort to the notation utilized in Refs.~\cite{parquet_conserving,Gievers_2022} that is we define the propagator bubbles 
\begin{align}
   \dot{\Pi}_s(x'_1, x'_2 | x_1, x_2) &= -\tfrac{1}{2} \diffsig \left[ G^{\Lambda}(x'_1 | x_1) G^{\Lambda}(x'_2 | x_2) \right] \notag \\ 
   \dot{\Pi}_t(x'_1, x'_2 | x_1, x_2) &= +\diffsig \left[ G^{\Lambda}(x'_1 | x_2) G^{\Lambda}(x'_2 | x_1) \right] \notag \\ 
   \dot{\Pi}_u(x'_1, x'_2 | x_1, x_2) &= -\diffsig \left[ G^{\Lambda}(x'_2 | x_2) G^{\Lambda}(x'_1 | x_1) \right] \,,
    \label{eq:bubbles}
\end{align} 
and two-particle contractions
\begin{align}
   [\Gamma^{\Lambda} \circ_s& \tilde{\Gamma}^{\Lambda}](x'_1, x'_2 | x_1, x_2) = \notag \\ 
   &\sum_{x_3, x_4} \G{x_3}{x_4}{x_1}{x_2} \Gt{x'_1}{x'_2}{x_3}{x_4} \notag \\ 
   [\Gamma^{\Lambda} \circ_t& \tilde{\Gamma}^{\Lambda}](x'_1, x'_2 | x_1, x_2) = \notag \\ 
   &\sum_{x_3, x_4} \G{x'_1}{x_4}{x_1}{x_3} \Gt{x_3}{x'_2}{x_4}{x_2} \notag \\ 
   [\Gamma^{\Lambda} \circ_u& \tilde{\Gamma}^{\Lambda}](x'_1, x'_2 | x_1, x_2) = \notag \\ 
   &\sum_{x_3, x_4} \G{x_4}{x'_2}{x_1}{x_3} \Gt{x'_1}{x_3}{x_4}{x_2} \,,
   \label{eq:vertex_contractions}
\end{align} 
such that
\begin{align}
    \diff \G{x'_1}{x'_2}{&x_1}{x_2} = [\Gamma^{\Lambda}_3 \bullet S^{\Lambda}](x'_1, x'_2 | x_1, x_2) \notag \\ 
    &+ [\Gamma^{\Lambda} \circ_s \dot{\Pi}_s \circ_s \Gamma^{\Lambda}](x'_1, x'_2 | x_1, x_2) \notag \\ 
    &+ [\Gamma^{\Lambda} \circ_t \dot{\Pi}_t \circ_t \Gamma^{\Lambda}](x'_1, x'_2 | x_1, x_2) \notag \\ 
    &+ [\Gamma^{\Lambda} \circ_u \dot{\Pi}_u \circ_u \Gamma^{\Lambda}](x'_1, x'_2 | x_1, x_2) \,.
    \label{eq:vertex_flow_general}
\end{align} 
So far, it may not be apparent to the reader why precisely these definitions of bubble functions and two-particle contractions are useful. For now, we will simply regard them as one specific way of grouping the diagrams on the right hand side of the vertex flow (see \cref{fig:flow_eq_general}) and postpone this particular discussion to \cref{subsec:truncation}.

A cumbersome property of the vertex flow is the appearance of the tadpole contracted three-particle vertex $\Gamma^{\Lambda}_3 \bullet S^{\Lambda}$. In other words, the computation of $\diff \Gamma^{\Lambda}$ requires knowledge about $\diff \Gamma^{\Lambda}_3$, which itself features a contribution $\Gamma^{\Lambda}_4 \bullet S^{\Lambda}$. More generally speaking, the flow of the $n$-particle vertex depends on all $\Gamma^{\Lambda}_m$ with $m \leq n + 1$, implying that the formally exact hierarchy of vertex ODEs cannot be solved without employing additional approximations, which we discuss in the next section. After having implemented such a \textit{truncation}, the flow equations for the action in Eq.~\eqref{eq:action} can be solved using the initial conditions \cite{Kopietz2010,Metzner2012}
\begin{align}
    \Sigma^{\Lambda \to \infty}(x'_1 | x_1) &= 0 \notag \\
    \Gamma^{\Lambda \to \infty}(x'_1, x'_2 | x_1, x_2) &= 4 V_{x'_1, x'_2| x_1, x_2} \notag \\
    \Gamma^{\Lambda \to \infty}_{n \geq 3}(x'_1, x'_2 | x_1, x_2) &= 0  \,.\label{eq:initalcond}
\end{align}
\begin{figure*}
    \centering
    \includegraphics[width = 0.9\textwidth]{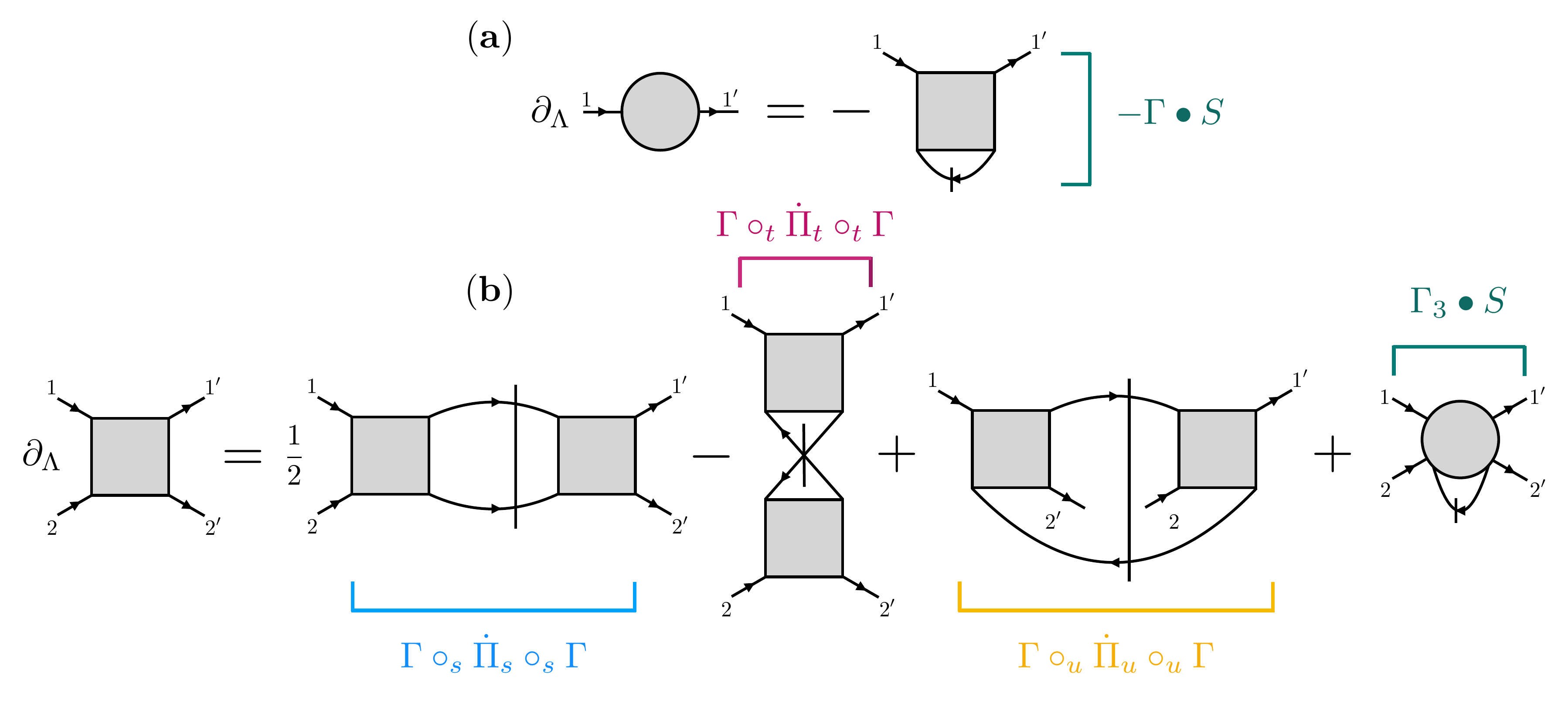}
    \caption{\textbf{Flow equations for the self-energy and two-particle vertex.} The self-energy flow shown in \textbf{(a)} simply consists of a vertex contracted with a single-scale propagator $S$. The contributions to $\Gamma$ in \textbf{(b)}, involve two vertices, which have their legs contracted with one of three propagator bubbles $\dot{\Pi}_c$. Dashed lines indicate partial derivatives $\diffsig G$. The three-particle vertex $\Gamma_3$ is difficult to compute (see \cref{subsec:truncation}) and requires some approximation.}
    \label{fig:flow_eq_general}
\end{figure*}

\subsection{Truncation of the flow equations}
\label{subsec:truncation}

The action considered in this review (see \cref{eq:action}) consists of a Gaussian part and a quartic interaction and we, thus, concern ourselves with approximations that truncate the flow equations beyond the two-particle level. In other words, the goal of this section is to present different ways of removing the expression $\Gamma^{\Lambda}_3 \bullet S^{\Lambda}$ from the right hand side of $\diff \Gamma^{\Lambda}$. To simplify the notation for the following discussions, we suppress the $\Lambda$-superscripts above all propagators and $n$-particle vertices and consider them implicitly as cutoff dependent. Furthermore, we dispense with writing out external arguments such as $G(x'_1 | x_1)$, since they can be reintroduced a posteriori without further complications. The flow of the vertex then reads
\begin{align}
    \diff \Gamma = \Gamma_3 \bullet S + \sum_c \Gamma \circ_c \dot{\Pi}_c \circ_c \Gamma \,.
\end{align}
The most simple approximation, the level-2 (L2) truncation, sets all $n$-particle vertices with $n \geq 3$ to zero, such that
\begin{align}
    \diff \Gamma \ \substack{(\text{L2}) \\ \approx} \sum_c \Gamma \circ_c \dot{\Pi}_c \circ_c \Gamma \,.
\end{align}
As long as the bare interaction $V$ is small if compared to, for example, the electronic bandwidth in itinerant fermion models, the L2 truncation can be justified in the $\Lambda \to \infty$ limit: contributions to $\Gamma_3$ are at least third order in $V$, whereas the vertex is $\mathcal{O}(V^2)$~\cite{Honerkamp2001,Platt2013}. 
During the flow, two scenarios are possible: 1) the vertex stays small and the L2 truncation remains well-controlled, or 2) $\Gamma$ flows to \textit{strong-coupling}, i.e., it becomes large and higher order vertices cannot be ignored, resulting in a breakdown of the flow. The latter usually happens when approaching a low-energy ordered phase \cite{Kopietz2010,Metzner2012} or when correlations mediated by a subclass of diagrams become particularly strong.

One of the downsides of the L2 truncation concerns the accuracy with which Ward identities are fulfilled \footnote{Ward identities are exact relations between vertices of different order that can be derived from conservation laws \cite{parquet_identities,Kopietz2010}}. More specifically, violations of the conservation law already set in at third order in $\Gamma$, thereby spoiling the robustness of the results obtained at the end of the RG flow. Moreover, the latter usually depend on the specific implementation of the regulator, complicating the analysis even further. A first attempt to improve the truncation of the flow equations, specifically with respect to the fulfilment of Ward identities, was made by Katanin in Ref.~\cite{Katanin2004} and amounts to the replacement of the partial derivative in $\dot{\Pi}_c$ by a full $\Lambda$-derivative: $\diffsig \to \diff$. Consequently, the vertex flow becomes
\begin{align}
    \diff \Gamma \ \substack{(\text{Kat.}) \\ \approx} \sum_c \Gamma \circ_c \diff \Pi_c \circ_c \Gamma \,,\label{eq:kataninsub}
\end{align}
where $\Pi_c$ is defined as in Eq.~\eqref{eq:bubbles} but without the partial derivative $\diffsig$. By expressing $G$ via Dysons's equation, one finds that
\begin{align}
    S^{\Lambda} = \diffsig G \to S_\text{kat}^{\Lambda} = \diff G &= \diff \left(G^{-1}_0 - \Sigma \right)^{-1} \notag \\ 
    &= \diffsig G + G \left(\diff \Sigma \right) G \,,
\end{align}
i.e., the propagator bubbles in the Katanin truncation feed back the self-energy flow into $\diff \Gamma$, augmenting the diagrams on the right-hand side (rhs) of the vertex flow by those $\mathcal{O}(\Gamma^3)$ diagrams without overlapping loops. If these additional diagrams are accounted for, single-channel FRG calculations\footnote{That is, ladder summations using only one of the terms $\Gamma \circ_c \diff \Pi_c \circ_c \Gamma$ on the rhs~of $\diff \Gamma$.} fulfill Ward identities exactly and, thus, self-consistency of the FRG approach is improved \cite{Katanin2004}. 

While the Katanin truncation reduces the systematic error induced by truncating three-particle and higher-order vertices (see Ref.~\cite{parquet_identities} for numerical results), it does not include all $\mathcal{O}(\Gamma^3)$ diagrams and violations of conservations laws will therefore likewise set in at third order in $\Gamma$. This issue was later addressed by Eberlein \cite{Eberlein_twoloop}, who proposed a scheme to systematically compute the missing third-order diagrams by first decomposing the vertex into three channels $g_c$ corresponding to the three different types of bubbles and contractions introduced in \cref{sec:generating_functionals} and subsequently inserting one-loop diagrams from channel $c$ into contractions with $c' \neq c$. In 2018, this idea was generalized by Kugler and von Delft in what is now called \textit{multiloop} FRG (MFRG) \cite{Kugler2018,parquet_conserving,MFRG_general}. This approximation mitigates many of the deficiencies mentioned above by instantiating an RG scheme that incorporates all diagrams of the so-called \textit{parquet} approximation (PA)~\cite{Diatlov-1957,Bychkov-1966,Roulet-1969,Abrikosov1965} - a set of coupled many-body relations that self-consistently connects the one- and two-particle level while maintaining an effective one-loop structure. Furthermore, the PA exactly incorporates Ward identities on the one-particle level \cite{parquet_conserving}, allowing for FRG calculations with an accuracy on par with QMC simulations in the weak to intermediate coupling regime \cite{Hille_quantitative}.

To present the general formulation of the MFRG flow, we adopt the language commonly used in the context of the PA and classify diagrams according to their two-particle reducibility in a particular channel~\cite{Roulet-1969}. A diagram is called \textit{two-particle reducible} (2PR) in the $s$-channel if it can be fully disconnected by cutting two parallel propagator lines. If however, those propagators point in opposite directions, we refer to them as $t$- or $u$-reducible. $t$-reducible diagrams differ from those reducible in the $u$-channel in the way external legs are assigned to the disconnected parts: in a $t$-reducible diagram the external legs lie on the same \textit{edge} of a fermionic vertex, whereas they lie on opposite \textit{corners} in a $u$-reducible term\footnote{The notion of "edges" and "corners" here refers to the diagrammatic representation of $\Gamma$ as a rectangular box.}. The total contribution of $c$ reducible diagrams to the vertex we denote by $g_c$. For simplicity, we drop the subscripts $c$ in the vertex contractions $\circ$ whenever the two-particle reducibility can already be deduced from the inserted bubble $\Pi_c$. To compute the multiloop ($m\ell$) flow of $g_c$, one first computes the respective Katanin ($1\ell$) diagrams, i.e.,
\begin{align}
    \diff g_c \overset{(1\ell)}{\approx} \Gamma \circ \diff \Pi_c \circ \Gamma \equiv \dot{g}^{(1\ell)}_c\,.
\end{align}
In a second step, one substitutes the one-loop flows in the complementary 2PR classes for the vertex to the left and right, while excluding the $\Lambda$-derivative in the bubble:
\begin{align}
    \diff g_c &\overset{(2\ell)}{\approx} \dot{g}_c^{(1\ell)} + \dot{g}_{\bar{c}}^{(1\ell)} \circ \Pi_c \circ \Gamma + \Gamma \circ \Pi_c \circ \dot{g}_{\bar{c}}^{(1\ell)} \notag \\
    &\equiv \dot{g}_c^{(1\ell)} + \dot{g}_c^{(2\ell)L} + \dot{g}_c^{(2\ell)R}\,.
\end{align}
Here, we introduced the left ($L$) and right ($R$) part of the $2\ell$ contribution $\dot{g}^{(2\ell)}_{c} \equiv \dot{g}_c^{(2\ell)L} + \dot{g}_c^{(2\ell)R}$ as well as the short-hand notation $g_{\bar{c}} = \sum_{c' \neq c} g_{c'}$. In a two-loop approximation, i.e., $\diff g_c \approx \dot{g}_c^{(1\ell)} + \dot{g}_c^{(2\ell)}$, the vertex flow contains all third-order diagrams as in the Eberlein construction, as well as some fourth order diagrams due to the insertion of $1\ell$-diagrams with Katanin bubbles. To construct the $m\ell$ diagrams for $m \geq 3$, we additionally need the central part
\begin{align}
    \dot{g}^{(m\ell)C}_c &= \dot{g}_{c}^{([m - 1]\ell)R} \circ \Pi_c \circ \Gamma \notag \\ 
    &= \Gamma \circ \Pi_c \circ \dot{g}_{c}^{([m - 1]\ell)L}\,,
\end{align}
such that $\dot{g}^{(m\ell)}_c = \dot{g}^{(m\ell)L}_c + \dot{g}^{(m\ell)C}_c + \dot{g}^{(m\ell)R}_c$. The multiloop flow in the $c$-channel is thus obtained as 
\begin{align}
    \diff g_c \overset{(m\ell)}{\approx} \sum_{n = 1}^{m} \dot{g}^{(n\ell)}_c \,,
\end{align}
from which the flow of the full vertex follows as $\diff \Gamma = \sum_c \diff g_c$. Finally, the central part of the $s$- and $u$-channel
\begin{align}
    \dot{g}^{C}_{\bar{t}} \overset{(m\ell)}{\approx} \sum_{n \geq 3}^{m} \dot{g}^{(n\ell)C}_{\bar{t}} \,,
\end{align}
can be used to add multiloop corrections $\dot{\Sigma}_{1(2)}$ to the self-energy flow
\begin{align}
    \diff \Sigma \overset{(m\ell)}{\approx} -\Gamma \bullet S + \dot{\Sigma}_1 + \dot{\Sigma}_2 \,,
\end{align}
where $\dot{\Sigma}_1 = \dot{g}^{C}_{\bar{t}} \bullet G$ and $\dot{\Sigma}_2 = \Gamma \bullet (G \times \dot{\Sigma}_1 \times G)$\footnote{$\times$ denotes an ordinary product.}. The additional terms in $\diff \Sigma$ are necessary to fully establish agreement of MFRG with the parquet approximation \cite{MFRG_general}. Since the derivatives $\diff \Pi_c$ in the Katanin bubbles already depend on the self-energy flow, the calculation of vertex and self-energy corrections can in principle be iterated until convergence is reached. 

One remarkable property of MFRG is the restoration of regulator independence at the end of the RG flow. For $\Lambda \to 0$, the multiloop equations precisely coalesce with the parquet approximation, which, as a general many-body relation, is insensitive to the type of regularization used throughout the RG flow.

Let us summarize the main aspects of this section. We have presented a general formulation of the functional renormalization group framework for interacting fermions with quartic interactions. In order to make the differential equations for the 1PI vertex functions soluble, an approximate truncation scheme is unavoidable. Our discussion introduced three commonly used truncation strategies: L2 truncation ($\Gamma_{n} = 0$ for $n \geq 3$), Katanin scheme ($\dot{\Pi}_c \to \diff \Pi_c$) and multiloop FRG, which adds parquet type diagrams to the flow of 2PR vertices and accounts for corrections to the self-energy. In the next section, we will occupy ourselves with the explicit implementations of these truncations within the pseudo-fermion FRG. For this purpose, we derive an efficient parameterization of the vertex functions based on the symmetries of the pseudofermion Hamiltonian presented in \cref{sec:auxiliary_fermions}. This will allow a compact representation of the bubble functions and vertex contractions which minimizes the numerical effort involved to compute them.

\section{Pseudo-fermion functional renormalization group}
\label{sec:PFFRG}

The reformulation of spin Hamiltonians in terms of pseudo-fermions, as introduced in \Sec{sec:operatormapping} allows for the implementation of the general fermionic FRG formalism from the previous section to treat pure spin systems. Here, we will introduce this fusion, the pseudo-fermion functional renormalization group (PFFRG), by translating the symmetries of the pseudo-fermion Green's functions derived in \Sec{sec:pseudofermionsymmetries} into an efficient parametrization of the vertex functions forming the basic building blocks for the FRG.

\subsection{Parameterization of the self-energy}
The self-energy is directly related to the one-particle Green's function according to Dyson's equation
\begin{equation}
  \begin{split}
  \Sigma(x'_1|x_1) &= G_0^{-1}(x'_1|x_1)-G^{-1}(x'_1|x_1) \\&= \i\omega_1 \delta_{x'_1, x_1} -G^{-1}(x'_1|x_1),\label{eqref:dyson}
  \end{split}
\end{equation}
where we have already used that kinetic terms vanish in the pseudo-fermion formulation and, thus, $G_0(x'_1|x_1) = \frac{1}{\i\omega_1}\delta_{x'_1, x_1}$ for the non-interacting propagator. As it is completely diagonal in all degrees of freedom, the symmetries of the one-particle Green's function listed in \tab{tab:onepartsymm} directly apply to the self-energy. 

Upon inspection, we note that $\U(1)$ symmetry guarantees locality of the self-energy in real space, which, in combination with translational symmetry of the lattice removes any spatial dependence\footnote{This is only strictly true for Archimedean lattices, for which all lattice sites are equivalent. For $n$ inequivalent types of lattice sites, one has to define $n$ different self-energies $\Sigma$.}. Similarly, time-translational invariance (TT) guarantees diagonality in frequency space.

Therefore, we find the intermediate parametrization
\begin{equation}\label{eq:paraself}
  \Sigma(x'_1|x_1) = \sum\limits_{\alpha=0,1,2,3}\Sigma^\alpha(\omega_1) \sigma^\alpha_{\mu'_1\mu_1} \delta_{i'_1i_1} \delta_{\omega'_1,\omega_1},
\end{equation}
where we have expanded the remaining spin structure in terms of Pauli matrices supplemented by a unit matrix $\sigma^0$.

The combination of hermiticity and time-reversal symmetry furthermore implies
\begin{equation}
\Sigma^\alpha(\omega^\noprime_{1}) \sigma^\alpha_{\mu'_{1}\mu^\noprime_{1}} = \mu'_{1} \mu_{1} \Sigma^\alpha(\omega_{1}) \sigma^\alpha_{\bar{\mu}_{1}\bar{\mu}'_{1}}. \label{eq:trhonepart}
\end{equation}
Realizing that
\begin{equation}
  \mu'_{1} \mu_{1} \sigma^\alpha_{\bar{\mu}_{1}\bar{\mu}'_{1}} = \mu'_{1} \mu_{1}\left(\sigma^{\alpha}_{\bar{\mu}'_{1}\bar{\mu}_{1}}\right)^* = \xi(\alpha)\sigma^\alpha_{\mu'_{1}\mu_{1}}
\end{equation}
with 
\begin{equation}
  \xi(\alpha) = \begin{cases}
    \phantom{-}1 & \alpha = 0\\
    -1 & \alpha \in \{1,2,3\}.
  \end{cases}
\end{equation}
immediately leads to the conclusion that only the $\alpha=0$ component of the self-energy is non-vanishing, removing any spin-dependence. Additionally, the combination of particle-hole conjugation implies that $\Sigma^0$ has to be antisymmetric in its frequency argument, while time-reversal symmetry renders it purely imaginary. Therefore, for the pseudo-fermion self-energy, we adopt the parametrization
\begin{equation}
  \Sigma(x'_1|x_1) = \i \gamma(\omega_1) \delta_{x'_1,x_1}\label{eq:selfenergy_final}
\end{equation}
with a real, antisymmetric function $\gamma$.

It is important to note that this simple structure heavily relies on time-reversal symmetry. Breaking it by, e.g., considering external magnetic fields would introduce both real- and imaginary parts and non-vanishing contributions from $\alpha=1,2,3$ in the self-energy parametrization of \eq{eq:paraself}.

\subsection{Spin and real space dependence of the two-particle vertex}
\label{sec:spindep}
Similar to the self-energy, we now aim at a simplified representation of the two-particle vertex, manifesting the symmetries of the pseudo-fermion Green's functions as summarized in \tab{tab:twopartsymm}. To this end, we first give the relation between the full two-particle Green's function and the corresponding vertex through the tree expansion~\cite{Negele2018}
\begin{equation}
  \begin{split}
  G_2(x'_1,&x'_2|x_1,x_2) \\= \sum\limits_{x'_3,x'_4,x_3,x_4} &G(x'_1|x'_3)G(x'_2|x'_4)\Gamma(x'_3,x'_4|x_3,x_4)\\
  &\times G(x_3|x_1)G(x_4|x_2) \\&+ G(x'_1|x_1) G(x'_2|x_2)\,.\label{eq:tree}
  \end{split}
\end{equation}

Due to the diagonality of $G$ in all indices, as discussed in the previous section, again all symmetries of the full two-particle Green's function $G_2$ directly carry over to the two-particle vertex $\Gamma$.

As in the case of the self-energy, we start form the local $\U(1)$ symmetry of the pseudo-fermions, which induces a bi-locality of the vertex function captured in the expansion
\begin{equation}
  \begin{split}
  \Gamma(x'_1,x'_2|x_1,x_2) = &\phantom{+} \Gamma_{=,i_1i_2}(x'_1,x'_2|x_1,x_2)\delta_{i'_{1}i_{1}}\delta_{i'_{2}i_{2}} \\&-  \Gamma_{\times,i_1i_2}(x'_1,x'_2|x_1,x_2)\delta_{i'_{1}i_{2}}\delta_{i'_{2}i_{1}}\, .\label{eq:bilocalparam}
  \end{split}
\end{equation}
Here, $\Gamma_=$ and $\Gamma_\times$ represent the vertex content, where site indices are constant across the equally numbered pairs of indices or swapped, respectively. Clearly, from crossing symmetry, the relation
\begin{equation}
  \Gamma_{\times,i_1i_2}(x'_1,x'_2|x_1,x_2) = \Gamma_{=,i_2i_1}(x'_1,x'_2|x_2,x_1)\, ,\label{eq:parallelcross}
\end{equation}
holds. Employing the space group symmetries of the lattice, we are able to project back one of the remaining site indices onto a single reference site, rendering the vertex only dependent on the difference vector between the sites $i_1$ and $i_2$. We will, however, not explicitly implement this fact in our notation, as this would complicate the flow equations discussed henceforth.

In numerical implementations, however, we will use this fact to approximate the vertex by neglecting vertex contributions, for which $\norm{\bm{R}_{i_1}-\bm{R}_{i_2}}<L$, i.e. we impose a maximum correlation length in some norm $\norm{\cdot}$. Effectively, this implements calculations in an infinite system, which avoids both boundary effects for calculations with open boundary conditions and finite momentum resolution imposed by periodic ones. Although the finite correlation length imposed will lead to broadened features in reciprocal space, in this way we are able to resolve magnetic phenomena incommensurate with the lattice. For details on the implementation see \cref{app:lattice}.

In the next step, we expand the spin dependence of the vertex in terms of Pauli matrices, leading to
\begin{equation}
  \Gamma_{=,i_1i_2}(x'_1,x'_2|x_1,x_2) = \Gamma^{\alpha\beta}_{i_1i_2}(\omega'_{1}, \omega'_{2}|\omega_{1}, \omega_{2}) \sigma^{\alpha}_{\mu'_{1}\mu_{1}} \sigma^\beta_{\mu'_{2}\mu_{2}}
\end{equation} 
with a similar expansion for $\Gamma_{\times,i_1i_2}$. Summation over $\alpha$ and $\beta$ is implied. We use a similar notation $g^{\alpha\beta}_{c,i_1i_2}(\omega'_{1}, \omega'_{2}|\omega_{1}, \omega_{2})$
for the $c$-channel contributions to the vertex.

In the special case of a Heisenberg Hamiltonian, which is spin-rotation invariant, this relation can be further simplified to
\begin{equation}
\begin{split}
  \Gamma_{=,i_1i_2}(x'_1,x'_2|x_1,x_2) &= \Gamma_{\mathrm{s},i_1i_2}(\omega'_{1}, \omega'_{2}| \omega_{1}, \omega_{2}) \sigma^\alpha_{\mu'_{1}\mu_{1}} \sigma^\alpha_{\mu'_{2}\mu_{2}}\\
  &+ \Gamma_{\mathrm{d},i_1i_2}(\omega'_{1}, \omega'_{2}| \omega_{1}, \omega_{2}) \delta_{\mu'_{1}\mu_{1}} \delta_{\mu'_{2}\mu_{2}},\label{eq:spindensparam}
  \end{split}
\end{equation} 
introducing the so-called spin- and density vertices $\Gamma_\mathrm{s}$ and $\Gamma_\mathrm{d}$, respectively. These two terms in \eq{eq:spindensparam} correspond to the only possible spin dependences of a two-particle vertex that obey the spin-rotation symmetry of a Heisenberg Hamiltonian.

The last symmetry we want to invoke at this point is a combination of the two particle-hole symmetries listed in \tab{tab:twopartsymm}, followed by a time-reversal transform, hermitian conjugation and another time-reversal operation. This sequence of symmetry transformations yields the relation
\begin{equation}
  \Gamma^{\alpha\beta}_{i_1i_2} = \xi(\alpha)\xi(\beta)\left(\Gamma^{\alpha\beta}_{i_1i_2}\right)^*,
\end{equation}
which significantly simplifies the analytic structure of the two-particle vertex functions. In particular, it indicates that the spin and density vertices $\Gamma_s$ and $\Gamma_d$ are purely real.

\begin{figure*}
  \centering
  \includegraphics[width=0.85\linewidth]{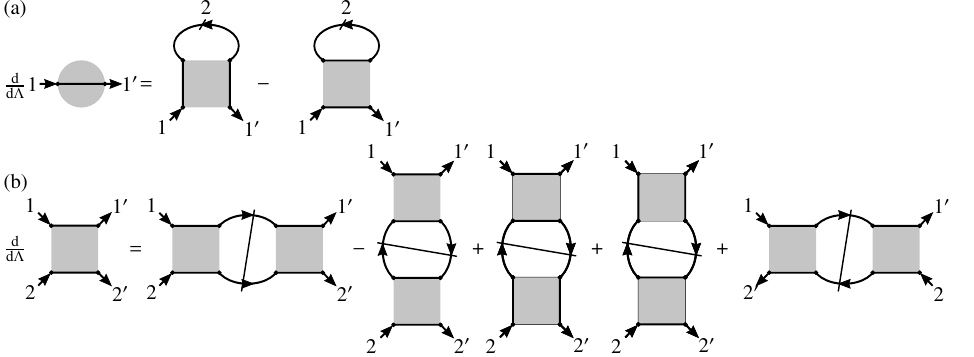}
  \caption{Diagrammatic representation of the multi-local PFFRG flow equations of the (a) self-energy $\Sigma$ [gray circle, \eq{eq:multilocalseflow}] and (b) two-particle vertex  $\Gamma_{=}$ [gray square, \eq{eq:multilocalflow}]. Along solid lines of the vertices, the site index remains constant. The arrows represent full Green's functions $G^\Lambda$, while slashed arrows are single scale propagators $S^\Lambda$. A slash crossing two propagators denotes a sum of the two possibilities of replacing one propagator by a single scale propagator. \label{fig:bilocalflow}}
\end{figure*}

As an intermediate result, we present the multilocal flow equations, obtained by inserting the bilocal parametrization of the two-particle vertex, \eq{eq:bilocalparam}, into \eq{eq:self_flow_general} and \eq{eq:vertex_flow_general}, respectively. Keeping in mind the locality of the self-energy, we find
\begin{equation}
  \begin{split}
  \dv{\Lambda} \Sigma^\Lambda(x_1|x_1) = &\sum\limits_{x_2} \Big(\Gamma_{\times,i_1i_1}^\Lambda(x_1,x_2|x_1,x_2)\\&- \sum\limits_j\Gamma_{=,i_1j}^\Lambda(x_1,x_2|x_1,x_2)\Big) S^\Lambda(x_2,x_2)\label{eq:multilocalseflow}
  \end{split}
\end{equation}
for the flow of $\Sigma$, as well as
\begin{equation}
  \begin{split}
    &\dv{\Lambda}\Gamma_{=,i_1i_2}^\Lambda(x'_1,x'_2|x_1,x_2) =
    \\&\phantom{-}\sum\limits_{x_3,x_4} \left[\Gamma_{=,i_1i_2}^\Lambda(x'_1,x'_2|x_3,x_4)\Gamma_{=,i_1i_2}^\Lambda(x_3,x_4|x_1,x_2)\right. \\
    &-\sum_j\Gamma_{=,i_1j}^\Lambda(x'_1,x_4|x_1,x_3)\Gamma_{=,ji_2}^\Lambda(x_3,x'_2|x_4,x_2) -\big(3\leftrightarrow4\big)\\
    &+\Gamma_{=,i_1i_2}^\Lambda(x'_1,x_4|x_1,x_3)\Gamma_{\times,i_2i_2}^\Lambda(x_3,x'_2|x_4,x_2) +\big(3\leftrightarrow4\big)\\
    &+\Gamma_{\times,i_1i_1}^\Lambda(x'_1,x_4|x_1,x_3)\Gamma_{=,i_1i_2}^\Lambda(x_3,x'_2|x_4,x_2) +\big(3\leftrightarrow4\big)\\
    &\left. +\Gamma_{=,i_1i_2}^\Lambda(x'_2,x_4|x_1,x_3)\Gamma_{=,i_1i_2}^\Lambda(x_3,x'_1|x_4,x_2) +\big(3\leftrightarrow4\big)\right]\\
    &\times G^\Lambda(x_3,x_3) S_\text{kat}^\Lambda(x_4,x_4)\\
    \end{split}\label{eq:multilocalflow}
\end{equation}
for the two-particle vertex. Note that we have already neglected the three-particle vertex terms and performed the Katanin substitution in these flow equations. Furthermore, we have explicitly incorporated the diagonality of the full and single-scale propagators in all their arguments. Using \eq{eq:parallelcross}, we can replace $\Gamma_\times$ by $\Gamma_=$, such that the two-particle vertex is only represented by $\Gamma_=$. In \fig{fig:bilocalflow}, we illustrate a diagrammatic representation of the multilocal flow equations.

Comparing \eq{eq:multilocalflow} to \eq{eq:vertex_flow_general}, there are a few notable features. Firstly, all but the third line in \eq{eq:multilocalflow} do not involve a site summation, rendering these terms bi-local. Secondly, the $t$-channel diagram contained in \eq{eq:vertex_flow_general} splits into three contributions: The third line in \eq{eq:multilocalflow} represents a RPA-like contribution, which involves a site summation. As this is the only term mixing correlations between different pairs of lattice sites, possibly generating longer-range correlations from initially short-ranged bare interactions. Therefore, we can expect it to be pivotal in the formation of long-range order, a notion we will put on more solid grounds in \Sec{sec:largeS}. The fourth and fifth lines in \eq{eq:multilocalflow}, originate from the intermixing of $\Gamma_\times$ and $\Gamma_=$ in the parameterization. 

The flow of the self-energy, \eq{eq:multilocalseflow}, also splits into two contributions, with the first term resembling a purely local Fock-style diagram, while the second term is a non-local Hartree contribution involving a sum over the lattice.

\subsection{Frequency parametrization}
\label{subsec:frequencyparam}

After discussing the implications of symmetries on the spin and real-space structure of the pseudo-fermion vertices, we now turn to an adequate treatment of the remaining frequency structure of $\Gamma^{\alpha \beta}_{i_1 i_2}$. As already mentioned in~\Sec{sec:pseudofermionsymmetries}, time translation invariance implies frequency conservation, so that the vertex has only three fermionic frequency arguments instead of four. In the early days of PFFRG, these were usually rewritten in terms of the three bosonic transfer frequencies
\begin{align}
  \omega_s &= \omega'_1 + \omega'_2 = \omega_1 + \omega_2\\
  \omega_t &= \omega'_1 - \omega_1 = \omega_2 - \omega'_2\\
  \omega_u &= \omega'_1 - \omega_2 = \omega_1 - \omega'_2\,,
  \label{eq:transfer_frequencies}
\end{align}
each of them associated with the energy exchanged during a scattering process in the corresponding 2PR channel. However, as pointed out in a seminal paper by Wentzell et al.~\cite{Wentzell2020}, this fully bosonic representation of the vertex function gives rise to complicated high-frequency asymptotics: if one of the transfer frequencies goes to infinity while the other two remain fixed, $\Gamma$ generally does not decay to zero but to a (non-vanishing) constant depending on the value of the other two frequencies.

\begin{figure}
  \centering
  \includegraphics[width=\columnwidth]{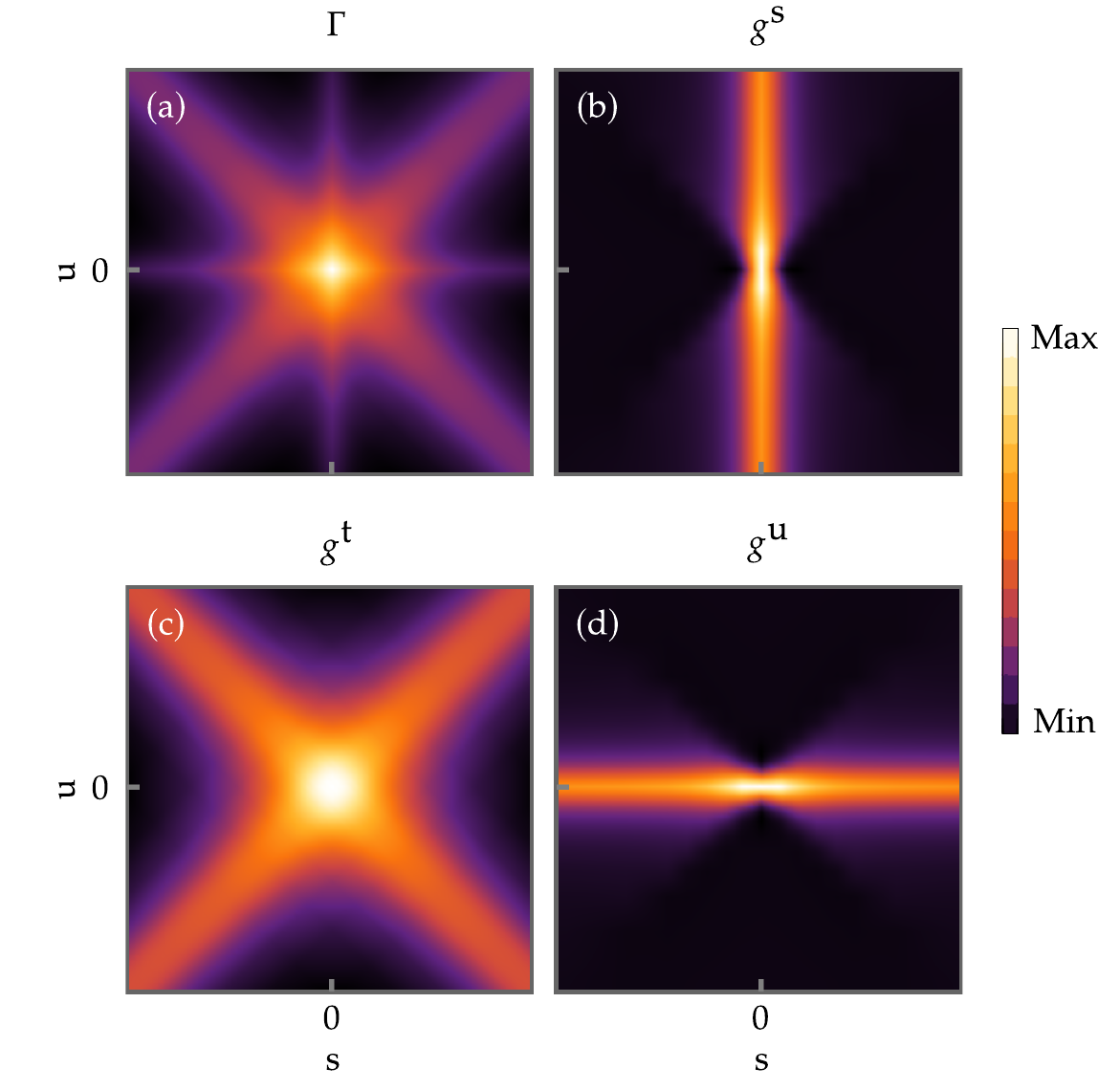}
  \caption{(a) Illustration of a full pseudo-fermion spin-vertex $\Gamma$ for the simple-cubic lattice antiferromagnet shown in the $t=0$ plane, showing characteristic double-cross structure. Its decomposition in the (b) $s$-, (c) $t$- and (d) $u$-channel contributions reveals the origin of the structure in the different channels. Due to this structure, the asymptotics are not well-defined in the transfer frequency parametrization, necessitating a parametrization for each channel individually.\label{fig:vertexdecomp}}
\end{figure}

As can be seen from~\fig{fig:vertexdecomp}, where an exemplary decomposition of the pseudofermion vertex $\Gamma$ into its 2PR contributions $g_c$ is shown, this constant arises from some residual, non-decaying contributions in the complementary channels $g_{c' \neq c}$. For example, if $s \to \infty$, all contributions from the $s$- and $t$-channel vanish (see~\fig{fig:vertexdecomp}(b) \& (d)), whereas the $u$-channel assumes some finite value that depends on $\omega_u$ (~\fig{fig:vertexdecomp}(c)). To model the vertex more accurately, Wentzell and coworkers introduced a mixed bosonic-fermionic frequency notation for each 2PR channel, grouping the diagrams contributing to $g_c(\omega_c, \nu_c, \nu'_c)$ into three types of kernel functions, i.e.
\begin{align}
  g_{c}(\omega_c,\nu_c,\nu_c') &= K_1^{c}(\omega_c) \notag \\ &+ K_2^{c}(\omega_c,\nu_c) + \bar{K}_2^{c}(\omega_c, \nu_c') \notag \\ &+ K_3^{c}(\omega_c,\nu_c,\nu_c') \,.
  \label{eq:kernelfuncs}
\end{align}
Most importantly, all $K$-functions decay to zero if one of their arguments is taken to infinity, which tremendously simplifies their numerical implementation. This decomposition can already be motivated from the lowest orders of perturbation theory, where so-called 'bubble' ($K_1$) and 'eye' ($K_2, \bar{K}_2$) diagrams govern the high-frequency structure~\cite{Wentzell2020}. 

\begin{figure}
  \centering
  \includegraphics[width=\columnwidth]{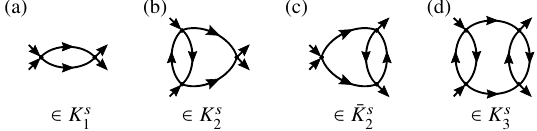}
  \caption{Exemplary diagrammatic contributions to the $s$-channel kernel functions $K_1^s$, $K_2^s$, $\bar{K}_2^s$, and $K_3^s$. For simplicity, here, we illustrate the bare vertex as a black dot. (a) $K_1^s$ diagram, where all external legs couple to the bare vertices. (b) and (c) are $K_2^s$ and $\bar{K}_2^s$ contributions, respectively, where the $\nu_s^\noprime$ or $\nu_s'$ dependencies arise from the additional loops. These diagrams are dubbed fish-eye diagrams. The nested three-loop diagram (d) contributes to the $K_3^s$ kernel. \label{fig:asymptoticdiagrams}}
\end{figure}

The rest functions $K_3$ then capture all diagrams which have an even more intricate loop structure and belong neither to $K_1$ nor to $K_2 / \bar{K_2}$, see~\fig{fig:asymptoticdiagrams}. More recent implementations of PFFRG~\cite{Kiese2020b, PFFRGSolver, Gresista2022, Thoenniss2020, Ritter2022} have adopted this strategy in order to improve numerical accuracy when integrating the FRG flow. Following Refs.~\cite{Kiese2020b, PFFRGSolver, Gresista2022}, the bosonic frequencies are defined as in~\eq{eq:transfer_frequencies}, whereas the remaining two fermionic frequencies are chosen as 
\begin{align}
  \nu_s &= (\omega_1-\omega_2)/2 &\nu'_s &= (\omega'_{2}-\omega'_{1})/2\notag\\
  \nu_t &= (\omega_1+\omega'_{1})/2 &\nu_t' &= (\omega_{2}+\omega'_{2})/2\notag\\
  \nu_u &= (\omega_1+\omega'_{2})/2 &\nu_u' &= (\omega'_{1}+\omega_{2})/2.\label{eq:nat_para}
\end{align}

\begin{figure*}
  \centering
  \includegraphics[width=0.85\linewidth]{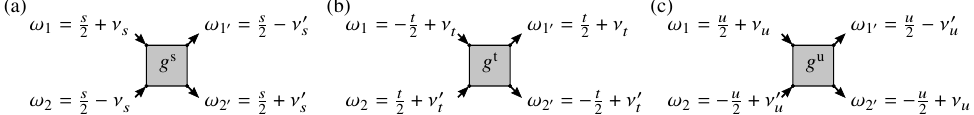}
  \caption{Natural frequency parametrization as used in the asymptotic parametrization scheme defined in \eq{eq:nat_para}. The diagrammatic $s$-, $t$-, and $u$-channels vanish in the limit of taking the corresponding bosonic transfer frequency $s/t/u$ to infinity. The shift by half the transfer frequency allows for a simpler implementation of symmetries. \label{fig:frequencyparam}}
\end{figure*}

The distribution of the frequencies to the external legs of a two-particle diagram is illustrated in \cref{fig:frequencyparam}. Please note, that this choice of fermionic frequencies is shifted with respect to Refs.~\cite{Li2016,Wentzell2020} to simplify symmetry relations.

Taking the appropriate high-frequency limits, the general symmetry relations in \tab{tab:twopartsymm} can straightforwardly be translated into the mixed frequency notation and assume are particularly simple form, see~\tab{tab:naturalsymm}. Most noteworthy, the $t$-channel decouples from the other two, while the $s$- and $u$-channels are coupled by inverting the fermionic frequencies. In total, these symmetries allow us to restrict the frequency domain of every channel to positive frequencies only, where, in addition, $\nu \geq \nu'$. We want to emphasize that those symmetries exchanging $\nu_c$ and $\nu'_c$ also imply that $K_2$ and $\bar{K}_2$ are equivalent for pseudo-fermionic systems and, thus, only one of them needs to be implemented. The full flow equations in this asymptotic frequency parametrization are presented in \Sec{app:floweqasymptoticfreq}.

\begin{table*}
  \setlength{\tabcolsep}{8pt}
  \setlength\extrarowheight{2pt}
  \centering
  \begin{tabular}{Ar}
  \toprule
    g^{s,\alpha\beta}_{i_1i_2}(\omega_s,\nu_s^\noprime,\nu_s') &=  g^{s,\beta\alpha}_{i_2i_1}(-\omega_s,\nu_s^\noprime,\nu_s') & (X $\circ$ TR $\circ$ H $\circ$ PH1 $\circ$ PH2)\\
    g^{s,\alpha\beta}_{i_1i_2}(\omega_s,\nu_s^\noprime,\nu_s') &= -\xi(\alpha)g^{u,\beta\alpha}_{i_2i_1}(\omega_s,-\nu^\noprime_s,\nu_s') &(PH2 $\circ$ X)\\
    g^{s,\alpha\beta}_{i_1i_2}(\omega_s,\nu_s^\noprime,\nu_s') &= -\xi(\beta)g^{u,\alpha\beta}_{i_1i_2}(\omega_s,\nu^\noprime_s,-\nu_s') &(PH2)\\
    g^{s,\alpha\beta}_{i_1i_2}(\omega_s,\nu_s^\noprime,\nu_s') &= g^{s,\beta\alpha}_{i_2i_1}(\omega_s,\nu_s',\nu^\noprime_s) &  (X $\circ$ TR $\circ$ H)\\
    \midrule
    g^{t,\alpha\beta}_{i_1i_2}(\omega_t,\nu_t^\noprime,\nu_t') &=  \xi(\alpha)\xi(\beta)g^{t,\alpha\beta}_{i_1i_2}(-\omega_t,\nu^\noprime_t,\nu_t') & (TR $\circ$ H)\\
    g^{t,\alpha\beta}_{i_1i_2}(\omega_t,\nu_t^\noprime,\nu_t') &= -\xi(\alpha)g^{t,\alpha\beta}_{i_1i_2}(\omega_t,-\nu^\noprime_t,\nu_t') &(PH1)\\
    g^{t,\alpha\beta}_{i_1i_2}(\omega_t,\nu_t^\noprime,\nu_t') &= -\xi(\beta)g^{t,\alpha\beta}_{i_1i_2}(\omega_t,\nu^\noprime_t,-\nu_t') &(PH2)\\
    g^{t,\alpha\beta}_{i_1i_2}(\omega_t,\nu_t^\noprime,\nu_t') &= g^{t,\beta\alpha}_{i_2i_1}(\omega_t,\nu_t',\nu^\noprime_t) &  (X $\circ$ TR $\circ$ H)\\
    \midrule
    g^{u,\alpha\beta}_{i_1i_2}(\omega_u,\nu_u^\noprime,\nu_u') &=  \xi(\alpha)\xi(\beta)g^{u,\beta\alpha}_{i_2i_1}(-\omega_u,\nu^\noprime_u,\nu_u') & (X $\circ$ TR $\circ$ H)\\
    g^{u,\alpha\beta}_{i_1i_2}(\omega_u,\nu_u^\noprime,\nu_u') &= -\xi(\beta)g^{s,\beta\alpha}_{i_2i_1}(\omega_u,-\nu^\noprime_u,\nu_u') &(X $\circ$ PH2)\\
    g^{u,\alpha\beta}_{i_1i_2}(\omega_u,\nu_u^\noprime,\nu_u') &= -\xi(\beta)g^{s,\alpha\beta}_{i_1i_2}(\omega_u,\nu^\noprime_u,-\nu_u') &(PH2)\\
    g^{u,\alpha\beta}_{i_1i_2}(\omega_u,\nu_u^\noprime,\nu_u') &= g^{u,\beta\alpha}_{i_2i_1}(\omega_u,\nu_u',\nu^\noprime_u) &  (TR $\circ$ H)\\
  \bottomrule
  \end{tabular}
  \caption{Symmetry relations of the channel contributions to the two-particle vertex functions for pseudo-fermion Hamiltonians expanded in terms of Pauli matrices in the natural frequency parametrization for the $s$-, $t$- and $u$-channels respectively, see \eq{eq:nat_para}. The rightmost column specifies the combinations of physical symmetries to realize these relations. TR denotes time-reversal, H hermitian conjugation, X crossing symmetry in both incoming and outgoing particles and PH1/2 is a particle-hole transformation for particle 1 or 2.\label{tab:naturalsymm}}
\end{table*}

\subsection{Choice of truncation}

In the multi-local flow equations, \eq{eq:multilocalflow}, we have already performed the Katanin-substitution to go beyond a conventional L2 truncation of the FRG equations. As already realized in the early days of PFFRG~\cite{Reuther2010}, the Katanin substitution is an essential modification to obtain meaningful results. Technically, the Katanin terms ensure the full feedback of the self-energy into the flow of the two-particle vertex. This creates damping effects for the onset of magnetic long range order that can be seen as an incorporation of quantum fluctuations: If a system flows into a magnetically ordered phase during the RG flow, the two-particle vertex increases. Since the self-energy flow is proportional to the two-particle vertex this also increases the self-energy. However, the self-energy, when fed back into the two-particle vertex via the additional Katanin terms, occurs in the {\it denominator} of the propagators. This suppresses the propagators in the flow equation of the two-particle and, hence, suppresses the two-particle vertex itself,  realizing a negative feedback loop that may suppress the initially assumed onset of magnetic order. In fact, without the Katanin terms, the PFFRG always predicts magnetic order and the obtained susceptibilities closely resemble those of a bare spin mean-field theory. The negative feedback loop of the Katanin terms is, therefore, essential to find any non-magnetic phases at all~\cite{Reuther2010}.

\begin{figure}[b]
\centering
\includegraphics[width=\columnwidth]{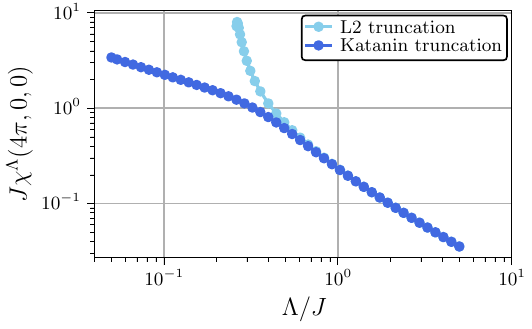}
\caption{Flow of the PFFRG spin-susceptibility at the pinch-point of the nearest-neighbor antiferromagnet on the Pyrochlore lattice. The L2 truncated flow features a divergence, indicative of long-range magnetic order, whereas the flow remains regular when the Katanin truncation is invoked.  \label{fig:pyroflow}}
\end{figure}

As a demonstration, in \fig{fig:pyroflow}, the flow of the spin-susceptibility of the Pyrochlore nearest-neighbor antiferromagnet (see \Sec{sec:pyro} for a discussion of the model) diverges in the L2 truncation, whereas it stays regular in the Katanin corrected case even for small RG scales $\Lambda$.

We will refrain from including multiloop corrections \cite{Thoenniss2020,Kiese2020b} in this review, as the qualitative results are not changed by these and the quantitative numerical improvements do not warrant the increased effort (see \Sec{sec:challenges} for details).

\subsection{Regulator}
\label{sec:regulators}

To complete the discussion of the PFFRG flow equations, we have to specify the regulator function $R(\Lambda)$ [defined before \eq{eq:diff_eff}] used to introduce the IR cutoff. While in itinerant systems, a cutoff in reciprocal space~\cite{Metzner2012}, as well as rescaling of the interactions~\cite{Honerkamp2004} or temperature itself~\cite{Honerkamp2001,Platt2013} are commonly used for performing the regularization, the formulation of PFFRG in real space at zero temperature calls for a different approach, which, most naturally, amounts to implementing the RG scale $\Lambda$ in frequency space. Here, we will discuss the two different implementations used in this context: the sharp step regulator already employed in earlier implementations of PFFRG~\cite{Reuther2010} and a recently introduced smoothened version~\cite{Kugler2018,Kiese2020b,Thoenniss2020}.

\subsubsection{Step regulator}
\label{sec:stepreg}
The most straightforward way to separate high- and low-energy degrees of freedom is by introducing a step-like regulator in frequency space of the form
\begin{equation}
  R(\omega,\Lambda) = \theta(|\omega|-\Lambda),
\end{equation}
where $\theta$ denotes the Heaviside step function. The scale-dependent bare propagator for a spin system in pseudo-fermionic representation is then given by
\begin{equation}
  \i G_0^\Lambda(\omega) = \frac{\theta(|\omega|-\Lambda)}{\omega},
\end{equation}
from which we immediately find the full propagator, by using Dyson's equation [\eq{eqref:dyson}], to be 
\begin{equation}
  \i G^\Lambda(\omega) = \frac{\theta(|\omega|-\Lambda)}{\omega+\gamma^\Lambda(\omega)}.
\end{equation}
The corresponding single-scale propagator [\eq{eq:singlescaledef}] can be calculated using Morris' lemma~\cite{Morris1994}, yielding
\begin{equation}
  \i S^\Lambda(\omega) = \frac{\delta(|\omega|-\Lambda)}{\omega+\gamma^\Lambda(\omega)},
\end{equation}
where $\delta$ denotes the Dirac-delta function. This form \emph{a posteriori} accounts for its name: the single-scale propagator in this formulation filters out exactly the frequency corresponding to the RG scale $\Lambda$. This considerably simplifies the right-hand side of the FRG equations in a L2 truncation, as it analytically replaces frequency integrations by a finite summation. 

However, invoking the Katanin substitution in \eq{eq:kataninsub} or performing a full multiloop scheme, this advantage is remidied due to integrations not containing the single-scale propagator. Furthermore, due to the non-analyticity of the regulator, the frequency dependence of the vertex functions shows characteristic kinks at $\Lambda$ dependent positions, which in a numerical implementation leads to an oscillating behavior of the RG flow, see e.g. \mRef{Reuther2010}.

\subsubsection{Smoothened frequency cutoff}
\label{sec:smoothreg}
To circumvent these numerical problems, in more recent implementations of PFFRG, a smooth regulator
\begin{equation}
  R(\omega,\Lambda) = 1-\e^{-\frac{\omega^2}{\Lambda^2}},\label{eq:smoothcutoff}
\end{equation}
is employed, which smears out the step at $|\omega| = \Lambda$ over a width of $\Lambda$. This regulator is similar to the so-called $\Omega$-flow used in itinerant fermion FRG~\cite{Giering2012}, in the sense that in \eq{eq:smoothcutoff} the suppression of the low-frequency region is done in a Gaussian shape, while in the $\Omega$-flow, this is done using a Lorentzian.

The bare propagator for the smooth cutoff is given by
\begin{equation}
  \i G_0^\Lambda(\omega) = \frac{1-\e^{-\frac{\omega^2}{\Lambda^2}}}{\omega},
\end{equation}
and, consequently, the full propagator reads as
\begin{equation}
  \i G^\Lambda(\omega) = \frac{1-\e^{-\frac{\omega^2}{\Lambda^2}}}{\omega+\gamma^\Lambda(\omega)}.
\end{equation}
Since, no discontinuities are present in this function, we can directly use \eq{eq:singlescaledef} to obtain the single scale propagator

\begin{equation}
  \i S^\Lambda(\omega) = \frac{2 \e^{-\frac{\omega^2}{\Lambda^2}}}{[\omega+\gamma^\Lambda(\omega)]^2} \frac{\omega^3}{\Lambda^3},
\end{equation}
which, as in the sharp cutoff case, features two peaks located symmetrically around $\omega=0$, but now at a frequency $\omega_p<\Lambda$, whereas we have $\omega_p=\Lambda$ in the case of a sharp cutoff.

\subsection{Initial conditions}
To close the PFFRG flow, we have additionally to specify the initial conditions [see \eq{eq:initalcond}] for our parameterization of the PFFRG. Since, the self-energy has to vanish at the beginning of the flow, we trivially find
\begin{equation}
  \gamma^{\Lambda\to\infty}(\omega) = 0.
\end{equation}
Antisymmetrizing the pseudo-fermion Hamiltonian in \eq{eq:heisenbergfermion}, we find for the two-particle vertex
\begin{equation}
  \Gamma^{\alpha\beta,\Lambda\to\infty}_{i_1i_2} = \frac{J^{\alpha\beta}_{i_1i_2}}{4}.\label{eq:pffrginitalcond}
\end{equation}

Interactions involving three or more spins would lead to a non-vanishing initial condition for the three-(or higher-)particle vertex, which is, although analytically possible to include, numerically not tractable due to the more involved frequency dependence.

\subsection{Susceptibilities}
\label{sec:susceptibilities}
As discussed in the previous sections, the FRG flow equations are formulated in terms of vertex functions. While these, in principle, contain the full physical information about the quantum mechanical state of a system, they are not physical observables and are therefore of limited use.

The simplest physical observable that can be straightforwardly calculated with PFFRG and which also allows for physically interpreting the system's quantum state is the {\it static} susceptibility or spin-spin correlator
\begin{equation}
  \chi^{\alpha\beta}_{ij}(\omega = 0) = \int\limits_0^\infty \dd \tau \ev{S_i^\alpha(\tau) S_j^\beta(0)}, \label{eq:realspacecorrel}
\end{equation}
where $\tau$ is the imaginary time and only contributions from $\alpha=\beta$ are finite for Heisenberg systems. Expressing the right-hand side of \eq{eq:realspacecorrel} in terms of pseudo-fermions, and using the tree expansion for the full two-particle Green's function [\eq{eq:tree}], we can express this quantity in terms of the self-energy and the two-particle vertex as 

\begin{equation}
  \begin{split}\label{eq:correlpffrg}
  &\chi^{\alpha\beta, \Lambda}_{ij}(\omega) = -\frac{1}{2} \frac{1}{2\pi} \int\dd\omega' G^\Lambda(\omega') G^\Lambda(\omega'+\omega) \delta_{ij}\delta_{\alpha\beta}\\ 
  &- \frac{1}{4} \left(\frac{1}{2\pi}\right)^2 \iint \dd\omega' \dd\omega'' G^\Lambda(\omega') G^\Lambda(\omega'+\omega) G^\Lambda(\omega'')\\
  &\times G^\Lambda(\omega''+\omega)\sum\limits_{\mu'_{1}\mu'_{2}\mu_{1}^\noprime\mu_{2}^\noprime} \Gamma^\Lambda(x'_{1},x'_{2}|x_1,x_2) \sigma^\alpha_{\mu_{1}^\noprime\mu'_{1}}\sigma^\beta_{\mu_{2}^\noprime\mu'_{2}}.
  \end{split}
\end{equation}
where $x'_{1}=(i,\omega'+\omega,\mu'_{1})$, $x'_{2}=(j,\omega'',\mu'_{2})$, $x_1=(i,\omega',\mu_{1}^\noprime)$, and $x_2=(j,\omega''+\omega,\mu_{2}^\noprime)$. Note that through its dependence on vertex functions, the correlator $\chi^{\alpha\beta,\Lambda}_{ij}(\omega)$ has acquired a $\Lambda$-dependence. While this expression is formulated for arbitrary frequencies $\omega$, it should be kept in mind that these are Matsubara frequencies defined on the imaginary frequency axes. Therefore, only the point $\omega=0$ corresponds to a physical quantity. Another physical observable can be obtained by integrating over frequencies in \eq{eq:realspacecorrel}, which yields the usual equal time (i.e., instantaneous) spin-spin correlator
\begin{equation}
    \langle S_i^\alpha S_j^\beta\rangle=\int \dd \omega \chi^{\alpha\beta}_{ij}(\omega).
\end{equation}
Since this additional frequency integration (which in numerical approaches is performed over a discrete mesh) introduces additional numerical errors, the static correlator in \eq{eq:realspacecorrel} is more often used in applications of the PFFRG.

By Fourier-transforming the static correlator into momentum space, one further obtains the static susceptibility
\begin{equation}
  \chi^\Lambda(\bm{k}) = \frac{1}{N} \sum\limits_{i,j} \e^{\i \bm{k}\cdot(\bm{r}_i-\bm{r}_j)} \chi^{\alpha\beta,\Lambda}_{ij}(\omega = 0).\label{eq:static_suscep}
\end{equation}
This quantity is of particular physical interest as it describes the magnetic response to an (infinitesimally small) static external magnetic field. Moreover, as it is defined in entire momentum space (i.e., it does not only correspond to the response to homogeneous magnetic fields but also to spatially varying ones) with this quantity one can identify the wave vectors of dominant magnetic fluctuations.

At this point, it is important to mention that, by definition, the PFFRG flow respects all symmetries of the initial spin Hamiltonian, in particular, $\chi^\Lambda(\bm{k})$ is invariant under the full space group of the underlying lattice. In magnetically ordered phases, however, time-reversal symmetry (and usually also lattice symmetries) are spontaneously broken. As the RG flow cannot enter phases with spontaneously broken symmetries, this will lead to an instability in the $\Lambda$-flow of $\chi^\Lambda(\bm{k})$ at a critical scale $\Lambda_c$, usually in the form of a divergence or kink. In this case, the wave vector $\bm{k}$ where $\chi^\Lambda(\bm{k})$ is maximal, provides the ordering wave vector of the respective magnetically ordered phase. Here, the implementation of the magnetic system as infinite with finite correlation length (see \cref{sec:spindep}) allows for accessing the precise location of incommensurate ordering vectors, e.g. for spin spirals, which would not be accurately resolvable with other boundary conditions~\cite{Reuther2011}. On the other hand, in spin liquid phases, where no spontaneous symmetry breaking occurs, a featureless flow of $\chi^\Lambda(\bm{k})$ down to the small $\Lambda$ limit is expected.

We note that, in principle one can also continue the flow into symmetry broken phases by including a suitable order parameter field~\cite{Salmhofer2004}. However, this requires an \emph{a priori} anticipation of the specific type of symmetry breaking to define this field. Furthermore, explicitly breaking symmetries on the Hamiltonian level may increase the numerical efforts enormously. To avoid both types of complications, applications of the PFFRG are usually performed without such symmetry breaking fields and, consequently, the flow has to be stopped when indications for a magnetic instability arise.

Through Kramers-Kronig relations, the static susceptibility $\chi^\Lambda(\bm{k})$ is also closely related to the dynamical spin structure factor $\mathcal{S}^{\alpha\beta}(\bm{k},\omega)$ that can be measured in neutron scattering experiments,
\begin{equation}
    \chi^{\Lambda=0}(\bm{k})=\int \text{d}\omega \frac{\mathcal{S}^{\alpha\beta}(\bm{k},\omega)}{\omega}.
\end{equation}
This correspondence allows one to perform direct theory-experiment comparisons based on the outcomes of PFFRG, which have been carried out very successfully in past applications~\cite{Balz2016,Chillal2020,Zivkovic2021}. Having numerical access to the system's momentum resolved spin fluctuations as contained in $\chi^\Lambda(\bm{k})$ is particularly important for magnetically disordered systems, such as quantum spin liquids. There, the precise distribution of signal in momentum space provides a superb characterization of the system's ground state magnetic properties which constitutes one of the key strengths of PFFRG.

In principle, by performing an analytic continuation of $\chi^{\alpha\beta,\Lambda}_{ij}(\omega)$  [\eq{eq:correlpffrg}] to the real frequency axis and additionally transforming this quantity to momentum space, one could even obtain the full {\it dynamical} spin structure factor $\mathcal{S}^{\alpha\beta}(\bm{k},\omega)$. However, an analytic continuation of numerical data is a long standing problem which so far has defied a satisfactory solution and, hence, this strategy of obtaining $\mathcal{S}^{\alpha\beta}(\bm{k},\omega)$ has until now not been further pursued.

\subsection{Single occupation constraint}
\label{sec:singleoccupation}
The pseudo-fermion representation necessitates the fulfilment of the single occupation per site contraint as given in \eqs{eq:numberconstraint}{eq:occupationconstraint} for \eq{eq:spinfermions} to be a faithful operator mapping.
As discussed in \Sec{sec:gaugefield}, a suitable $\SU(2)$ gauge field will act as a Lagrange multiplier enforcing the constraint.
In FRG, however, the inclusion of such a non-Abelian field complicates the flow equations further~\cite{Roscher2019} and, therefore, this has not been pursued to date.

At finite temperature, the inclusion of an imaginary chemical potential $\mu_{\mathrm{PF}}$, as initially put forward by Popov and Fedotov~\cite{popov_functional-integration_1988} projects out the contributions from unphysical pseudo-fermion sectors, as discussed further in \Sec{sec:finiteT}. In the limit $T\to0$, $\mu_{\mathrm{PF}}$ vanishes. As this limit, however, does not commute with the path integral, a vanishing chemical potential does only guarantee a fulfillment of the constraints on average [i.e. $\langle\cd_{i\downarrow}\cc_{i\downarrow}+\cd_{i\uparrow}\cc_{i\uparrow}\rangle=1$ instead of \eq{eq:numberconstraint}], as it implies half-filling of the particle-hole symmetric pseudo-fermionic system.

This average constraint was used in all $T=0$ implementations of PFFRG to date, motivated by initial studies that present physically correct results when compared to an exact implementation of the Popov-Fedotov scheme~\cite{Reuther2011,Roscher2019}. Another systematic treatment of the single occupation can be reached by implementing a level-repulsion term

\begin{equation}
  \mathcal{H}_\text{LR} = -A \sum_i \bm{S}_{i}^2,\label{eq:levelrepulsion}
\end{equation}
which for $A>0$ energetically favors the physical $S=1/2$ states, gapping out the unphysical $S=0$ sector of the pseudo-fermions, in the limit $A\to\infty$ leading to an exact fulfillment of the constraint. Large values of $A$ however spoil numerical stability, of the FRG flow, but up to this point it has been shown that the flow remains essentially unaffected by the level-repulsion~\cite{Kiese2020}.
Recent studies on small clusters, however, suggest that small size systems can be found, in which the constraint violation can spoil the results of PFFRG~\cite{schneiderTamingPseudofermion2022}.

\subsection{Generalizations}
\subsubsection{Arbitrary spin-length $S$}
\label{sec:largeS}

Although most interesting from the perspective of quantum fluctuations, the $S=1/2$ case of spin operators treated by pseudo-fermions is not generally applicable to model Hamiltonians of real materials, which often feature higher spin $S$ moments. The obvious way to extend \eqref{eq:spinfermions} would be to replace the Pauli-matrices $\sigma$ with their higher-spin counterparts, leading to a pseudo-fermionic representation comprising $2S+1$ flavors of particles per site~\cite{Liu-2010}. The occupation constraint then calls for a $1/(2S+1)$ filling to achieve one-particle per-site on average. The corresponding chemical potential is, however, not \emph{a priori} known and nor is particle-hole symmetry present in such a framework, thereby complicating the implementation of such a scheme.

Therefore, for PFFRG, as an alternative route, it has been put forward in \mRef{Baez2017} to introduce $2S$ replicas of spin $S=\nicefrac{1}{2}$ operators per site to express a single spin-$S$ operator at site $i$ as
\begin{equation}
  \bm{S}_i = \sum\limits_{\kappa=1}^{2S} \bm{S}_{i,\kappa}, 
\end{equation}
with $\kappa$ enumerating the different replicas. Introducing the pseudo-fermion mapping,~\eq{eq:spinfermions}, for each constituent spin, we find an alternative pseudo-fermionic representation
\begin{equation}
  \bm{S}_i^\alpha =  \frac{1}{2} \sum\limits_{\kappa,\mu',\mu} \cd_{\vphantom{\nu}\mu'\kappa} \sigma^\alpha_{\mu'\mu} \cc_{\mu\kappa},\label{eq:spinsfermions}
\end{equation}
now subject to the constraint that at each lattice point, the system has to be at half-filling and simultaneously the total spin length must be maximized.

While the first condition can again be implemented by means of an average projection scheme, the second needs a bit more care: In addition to the physical sector with spin-length $S$, we have introduced $S$~$(S-1/2)$ unphysical sectors with lower spin for $2S$ being even (odd). To minimize their contributions in the calculations, a modified version of \eq{eq:levelrepulsion} as a level-repulsion term~\cite{Baez2017}
\begin{equation}
  \mathcal{H}_\text{LR} = -A \sum_i \left(\sum\limits_{\kappa=1}^{2S} \bm{S}_{i,\kappa}\right)^2 \label{eq:energypenalty}
\end{equation}
is added to the Hamiltonian.

For $A>0$, this will energetically favor the case where the maximal spin length $S$ per site is achieved, while gapping out sectors with lower spin value. In practical calculations, however, the inclusion of this term has been shown to have negligible effects, as the spin replicas already tend to form the largest spin length multiplets~\cite{Baez2017}.

\subsubsection{Flow equations at arbitrary $S$}
The modifications needed to implement the replica scheme in the PFFRG flow equations \eqs{eq:multilocalseflow}{eq:multilocalflow} are in the form of additional factors, which do not change the general structure of the equations.

Firstly, we note that in \eq{eq:spinsfermions} every site index $i$ is accompanied by an additional flavor index $\kappa$. As the $\U(1)$ gauge symmetry of the pseudo-fermions (cf. \Sec{sec:pseudofermiongaugesymm}) acts on every replica of the $S=\nicefrac{1}{2}$ fermions separately, we find a locality not only in the site index as discussed in \Sec{sec:localU1}, but also for the flavor index.

Secondly, considering the initial conditions of the vertex in \eq{eq:pffrginitalcond}, we see that these are agnostic to the flavor index. Combined with the flavor index locality, this leads to the vertex function staying completely independent of the flavor index during the flow.
Therefore, any summation over flavor indices is trivially carried out, contributing a factor of $2S$ in the flow equations wherever there is an internal site summation, due to the intimate connection between site and flavor indices discussed above. Therefore, the only changes to the flow equations are factors $2S$ for the site summation in \eq{eq:multilocalseflow} and the RPA-like contribution in \eq{eq:multilocalflow}.

Since, for increasing $S$, quantum fluctuations become less pronounced and the tendency towards classical long-range order is enhanced, this term can be identified as the one inducing such a phase transition.

\subsubsection{Equivalence of $S\to\infty$ to Luttinger-Tisza}

In the limit $S\to\infty$, this notion becomes particularly clear~\cite{Baez2017}. The only surviving term in the two-particle vertex flow then is the non-local RPA contribution in the $t$-channel. Therefore, the only frequency structure of the initially frequency independent vertex will be in the corresponding transfer frequency.

Introducing the shorthand notation
\begin{equation}
  \tilde{\Gamma}^{s/d,\Lambda}_{ij}(t) = \frac{1}{2S} \Gamma^{s/d,\Lambda}_{ij}(s,t,u),\label{eq:vertexshort}
\end{equation}
for the spin and density vertices introduced in \Sec{sec:spindep}, the flow equation for the latter simplifies to 
\begin{equation}
  \begin{split}
  \tilde{\Gamma}^{d,\Lambda}_{i_1i_2}(t) &= \frac{1}{\pi} \int\dd\omega \sum\limits_j \tilde{\Gamma}^{d,\Lambda}_{i_1j_{\vphantom{2}}}(t)\tilde{\Gamma}^{d,\Lambda}_{j_{\vphantom{2}}i_2}(t) \\ \times&\left(S_\text{kat}^\Lambda(\omega)G^\Lambda(\omega+t) + (\omega \leftrightarrow \omega+t)
  \right),
  \end{split}
\end{equation}
which stays finite due to the rescaling by $2S$ in \eq{eq:vertexshort} and decouples from the spin vertex flow. Therefore, the initially vanishing density vertex does not become finite during the flow. Similarly, the self-energy will only couple to the density vertex and therefore identically vanish.

Hence, the flow equation for the remaining spin vertex assumes a tractable form, when using the step-like regulator introduced in \Sec{sec:stepreg}
\begin{equation}
  \begin{split}
  \tilde{\Gamma}^{s,\Lambda}_{i_1i_2}(t) &= \frac{1}{\pi} \int\dd\omega \sum\limits_j \tilde{\Gamma}^{s,\Lambda}_{i_1j_{\vphantom{2}}}(t)\tilde{\Gamma}^{s,\Lambda}_{j_{\vphantom{2}}i_2}(t) \\ \times &\left(\frac{\delta(|\omega|-\Lambda)}{\omega}\frac{\theta(|\omega+t|-\Lambda)}{\omega+t} + 
  (\omega \leftrightarrow \omega+t)
  \right),
  \end{split}
\end{equation}
which allows for an anlytical solution of the frequency integration. After Fourier transformation of the spatial dependence, assuming a Bravais-lattice for convenience, the flow equation reads
\begin{equation}
  \tilde{\Gamma}^{s,\Lambda}(\bm{k},t) = \frac{2}{\pi\Lambda(\Lambda+t)}  \tilde{\Gamma}^{s,\Lambda}(\bm{k},t)^2,\label{eq:largesspin}
\end{equation}
which is amenable to an analytic solution~\cite{Buessen2018a}
\begin{equation}
  \tilde{\Gamma}^{s,\Lambda}(\bm{k},t) = \frac{J(\bm{k})/4}{1+\frac{J(\bm{k})}{2\pi\Lambda} \ln(1+\frac{t}{\Lambda})},
\end{equation}
where $J(\bm{k})$ is the Fourier transform of the bare Heisenberg interaction.

This flow features a leading divergence at frequency $t=0$ for 
\begin{equation}
  \Lambda_c = -\frac{\min_{\bm{k}} J(\bm{k})}{2\pi},\label{eq:lambdacritclass}
\end{equation}
i.e., the spin vertex diverges at the point in reciprocal space, where the Fourier transform of the initial interaction is most negative. Following \eq{eq:static_suscep} this feature will also appear in the static susceptibility, implying long-range order governed by this wave-vector.

Therefore, PFFRG in the $S\to\infty$ limit is equivalent to the Luttinger-Tisza method~\cite{Luttinger1946,Lyons1960,Kaplan2007}, where the same finding is true. This means, on Bravais lattices, PFFRG reproduces the exact classical ground-state~\cite{Baez2017}, whereas it is equivalent to a classical $\mathcal{O}(N\to\infty)$ mean-field in all other cases~\cite{Muller2023}.

In case of a multi-site basis, the minimum in \eq{eq:lambdacritclass} has to be taken over the eigenvalues of the matrix valued Fourier transform in sublattice space.

\subsubsection{Symmetry enhanced $\SU(N)$}

A second generalization of the pseudo-fermion approach is designed to enhance quantum fluctuations in contrast to the large-$S$ generalization in the previous section which approached classical behavior. To this end, the $\SU(2)$ symmetry group of spins is promoted to $\SU(N)$~\cite{Buessen2018a,Roscher2018a}, effectively allowing for more quantum degrees of freedom to the spin operators. This generalization is not uniquely defined, and several implementations with possibly different ground-state properties exist~\cite{Arovas1988}, but all have in common that quantum fluctuations are enhanced for $N>2$, rendering them dominant in the $N\to\infty$ limit.

Following \Refs{Buessen2018a}{Roscher2018a}, we introduce the generalization by introducing the generators $T^\alpha$ of $\SU(N)$ in the fundamental representation of this group, with $\alpha \in {1,2,\dots,N^2-1}$. These hermitian, traceless $N\times N$ matrices follow the $\mathfrak{su}(N)$ Lie-algebra
\begin{equation}
  \commutator{T^\alpha}{T^\beta} = \i \sum\limits_{\gamma=1}^{N-1} f_{\alpha\beta\gamma} T^\gamma,
\end{equation}
where $f$ are the structure constants of the group. Replacing the Pauli matrices in \eq{eq:spinfermions}, we find the fermionic decomposition of $\SU(N)$ spins to be\footnote{For the $\SU(2)$ case, $T^\alpha = 1/2 \sigma^\alpha$.}
\begin{equation}
  S^\alpha = \sum\limits_{\mu,\nu=1}^N \cd_\mu T^\alpha_{\mu\nu} \cc_\nu,
\end{equation}
where we have introduced $N$ flavors of pseudo-fermions. To render the operator mapping exact, we additionally have to introduce the half-filling per site constraint 
\begin{equation}
  \sum\limits_{\mu=1}^{N} \cd_\mu\cc_\mu = \frac{N}{2},\label{eq:sunconstraint}
\end{equation}
which immediately constrains this generalization to even $N$. In the $T\to0$ limit, \eq{eq:sunconstraint} can be treated by an average projection scheme, as discussed in \cref{sec:singleoccupation}. At finite temperatures, a Popov-Fedotov like scheme, employing imaginary chemical potentials is possible, however, it necessitates introducing distinct potentials for every fermion flavor~\cite{Kiselev2001}.

Employing a parameterization in terms of spin and density like vertex components 
\begin{equation}
  \begin{split}
  \Gamma_{=,i_1i_2}(1',2';1,2) &= \Gamma_{\mathrm{s},i_1i_2}(\omega'_{1}, \omega'_{2}| \omega^\noprime_{1}, \omega^\noprime_{2}) T^\alpha_{\mu'_{1}\mu^\noprime_{1}} T^\alpha_{\mu'_{2}\mu^\noprime_{2}} \\ &+ \Gamma_{\mathrm{d},i_1i_2}(\omega'_{1}, \omega'_{2}| \omega^\noprime_{1}, \omega^\noprime_{2}) \delta_{\mu'_{1}\mu^\noprime_{1}} \delta_{\mu'_{2}\mu^\noprime_{2}},\label{eq:spindensparamN}
  \end{split}
\end{equation} 
for Heisenberg-like interactions independent of $\SU(N)$ components, we immediately see that the general structure of the flow equations will remain unchanged by this generalization, which amounts to a mere change of prefactors. For the full set of $\SU(N)$ equations we refer the reader to \Refs{Buessen2018}{Buessen2019b}. In contrast, the symmetries of Greens' and subsequently vertex functions discussed in \Sec{sec:pseudofermionsymmetries} do not all survive the generalization. By promoting $\SU(2)$ to $\SU(N)$, the pseudo-fermion mapping naturally looses its $\SU(2)$ gauge symmetry. While the $\U(1)$ subgroup still remains intact, rendering the Green's functions multilocal, local particle-hole transformations based on the $\mathds{Z}_2$ subgroup of $\SU(2)$ are no longer present. Inspecting \tab{tab:naturalsymm}, this especially affects the mapping between $s$- and $u$-channel, which become independent for $N>2$.

Indeed, in the $N \to \infty$ limit, the only contribution to the two-particle vertex is the $u$-channel diagram of the spin vertex, which does not generate non-local terms. Thus, the susceptibility will remain finite throughout the whole RG flow, while the vertex itself will diverge, signalling a transition into a spin rotationally invariant ordered state (e.g., a valence bond crystal or a spin liquid), but not a magnetically ordered phase~\cite{Buessen2018a,Roscher2018a}. This reproduces the analytical mean-field results for $N\to\infty$, which are exact in this limit. The Katanin truncation is a vital ingredient in making this connection, \emph{a posteriori} rationalizing the necessity to include these corrections to obtain magnetically disordered ground-states.

\section{Extension to finite temperature}
\label{sec:finiteT}

\subsection{Motivation}
 \label{sec:finiteT-Motivation}

There are several reasons to study quantum spin Hamiltonians like \eqref{eq:mod-ham} also at finite temperatures $T>0$: 
(i) First, this is required if one desires quantitative modelling of experiments which are never conducted at $T=0$. Note, however, that the assumption of thermal equilibrium is mostly appropriate for solid-state applications, whereas experiments on cold-atom implementations of spin systems often operate with quench protocols and thermalization is not necessarily guaranteed within the available timescales. 
(ii) Second, from a theoretical point of view it is significant that spin $S=1/2$ Hamiltonians like the spin model \eqref{eq:mod-ham} do not come with a small parameter and this does not change if spins are represented by fermionic partons. However, it is well known that the smallness of the parameter $J/T$ can control perturbative expansions \cite{auerbachInteractingElectrons1994,rohringerRevModPhys.90.025003}, for example the high-temperature series expansion for static properties \cite{elstnerFiniteTemperature1993} or the pseudo-fermionic diagrammatic Monte Carlo technique \cite{kulaginBoldDiagrammatic2013,kulaginBoldDiagrammatic2013a,Huang2016}. Recently, this type of control has also been implemented for the PFFRG, see \cref{sec:Popov}. Finally, from a practical point of view, finite temperatures are associated with discrete Matsubara frequencies which are easier to handle numerically than the continuous frequencies at $T=0$.
(iii) Third, on top of quantum fluctuations present at $T=0$, turning to $T>0$ switches on thermal fluctuations which might have interesting consequences especially when they are competing. For example, it is well known that in one- and two-spatial dimensions thermal fluctuations melt any ordered phase with a spontaneously broken continuous symmetry \cite{merminAbsenceFerromagnetism1966}. For discrete symmetries or in three spatial dimensions, finite critical temperatures $T_c$ mark the boundary between a fluctuation dominated disordered regime and the ordered regime which survives to finite $T$. Moreover, if a zero temperature quantum phase transition \cite{sachdevQuantumPhase2011} driven by quantum fluctuations is present, the competition of quantum- and thermal fluctuations in the vicinity of the critical point lead to non-trivial scaling and power-laws with respect to $T$ that allow to infer information on the experimentally inaccessible limit at $T=0$, see Ref.~\cite{scheieWitnessingQuantum2023} for a recent experimental example on the triangular lattice. In summary, it is a highly relevant task to access static and dynamic properties of spin systems also at finite temperature. In the following, we review the role of pseudo-fermionic FRG-based methods in this endeavor. 

\subsection{Popov-Fedotov trick for PFFRG}
\label{sec:Popov}
As emphasized in \cref{sec:auxiliary_fermions}, the PF representation of spin operators which forms the basis of PFFRG, introduces unphysical states in the Hilbert space. It has been argued that these additional $S=0$ states reside at excited energies above the ground state and at $T=0$ cannot thus affect the physical observables computed via the PF representation with PFFRG. On the one hand, recently found explicit counterexamples \cite{schneiderTamingPseudofermion2022} question this lore at $T=0$. On the other hand, for $T>0$, where unphysical excited states are thermally populated, it is clear that PFFRG is certainly inapplicable and additional measures must be taken in order to project out the non-magnetic spin-$0$ PF states. 

One way to achieve this is through the so-called \textit{Popov-Fedotov trick} \cite{popov_functional-integration_1988}: An imaginary valued chemical potential $\mu_{\text{Popov}}$ is added to the PF Hamiltonian \cref{eq:heisenbergfermion}, i.e., $\widetilde{\mathcal{H}} \to \widetilde{\mathcal{H}}_{\text{Popov}} = \widetilde{\mathcal{H}} + \mu_{\text{Popov}} \sum_j(n_{j,\uparrow} + n_{j,\downarrow} - 1)$, where $n_{j,\sigma} = c^{\dag}_{j,\sigma} c_{j,\sigma}$ is the PF density at site $j$. The value of $\mu_{\text{Popov}} = \i \pi T/2$ is then chosen such that the unphysical contributions to the partition function cancel out when calculating the partition function. Thermal expectation values are also cleared of non-magnetic contributions and should therefore resemble their counterparts in the original spin Hamiltonian.

While the Popov-Fedotov trick is routinely employed in PF based diagrammatic Monte Carlo \cite{kulaginBoldDiagrammatic2013,kulaginBoldDiagrammatic2013a,Huang2016}, the numerical implementation of the PF trick in PFFRG was pioneered only recently in Ref.~\cite{schneiderTamingPseudofermion2022} and comes at a price. Since $\mu_{\text{Popov}}$ is purely imaginary, it will change sign under the anti-unitary time-reversal transformation and also hermitian conjugation. Furthermore, the term involving $\mu_{\text{Popov}}$ is also odd under the global particle-hole symmetry. Fortunately, pairwise combinations of the above-mentioned transformations remain a symmetry of the PF Hamiltonian $\widetilde{\mathcal{H}}_{\text{Popov}}$ and most symmetry constraints of the vertices discussed above remain intact \cite{schneiderTamingPseudofermion2022}. The numerical effort for using the Popov-Fedotov trick within PFFRG increases the runtime and memory requirements by roughly a factor of four.

An important qualitative difference to $T=0$ PFFRG is found in the behavior of the flow. For $T>0$, the flowing quantities are generally smooth and convergent as $\Lambda \to 0$ for any parameter choice. This is related to the presence of a finite smallest Matsubara frequency $\pi T$ which avoids the zero-temperature pole in the bare fermionic propagator $~1/\i\omega_n$. As a consequence, at $T>0$, the PFFRG with the Popov-Fedotov trick generally produces quantitative results for desired properties which are expected to be  accurate for large $T/J$. In \cref{fig:HeisenbergDimerPopov}(top), these end-of-flow results for the static spin correlation functions for the AFM Heisenberg dimer $\mathcal{H}=J \mathbf{S}_1 \cdot \mathbf{S}_2$ are presented (orange markers) and compared to exact results (solid black lines). Most importantly, the FRG data is in quantitative agreement with the exact results for intermediate and high temperatures ($T/J \gtrsim 0.3$). If the Popov-Fedotov trick is left out, the application of both the ED as well as FRG approach to $\widetilde{\mathcal{H}}$ (dashed black lines and blue symbols, respectively) disagree with the true spin results due to the contribution of unphysical states. As a signature of the contribution of unphysical states, the equal-time correlator $C_{i} = 4\langle \tilde{S}^{z}_i \tilde{S}^{z}_i \rangle$ (bottom panel) should equal unity in an exact representation of the spin-$1/2$ algebra \cite{Ritter2022}. Without the Popov-Fedotov trick, $C_{i}$ approaches $1/2$ at high-temperatures where both physical and unphysical states contribute equally to the partition function. In contrast, once the PF trick is employed, $C_{i}$ correctly converges to one at least for large enough $T$. At low $T$, where the truncation of the FRG flow equation degrades the quality of the results, the spin constraint ceases to remain exactly fulfilled and, instead, decreases from one. Simultaneously, the PFFRG results for the dimer correlation functions start to deviate from the ED result. This issue will be further discussed in \cref{subsec:lowTProblem}.

\begin{figure}
    \centering
    \includegraphics[width = \linewidth]{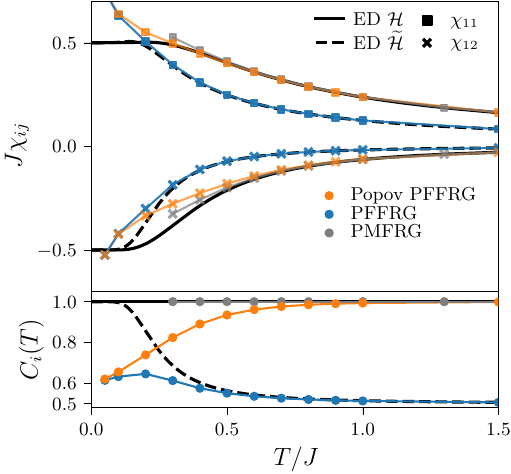}
    \caption{\textbf{Finite-$T$ results for the Heisenberg dimer.} Black lines indicate exact results for the unconstrained pseudofermion Hamiltonian ($\widetilde{\mathcal{H}}$) and the original spin model ($\mathcal{H}$). The equal-time correlator $C_{i} = 4\langle S^{z}_i S^{z}_i \rangle$ should equal unity in an exact representation of the spin-$\tfrac{1}{2}$ algebra. Data reproduced from \cite{schneiderTamingPseudofermion2022} and \cite{Niggemann2020}.}
    \label{fig:HeisenbergDimerPopov}
\end{figure}

\subsection{Pseudo-Majorana FRG}
\label{subsec:PMFRG}
\subsubsection{Pseudo-Majorana representation for spin $1/2$} \label{sec:pm-representation}
Instead of relying on the Popov-Fedotov trick to project out unphysical states, it is also possible to work with an alternative fermionic representation of spin $S=1/2$ devoid of unphysical states altogether. This however requires real Majorana fermions ($\eta$) instead of complex Abrikosov fermions ($c$,$c^\dagger$). 
The pseudo-Majorana (PM) representation \cite{Martin1959,TsvelikMajorana,FuMajorana,schaden2023bilinear} for spin $S=1/2$ operators at site $j$ is defined in terms of three Majorana fermion operators $\eta_j^\alpha$ with $\alpha={x,y,z}$,
\begin{equation}
    \bar{S}^x_j = -\i \eta^y_j \eta^z_j \text{,} \quad  \bar{S}^y_j = -\i \eta^z_j \eta^x_j \text{,} \quad  \bar{S}^z_j = -\i \eta^x_j \eta^y_j.\label{eq:MajoranaRep}
\end{equation}
The Majorana operators fulfil anti-commutation relations $\{\eta^\alpha_i,\eta^\beta_j\} = \delta_{ij}\delta_{\alpha \beta}$ from which we derive the normalization $(\eta_j^\alpha)^2=1/2$. Furthermore, it holds $(\eta_j^\alpha)^\dagger = \eta_j^\alpha$. From these rules it follows that the spin algebra is fulfilled by the operators $\bar{S_j^\alpha}$ and the spin length is $S=1/2$. As a consequence, no eigenstates at energies different from the spin eigenstates can appear.

As two Majoranas $\eta,\eta^\prime$ can be combined to form a complex fermion via $c=(\eta-\i\eta^{\prime})/\sqrt{2},\;c^\dagger=(\eta+\i\eta^{\prime})/\sqrt{2}$, the Hilbert space dimension per Majorana is $\sqrt{2}$ and the total dimension of the PM Hilbert space for $N$ spins is thus  $2^{3N/2}$, enlarged in comparison to the dimension $2^N$ of the spin system by a factor $2^{N/2}$. This factor is explained by an unphysical degeneracy of each spin-system's eigenstate required by the local $Z_2$ gauge symmetry of \cref{eq:MajoranaRep}, $\eta^\alpha_j \rightarrow - \eta^\alpha_j$ for all $\alpha$. More precisely, consider the operator $\Theta_{j} = \i \eta^x_j \eta^y_j \eta^z_j$ which commutes with all PM spin operators $\bar{S}_{j^\prime}^{\alpha}$, for both $j=j^\prime$ and $j\neq j^\prime$. Thus, the set $\{\Theta_1,\Theta_2,...,\Theta_N\}$ contains $N$ constants of motion. However, as this set consists of Majorana operators, it is the fermion parities $p_{(i,j)}\equiv2\i\Theta_{i}\Theta_{j}=\pm1$ for arbitrary but fixed pairing of sites $(i,j)$ that provide the $N/2$ eigenvalues $\pm1$ which distinguish the $2^{N/2}$ degenerate states. 

The unphysical degeneracy of eigenstates in the PM representation may seem problematic at first sight. However, for a correlation-function based method like the FRG, it is almost invisible. Indeed, consider the relation between the physical partition function $\mathcal{Z}$ and the pm partition function, $\mathcal{\bar{Z}}=2^{N/2}\mathcal{Z}$, by the aforementioned degeneracy. Thus, the free energies per site 
\begin{equation}
f=\bar{f}+\frac{T}{2}\log\left(2\right)   \label{eq:pm-f} 
\end{equation}
are simply related by a temperature dependent but trivial offset.

An expectation value of an arbitrary (time-evolved or composite) spin operator denoted by $O$ in a state $\rho$ of the spin system is defined as follows, 
\begin{equation}
    \left\langle O\right\rangle =\frac{\mathrm{tr}O\rho}{\mathrm{tr}\rho}=\frac{1}{\sum_{\sigma}\left\langle \sigma|\rho|\sigma\right\rangle }\sum_{\sigma,\sigma^{\prime}}\left\langle \sigma|O|\sigma^{\prime}\right\rangle \left\langle \sigma^{\prime}|\rho|\sigma\right\rangle,
\end{equation} 
where $ | \sigma \rangle$ denotes a basis of the spin system. In the PM representation, the Hilbert space is enlarged by a factor $2^{N/2}$ that encodes the different parity configurations $p_{(i,j)}$. However, the thermal PM state $\e^{-\beta\bar{H}}$ takes the form $\bar{\rho}=\rho\otimes1_{p}$ since $\bar{H}$ commutes with all $p_{(i,j)}$. The same is true for non-thermal states that only depend on the PM spin operators $\{\bar{S}_{i}^{\alpha}\}_{i,\alpha}$. We split off the trace over parity configurations $\mathrm{tr}=\mathrm{tr}_{\sigma}\mathrm{tr}_{p}$ and note that spin observables in pm representation $\bar{O}$ also do not depend on the parity configuration. We use $\mathrm{tr}_{p}\bar{\rho}=2^{N/2}\rho$ which cancels between numerator and denominator to conclude that any expectation values and correlation functions take their physical value in the PM representation \cite{Shnirman2003,Schad2015,Schad2016}
\begin{equation}
\left\langle \bar{O}\right\rangle =\frac{\mathrm{tr}_{\sigma}\mathrm{tr}_{p}\bar{O}\bar{\rho}}{\mathrm{tr}_{\sigma}\mathrm{tr}_{p}\bar{\rho}}=\frac{\mathrm{tr}_{\sigma}O\mathrm{tr}_{p}\bar{\rho}}{\mathrm{tr}_{\sigma}\mathrm{tr}_{p}\bar{\rho}}=\left\langle O\right\rangle .    
\end{equation}
We conclude this exposition on the PM representation by writing the Heisenberg spin Hamiltonian in PM representation, where the sum is over bonds,
\begin{equation}
    \bar{H}=\sum_{(ij)}\left(-J_{ij}\right)\left(\eta_{i}^{y}\eta_{i}^{z}\eta_{j}^{y}\eta_{j}^{z}+\eta_{i}^{x}\eta_{i}^{z}\eta_{j}^{x}\eta_{j}^{z}+\eta_{i}^{x}\eta_{i}^{y}\eta_{j}^{x}\eta_{j}^{y}\right). \label{eq:H_Heisenberg_pm}
\end{equation}
Similar to the PF representation, we arrived at a purely interacting fermionic Hamiltonian, now written in terms of Majorana operators. In the following, we present the diagrammatic PMFRG approach to this problem. To keep the notation light, we limit ourselves to Heisenberg systems, but generalizations to other bi-linear spin couplings, e.g.~of XXZ-type \cite{sbierskiMagnetismTwodimensional2023}, are straightforward.

\subsubsection{Pseudo-Majorana correlators and one-line irreducible vertices} \label{sec:pm-correlators}
We define the temporal Fourier transform of the two-point imaginary time-ordered PM correlation functions \cite{Niggemann2021}
\begin{eqnarray}
    G(1,2)&=&\int_{0}^{\beta}\mathrm{d}\tau_{1,2} \e^{\i\omega_{1}\tau_{1}+\i\omega_{2}\tau_{2}}\left\langle \mathrm{T}_\tau \eta_{j_{1}}^{\alpha_{1}}(\tau_{1})\eta_{j_{2}}^{\alpha_{2}}(\tau_2)\right\rangle \\
    &\equiv&\beta\delta_{\omega_{1}+\omega_{2},0}G_{j_{1}j_{2}}^{\alpha_{1}\alpha_{2}}(\omega_{2}) 
\end{eqnarray}
where $1 \equiv (j_{1}, \alpha_{1}, \omega_{1})$. The four-point function is defined analogously, $G_{j_{1}j_{2}j_{3}j_{4}}^{\alpha_{1}\alpha_{2}\alpha_{3}\alpha_{4}}(\omega_{1},\omega_{2},\omega_{3})\!=\!\int_{0}^{\beta}\!\!\mathrm{d}\tau_{1,2,3}\e^{\i\omega_{1}\tau_{1}+\i\omega_{2}\tau_{2}+\i\omega_{3}\tau_{3}} \!\! \left\langle \! \mathrm{T}_\tau \eta_{j_{1}}^{\alpha_{1}}(\tau_{1})\eta_{j_{2}}^{\alpha_{2}}(\tau_{2})\eta_{j_{3}}^{\alpha_{3}}(\tau_{3})\eta_{j_{4}}^{\alpha_{4}}\!\right\rangle$. Here, $\omega_1$ is shorthand for $\omega_{m_1}= \pi T(2m_1+1),\,m_1\in\mathbb{Z}$. The Heisenberg imaginary time evolution is $\eta(\tau)= \e^{\bar{H}\tau}\eta \e^{-\bar{H}\tau}$ and the Fourier transform convention is such that the $n$ frequencies of the $n$-point vertex sum to zero.

An important consequence of the PM representation is that we are dealing with only one type of operator, e.g.,~$\eta$ instead of $c,c^\dagger$. This means that for imaginary time-ordered correlation functions any pair of index tuples $1,2,...$ can be exchanged for a minus sign. This antisymmetry, in conjunction with the hermitian conjugation property $\left\langle O\right\rangle =\left\langle O^\dagger \right\rangle ^{\star}$ ensures that $G(1,2)\equiv -\i g(1,2) \in \i\mathbb{R}$ and $G\left(1,2,3,4\right)\in\mathbb{R}$.

The \emph{local} $\mathbb{Z}_2$ gauge symmetry of the PM representation $\eta_{i}^{\alpha}\rightarrow-\eta_{i}^{\alpha}\;\forall \alpha\in\{x,y,z\}$ ensures that each non-vanishing PM correlator is bi-local which means that each site-index appears at least twice. In the PM representation, time-reversal symmetry is implemented as complex conjugation $\i\rightarrow -\i$ which indeed flips all spin operators in Eq.~\eqref{eq:MajoranaRep} and implies that $g(1,2)$ and $G(1,2,3,4)$ are invariant under multiplying all frequency arguments by $-1$. Spin rotation symmetries act on the vector of Majorana operators $(\eta^x,\eta^y,\eta^z)^\mathrm{T}$ as on the vector of spin operators $(S^x,S^y,S^z)^\mathrm{T}$ \cite{FuMajorana}. If present, these symmetries further restrict the flavor combinations $\alpha_{1,2,...}$ of non-vanishing correlation functions, for details see Ref.~\cite{Niggemann2021}. For the Heisenberg case, the above symmetries lead to the following parameterization of the tw{o-point function,
\begin{equation}
g(1,2)=\beta\delta_{\omega_1+\omega_2,0}\delta_{j_1,j_2}g_{j_1}(\omega_2),\;g_{j}(\omega)=\frac{1}{\omega+\gamma_{j}(\omega)}
\end{equation}
with $\gamma_{j}(\omega)=-\gamma_{j}(-\omega)\in\mathbb{R}$.
The connected part of the four-point correlation functions $G(1,2,3,4)$ are linked to the one-line irreducible vertices $\Gamma(1,2,3,4)$ by the tree expansion. For the Heisenberg case, there are only three independent non-zero vertices, which, without loss of generality, can be defined as
\begin{align}
V_{ij}^{a}(s,t,u)&=&\beta\delta_\omega\Gamma(ix\omega_{1},ix\omega_{2},jx\omega_{3},jx\omega_{4}),\\
V_{ij}^{b}(s,t,u)&=&\beta\delta_\omega\Gamma(ix\omega_{1},ix\omega_{2},jy\omega_{3},jy\omega_{4}),\\
V_{ij}^{c}(s,t,u)&=&\beta\delta_\omega\Gamma(ix\omega_{1},iy\omega_{2},jx\omega_{3},jy\omega_{4}),
\end{align}
where $\delta_\omega \equiv \delta_{\omega_{1}+\omega_{2}+\omega_{3}+\omega_{4},0}$ and we defined the bosonic frequencies as $s=\omega_{1}+\omega_{2}$, $t=\omega_{1}+\omega_{3}$ and $u=\omega_{1}+\omega_{4}$. The most important frequency symmetries of the $V$, which allow to focus on non-negative $s,t,u$ are
\begin{eqnarray}
V_{ij}^{a,b,c}\left(s,t,u\right)&=&V_{ij}^{a,b,c}\left(-s,t,u\right),\\
V_{ij}^{a,b,c}\left(s,t,u\right)&=&V_{ji}^{a,b,c}\left(s,-t,u\right),\\
V_{ij}^{a,b,c}\left(s,t,u\right)&=&V_{ji}^{a,b,c}\left(s,t,-u\right),
\end{eqnarray}
and the exchange symmetries between $u$ and $t$ further simplify the numerical effort,
\begin{eqnarray}
V_{{\color{blue}i}j}^{a,b}\left(s,t,u\right)&=&-V_{ij}^{a,b}\left(s,u,t\right),\\
V_{{\color{blue}i}j}^{c}\left(s,t,u\right)&=&\left[-V_{ij}^{a}+V_{ij}^{b}+V_{ij}^{c}\right]\left(s,u,t\right).
\end{eqnarray}
As in PFFRG, spatial symmetries, both point-group and translation, drastically reduce the number of vertices in an actual calculation. 

Finally, we provide the main observables of interest for a spin system in terms of PM quantities. First, we remind the reader that the free energy in \cref{eq:pm-f} provides access to thermodynamic quantities like the energy per site $e =  f - T \frac{df}{dT} $, the specific heat $c = \frac{de}{dT}$ and even the entropy $s = (e -f) /T $, which is often challenging to obtain from other methods. Moreover we write the dynamic spin susceptibility on the Matsubara axis in terms of PM propagators and vertices, again for the Heisenberg case,
\begin{align}
\chi_{ij}^{zz}(\i\nu)=	\int_{0}^{\beta}d\tau\,\e^{\i\nu\tau}\left\langle S_{i}^{z} (\tau)S_{j}^{z}\right\rangle \\
	=\delta_{ij}T\sum_{\omega}g_{i}\left(\omega\right)g_{i}(\omega-\nu)  \label{eq:pm-chi}\\ 
	+	\, T^{2}\,\sum_{\omega,\omega^{\prime}} \, g_{i}\left(\omega\right)g_{i}\left(\omega-\nu\right)g_{j}\left(\omega^{\prime}+\nu\right)g_{j}\left(\omega^{\prime}\right)\nonumber\\
 \times V_{ij}^{c}\left(\nu,\omega-\omega^{\prime}-\nu,\omega+\omega^{\prime}\right).\nonumber
\end{align}

\subsubsection{Pseudo-Majorana path integral and FRG flow equations} 
Given the fact that Majorana operators and complex fermionic operators are related by a unitary rotation it is not surprising that there exists a Grassmann field path integral for the partition function of interacting Majorana systems. Calling the field $\zeta_a$, its action reads \cite{ shankarEqualityBulk2011, Nilsson2013, schadMajoranaRepresentation2016}
\begin{equation}
    S[\{\zeta_{a}\}_{a}]=\int_{0}^{\beta}d\tau\,\Bigl(\sum_{a}\frac{1}{2}\zeta_{a}(\tau)\partial_{\tau}\zeta_{a}(\tau)+H\left[\zeta_{a}(\tau)\right]\Bigr), 
\end{equation}
where, just like in the complex fermion case, each occurrence of $\eta_a$ in the Hamiltonian $H$ is to be replaced by $\zeta_a(\tau)$. With a sufficiently general formulation of the FRG at hand \cite{Kopietz2010}, it is a straightforward task to derive flow equations for the interaction correction of the free energy, self-energy and 4-point vertex \cite{Niggemann2021}. Since there is just a single field-type ($\zeta$), the corresponding diagrams do not feature incoming and outgoing lines and all vertices are fully antisymmetric under exchanges of all legs. Likewise, the $s,t$ and $u$ channels in the vertex flow equations are related to each other by simple permutations of indices.

In the following we will express these flow equations directly in terms of the PM quantities defined in the previous \cref{sec:pm-correlators}. In analogy to PFFRG, we use a multiplicative frequency cutoff, which in light of the discrete Matsubara frequencies is chosen to be smooth, $G_0(\omega) \rightarrow G_0^\Lambda(\omega) = G_0(\omega) \vartheta_\Lambda(\omega)$. A standard choice is a Lorentzian, $\vartheta_{\Lambda}(\omega)=\omega^{2}/(\omega^{2}+\Lambda^{2})$ with $\partial_{\Lambda}\vartheta_{\Lambda}(\omega)=-2\omega^{2}\Lambda\left(\omega^{2}+\Lambda^{2}\right)^{-2}$. 

We first give the initial conditions for $\Lambda \rightarrow \infty$. The free PM energy per site $\bar{f}$ starts at its non-interacting value $-3T/2\,\mathrm{log}2$ and the vertex $V^c_{ij}(s,t,u)$ initially is $-J_{ij}$. All other vertices and the self energy vanish initially. The flow equations for the PM free energy per site and the self-energy are
\begin{eqnarray}
    \partial_{\Lambda}\bar{f}^\Lambda&=&\frac{-3T}{N}\sum_{k}\sum_{\Omega>0}\frac{\partial_{\Lambda}\vartheta_{\Lambda}(\Omega)}{\vartheta_{\Lambda}(\Omega)}\gamma_{k}^{\Lambda}(\Omega)g_{k}^{\Lambda}(\Omega),\label{eq:pm-f-flow}\\
    \partial_{\Lambda}\gamma_{i}^{\Lambda}(\omega)&=&-T\sum_{\Omega>0}\sum_{k}\left[g_{{k}}^{\Lambda}(\Omega)\right]^{2}\frac{\Omega\partial_{\Lambda}\vartheta_{\Lambda}\left(\Omega\right)}{\vartheta_{\Lambda}^{2}\left(\Omega\right)}\\
    &&\times\left[V_{ki}^{a,\Lambda}+2V_{ki}^{b,\Lambda}\right]\left(0,\Omega-\omega,\Omega+\omega\right).\nonumber
\end{eqnarray}
The flows of the 4-point vertices are provided by
\begin{eqnarray}
    \partial_{\Lambda}V_{ij}^{a,\Lambda}\left(s,t,u\right)	&=&X_{ij}^{a,\Lambda}\left(s,t,u\right)-\tilde{X}_{ij}^{a,\Lambda}\left(t,s,u\right)\\
    &&+\tilde{X}_{ij}^{a,\Lambda}\left(u,s,t\right),\nonumber\\
\partial_{\Lambda}V_{ij}^{b,\Lambda}\left(s,t,u\right)	&=&X_{ij}^{b,\Lambda}\left(s,t,u\right)-\tilde{X}_{ij}^{c,\Lambda}\left(t,s,u\right)\\
&&+\tilde{X}_{ij}^{c,\Lambda}\left(u,s,t\right),\nonumber\\
\partial_{\Lambda}V_{ij}^{c,\Lambda}\left(s,t,u\right)	&=&X_{ij}^{c,\Lambda}\left(s,t,u\right)-\tilde{X}_{ij}^{b,\Lambda}\left(t,s,u\right)\\
&&+\tilde{X}_{ij}^{d,\Lambda}\left(u,s,t\right).\nonumber
\end{eqnarray}
Here, the objects denoted by $X$ and $\tilde{X}$ are bubble functions. In the case of $X$, they are defined by
\begin{eqnarray}
& &X_{ij}^{a,\Lambda}\left(s,t,u\right)=T\sum_{\Omega,k}\Pi_{k}^{\Lambda}(s,\Omega)\\
&\times&V_{ki}^{a,\Lambda}\left(s,\Omega+\omega_{1},\Omega+\omega_{2}\right)V_{kj}^{a,\Lambda}\left(s,\Omega-\omega_{3},\Omega-\omega_{4}\right)\nonumber\\
&+&2\times\left(a\rightarrow b\right),\nonumber
\end{eqnarray}
and
\begin{eqnarray}
& &X_{ij}^{b,\Lambda}\left(s,t,u\right)=T\sum_{\Omega,k}\Pi_{k}^{\Lambda}(s,\Omega)\\
&\times&V_{ki}^{a,\Lambda}\left(s,\Omega+\omega_{1},\Omega+\omega_{2}\right)V_{kj}^{b,\Lambda}\left(s,\Omega-\omega_{3},\Omega-\omega_{4}\right)\nonumber\\
&+&\left(a\leftrightarrow b\right)+\left(a\rightarrow b\right),\nonumber
\end{eqnarray}
and
\begin{eqnarray}
& &X_{ij}^{c,\Lambda}\left(s,t,u\right)=T\sum_{\Omega,k}\Pi_{k}^{\Lambda}(s,\Omega)\\
&\times&V_{ki}^{c,\Lambda}\left(s,\Omega+\omega_{1},\Omega+\omega_{2}\right)V_{kj}^{c,\Lambda}\left(s,\Omega-\omega_{3},\Omega-\omega_{4}\right)\nonumber\\
&+&\left(\omega_{1}\leftrightarrow\omega_{2},\omega_{3}\leftrightarrow\omega_{4}\right),\nonumber
\end{eqnarray}
where $\Pi_{k}^{\Lambda}(s,\Omega)=\dot{g}_{k}^{\Lambda}\left(\Omega\right)g_{k}^{\Lambda}\left(\Omega+s\right)$. The single-scale propagator is $\dot{g}_{j}^{\Lambda}(\omega)=\left(g_{j}^{\Lambda}(\omega)\right)^{2}\left[\omega\frac{\vartheta_{\Lambda}^{\prime}\left(\omega\right)}{\vartheta_{\Lambda}^{2}\left(\omega\right)}-{\partial_{\Lambda}\gamma_{j}^{\Lambda}\left(\omega\right)}\right]$. The second term in the brackets represents the Katanin truncation \cite{Katanin2004}. The local bubble functions $\tilde{X}_{ii}$ are by definition equivalent to the local $X_{ii}$. They are given as $\tilde{X}_{ii}^{\mu,\Lambda}\left(s,t,u\right)\equiv X_{ii}^{\mu,\Lambda}\left(s,t,u\right)$ for $\mu=a,b,c$ and $\tilde{X}_{ii}^{d,\Lambda}\left(s,t,u\right)\equiv-X_{ii}^{c,\Lambda}\left(s,u,t\right)$. In the nonlocal case, they are defined by
\begin{eqnarray}
& &\tilde{X}_{a;ij}^{\Lambda}\left(s,t,u\right)=T\sum_{\Omega}\Pi_{ij}^{\Lambda}(s,\Omega)\\
&\times&V_{a;ji}^{\Lambda}\left(\Omega+\omega_{1},s,\Omega+\omega_{2}\right)V_{a;ji}^{\Lambda}\left(\Omega-\omega_{3},s,\Omega-\omega_{4}\right)\nonumber\\
&+&2\times\left(a\rightarrow c\right),\nonumber
\end{eqnarray}
and
\begin{eqnarray}
& &\tilde{X}_{b;ij}^{\Lambda}\left(s,t,u\right)=T\sum_{\Omega}\Pi_{ij}^{\Lambda}(s,\Omega)\\
&\times&V_{a;ji}^{\Lambda}\left(\Omega+\omega_{1},s,\Omega+\omega_{2}\right)V_{c;ji}^{\Lambda}\left(\Omega-\omega_{3},s,\Omega-\omega_{4}\right)\nonumber\\
&+&\left(c\leftrightarrow a\right)+\left(a\rightarrow c\right),\nonumber
\end{eqnarray}
and
\begin{eqnarray}
& &\tilde{X}_{c;ij}^{\Lambda}\left(s,t,u\right)=T\sum_{\Omega}\Pi_{ij}^{\Lambda}(s,\Omega)\\
&\times&V_{b;ji}^{\Lambda}\left(\Omega+\omega_{1},\Omega+\omega_{2},s\right)V_{b;ji}^{\Lambda}\left(\Omega-\omega_{3},\Omega-\omega_{4},s\right)\nonumber\\
&+&\left(b\rightarrow c\right),\nonumber
\end{eqnarray}
and
\begin{eqnarray}
& &\tilde{X}_{d;ij}^{\Lambda}\left(s,t,u\right)=T\sum_{\Omega}\Pi_{ij}^{\Lambda}(s,\Omega)\\
&\times&V_{b;ji}^{\Lambda}\left(\Omega+\omega_{1},\Omega+\omega_{2},s\right)V_{c;ji}^{\Lambda}\left(\Omega-\omega_{3},\Omega-\omega_{4},s\right)\nonumber\\
&+&\left(b\leftrightarrow c\right).\nonumber
\end{eqnarray}
In these equations, $\Pi_{ij}^{\Lambda}(s,\Omega)=\dot{g}_{i,\Lambda}\left(\Omega\right)g_{j,\Lambda}\left(\Omega+s\right)+\dot{g}_{j,\Lambda}\left(\Omega+s\right)g_{i,\Lambda}\left(\Omega\right)$.
The symmetries of the $X$-terms under frequency flip or exchange are the same as for the vertices $V$. For the $\tilde{X}$-terms, they are slightly different. For $\mu\in\{a,b,c,d\}$ it holds that $\tilde{X}_{ij}^{\mu,\Lambda}\left(s,t,u\right)=\tilde{X}_{ji}^{\mu,\Lambda}\left(-s,t,u\right)=\tilde{X}_{ij}^{\mu,\Lambda}\left(s,-t,u\right)=\tilde{X}_{ji}^{\mu,\Lambda}\left(s,t,-u\right)$ and we also have $\tilde{X}_{ij}^{d,\Lambda}\left(s,t,u\right)=\left[\tilde{X}_{ij}^{a,\Lambda}-\tilde{X}_{ij}^{b,\Lambda}-\tilde{X}_{ij}^{c,\Lambda}\right]\left(u,t,s\right)$.

This concludes the description of the PMFRG flow equations for the most elementary case of a Heisenberg spin $S=1/2$ system. We emphasize again that the numerical implementation is greatly simplified by the fact that Matsubara frequencies are discrete and we usually use a frequency box with $N_\omega=20...60$ positive frequencies for the self-energy and the vertices. Of course, for a given $T/J$, one should check that the results are converged with respect to $N_\omega$. However, in practice, this is never a problem, since it turns out that the truncation of the FRG hierarchy of flow equations is usually the main bottleneck. This can be seen in the example of \cref{fig:HeisenbergDimerPopov} which shows the PMFRG results (grey symbols) for the static spin susceptibility of the AFM Heisenberg dimer. The deviations from the exact results (black lines) starting below $T/J\simeq0.4$ are not caused by any approximation in solving the flow equations but are only due to the neglect of the flow of higher-order vertices, i.e.~the truncation of the flow equation hierarchy.  

\subsubsection{Perturbative control and intrinsic consistency checks}
As mentioned in \cref{sec:finiteT-Motivation} under point (ii), a crucial technical advantage in working at $T>0$ is the ability to control the PMFRG with the smallness of the parameter $J/T$. For example, the contribution of the neglected six-point vertex scales as $\Gamma^6 \sim \mathcal{O}\left(J^3/T^2\right)$ in the standard one-loop truncation, which means that the four-point vertex $V(s,t,u)$ is accurate up to $\mathcal{O}(J^3/T^{2})$. Checking this scaling explicitly is an excellent test of any PMFRG implementation. However, since the (PM)FRG is beyond plain perturbation theory by incorporating infinite-order resummations, the question is how to decide down to which $T/J$ the results can be trusted. Ideally, in the absence of an exact solution as for the dimer case, some method-intrinsic quality check, similar to the check of the spin-magnitude in PFFRG \cite{Thoenniss2020} or similar developments in the Hubbard model community \cite{rohringerRevModPhys.90.025003}, is desired.

Generally speaking, any quantity that can be computed alternatively from the two-point and the four-point functions $\gamma$ and $V$, respectively, can be used for such an internal consistency check. Consider, for instance, the energy per site which is defined as
\begin{align}
    e &= \frac{1}{N} \langle H \rangle \nonumber = \frac{1}{N} \sum_{(i,j)} J_{ij} \langle \bm{S}_i\cdot \bm{S}_j \rangle. \label{eq:EnergyPerSite}
\end{align}
On the one hand, the equal-time spin correlators $\langle \bm{S}_i\cdot \bm{S}_j \rangle$ can be computed from the PM four-point vertex $V$ via a Matsubara sum of \cref{eq:pm-chi}. On the other hand, as discussed above, the energy can also be computed from the free energy $f$ that flows according to \cref{eq:pm-f-flow} where only $\gamma$ enters. Without the truncation of the hierarchy of flow equations, the exact vertices would be obtained and the two approaches have to produce the same result. In practical terms, this statement can be used in reverse to conclude that an observed consistency of both results for the energy signals that the truncation of the flow equations is innocuous. Heuristically, our experience with exactly solvable models indeed supports this point of view and leads us to discard PMFRG results once the difference grows beyond $\simeq 5\%$. Formally, however, the conclusion is not quite correct as even a perfect consistency between 2- and 4-point vertices would not be a guarantee for the exactness of a many-body calculation as such a consistency is a general feature of a conserving approximation. 

An alternative consistency check specific for the PM case can be derived from the local constant of motion $\Theta_i = -2\i \eta^x_i \eta^y_i \eta^z_i$, introduced in \cref{sec:pm-representation}. Since $\bar{S}^\alpha_i = \Theta_i \eta^\alpha_i $ and $ \Theta^2_i = \frac{1}{2}$, we may write the local spin-spin correlator as $\langle \bar{S}^z_i(\tau) \bar{S}^z_i(0) \rangle = \langle \Theta_i \eta^z_i (\tau) \Theta_i \eta^z_i(0) \rangle = \frac{1}{2} \langle  \eta^z_i(\tau) \eta^z_i(0) \rangle$. Therefore, for the local case $i=j$, the (static) susceptibility of \cref{eq:pm-chi} can be alternatively computed via
\begin{equation}
\chi_{jj}^{zz}(\i\nu=0)=\sum_{n}\frac{1}{\pi\left(2n+1\right)}g_{j}^{z}\left(\omega_{n}\right) \label{eq:pm-theta^2Check}
\end{equation}
which only involves the self-energy. Again, if the FRG truncation matters, one expects a sizeable difference between \cref{eq:pm-chi} and \cref{eq:pm-theta^2Check}.

\subsection{Pseudo-fermions versus pseudo-Majoranas}
We have now discussed two pseudo-particle FRG methods where the spin operators are either expressed in terms of pseudo-fermions (PFFRG) or in terms of pseudo-Majorana operators (PMFRG). We view both approaches as complementary techniques with distinct strengths and weaknesses. The PFFRG (without an implementation of the Popov-Fedotov trick) allows for a meaningful application only at $T=0$, which, in fact, is the physically most interesting situation in many applications. On the other hand, there is usually no control parameter for a quantum system at $T=0$ that guarantees the accuracy of PFFRG results. Therefore, one can at most expect qualitatively correct results. The PMFRG allows an application at finite temperatures and introduces such a control parameter, namely $J/T$. For $J/T\ll 1$ results are strictly error controlled in a perturbative sense. However, reaching the physically most interesting regime where $J/T\gtrsim 1$ while keeping results quantitatively correct, may be unfeasible. Of course, the PMFRG can also be applied at $T=0$, in which case, however, one gives up the advantage of error control for which it was designed. The choice between both methods, PFFRG and PMFRG, therefore, crucially depends on the specific needs, quantitative results at elevated $T$ or more qualitative insights into $T=0$ properties.

It is also worth discussing the differences between PFFRG {\it with} the Popov-Fedotov trick and the PMFRG. Both approaches represent two different strategies of avoiding the unphysical states in the conventional PFFRG. While the Popov-Fedotov trick can be implemented without many changes in existing PFFRG codes, the PM representation has a number of advantages: First, even with truncations in the hierarchy of the flow equations, the PM representation fulfills the spin-length constraint $S=1/2$ exactly by construction, while this is not the case with the Popov-Fedotov trick. Technically, this is rooted in the frequency symmetries of the PM vertex function which cancels its contribution to $C_i$ so that only the bare bubble contributes unity. A second advantage of the PM approach is that it remains well-defined for $T=0$, where the Popov-Fedotov trick is inapplicable. Third, the Majorana Hamiltonian is again Hermitian, allowing for a numerically more feasible implementation.\\


\section{Applications of PFFRG and PMFRG}
\label{sec:applications}

\subsection{Magnetic order}

\begin{figure}
  \centering
  \includegraphics[width = \linewidth]{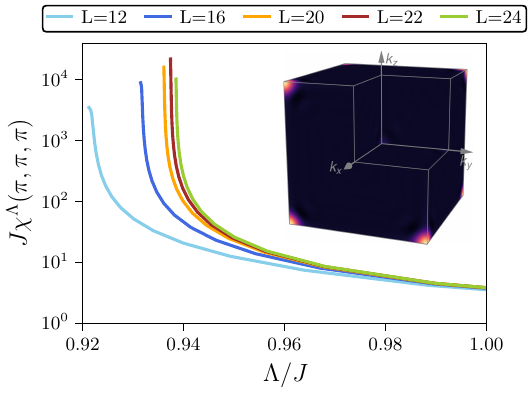}
  \caption{Nearest-neighbor antiferromagnet on simple cubic lattice with $J=1$: PFFRG flow of the spin susceptibility at the dominant wave-vector $\bm{k}_N = {(\pi,\pi,\pi)}$ for different maximal correlation lengths $L$ taken into account. Increasing $L$ sharpens the flow breakdown, which happens at $\Lambda_c/J \approx 0.939$. Inset: Susceptibility profile in reciprocal space showing Bragg peaks at the corners of the Brillouin characteristic for the N\'eel ordered state.}  
  \label{fig:pffrg_cubic}
\end{figure}

One of the most elementary characteristics of (quantum) magnets is their potential tendency to develop magnetic long-range order breaking the global spin rotation symmetry, e.g.~$\SU(2)$ for a Heisenberg model. The PFFRG and PMFRG are perfectly suited to detect magnetic ordering since they provide access to the spin-spin correlation function which is related to the four-fermion correlator. Physically, this object takes the role of a spin-susceptibility which is expected to peak at a magnetic ordering transition. Before we embark into details, we remind the reader of an important no-go theorem by Mermin-Wagner \cite{merminAbsenceFerromagnetism1966} which forbids the spontaneous breaking of a continuous symmetry [as, e.g.~$\SU(2)$ or $\U(1)$] at non-zero temperature for short-range interacting systems below three dimensions. In PFFRG, the parameter $\Lambda$, however, plays a similar role as the temperature $T$, since both act like a low-energy cutoff (the latter via the existence of a minimal Matsubara frequency). Although this implies that there should be no sign of order in the PFFRG flow of the spin susceptibility for two-dimensional systems, still a divergence is found for non-vanishing RG scale. In contrast to the flow breakdown in three dimensions, which is determined by the phase transition of the system, in the two-dimensional case it is an artefact of the truncation of the flow equations.

In practice, the detailed protocol to detect magnetic order is different for PFFRG and PMFRG. As the former focuses on $T=0$, only magnetic ordering tendencies in the the ground state can be assessed. In the following, for concreteness we focus on the nearest-neighbor antiferromagnet (NNAF) on the simple cubic lattice. Historically this was the first three-dimensional system investigated with PFFRG~\cite{Iqbal2016b}. In the classical $S\to\infty$ limit, the spins form a N\'eel ordered ground state, where neighboring spins align antiparallel to each other. Although not an eigenstate of the quantum model, the dominant correlations of this state carry over to the spin $1/2$ model, where Bragg peaks appear at the corners of the Brillouin zone, i.e.~at $\bm{k}_N = (\pi,\pi,\pi)$ and symmetry related points~\cite{Sandvik1998}. 

The PFFRG flow of the maximum of the static susceptibility for this system in \fig{fig:pffrg_cubic} shows a smooth behavior up to a spike at $\Lambda_c$. This signals a long-range ordered ground state, the formally expected divergent susceptibility  is regularized by both the finite size of the numerically treated correlations and the finite frequency resolution. Increasing the former leads to an increasingly faster and earlier divergence, which converges towards $\Lambda_c/J \approx 0.939$. Evaluating the static susceptibility just before this peak in reciprocal space indeed reveals the expected Bragg peaks on the (equivalent) corners of the Brillouin zone (see inset of \fig{fig:pffrg_cubic}). Note that the susceptibility peak could also be regularized by a small but finite magnetization which would then grow below $\Lambda_c$. However, so far this has not been implemented in PFFRG due to the ensuing breaking of time-reversal and spin-rotation symmetries, see the discussion below in \cref{subsec:magneticFields}.

As the PMFRG focuses on finite temperatures, it is complementary to PFFRG also in the search for magnetic order. However, as PMFRG flows converge in the temperature regime where the method is controlled we can expect quantitative results for critical temperatures which are to be extracted from \emph{end-of-flow} PMFRG results. To discuss the required workflow that applies to continuous phase transitions we return to the cubic lattice NNAF, where error controlled quantum Monte Carlo determined $T_c/J=0.946(1)$~\cite{Sandvik1998}. This was done for a finite-size system with periodic boundary conditions using the concept of finite-size scaling of the magnetic susceptibility at the ordering wave-vector~\cite{sandvikComputationalStudies2010}. As periodic boundary conditions are inconvenient for PMFRG which preferably is applied to infinite and translational invariant lattices, it was shown in Ref.~\cite{Niggemann2021a} that scaling in the vertex cutoff length $L$ can be used as a practical alternative. Other than that, the finite-size scaling program continues in a standard way: One possibility \cite{Niggemann2021a} is to determine the critical temperature by the pure power law-scaling of the N\'eel susceptibility, $\chi_N(T=T_c,L)/L^2 \sim L^{-\eta}$. Here, $\eta$ is the anomalous dimension, known to take on the small value $\eta=0.035$ for the \emph{classical} three-dimensional Heisenberg universality class that governs the finite-temperature magnetic transition even for quantum $S=1/2$ models. Another possibility is to use the so called correlation ratio to determine the ratio of magnetic correlation length $\xi$ and $L$,
\begin{equation}
    \xi/L=\frac{1}{2\pi}\sqrt{\frac{\chi(\bm{k}_{N})}{\chi(\bm{k}_{N}+\frac{2\pi}{L}\mathbf{e}_{x})}-1}.
    \label{eq:correlationRatio}
\end{equation}
This relation essentially relates the sharpness of the momentum-space susceptibility peak at the ordering wave-vector to the correlation length of the infinite system, see Ref.~\cite{sandvikComputationalStudies2010} for a detailed discussion. The critical temperature is reached when $\xi\rightarrow\infty$, e.g.~when $\xi/L$ does not depend on $L$. From the data in \cref{fig:cubicPMFRG} we read off $T_c=0.92J$, less than $3\%$ away from the error controlled calculation. Similar accuracies for critical temperatures were obtained in other models for which an error controlled QMC benchmark exists, namely the AFM $J_1-J_3$ cubic lattice model, the FM Heisenberg model on the pyrochlore lattice \cite{Niggemann2021a} or the square-lattice FM long-range dipolar XY-model \cite{sbierskiMagnetismTwodimensional2023}.

Going back to the scaling data of the cubic lattice NNAF (\cref{fig:cubicPMFRG}) it is also possible to attempt a scaling collapse of the data by changing the horizontal axes to $L|T-T_c|^\nu$. The collapse \cite{Niggemann2021a} is almost perfect with $\nu=0.71$ the known correlation length critical exponent of the three-dimensional Heisenberg universality class, while the mean-field value $\nu_\mathrm{MF}=1/2$ is clearly rejected. On the other hand it is well known that the underlying field theory of the Wilson-Fisher fixed point requires interactions in the order parameter. Such four-spin terms are generically absent in the truncation of the PMFRG flow equations \footnote{In the context of spin-fRG of \cref{subsec:SpinFRG}, the flow around the Wilson-Fisher fixed point can be constructed from a suitable cutoff scheme, see Ref.~\cite{tarasevychSpinFunctional2022}.}, although the bi-local terms are included via a non-trivial identity of the pseudo-Majorana representation \cite{Niggemann2021a}. In conclusion, the quantitative accuracy of the critical exponents obtained by a scaling collapse could possibly be accidental in nature or have a reason beyond current understanding.

\begin{figure}
  \centering
  \includegraphics[width = \linewidth]{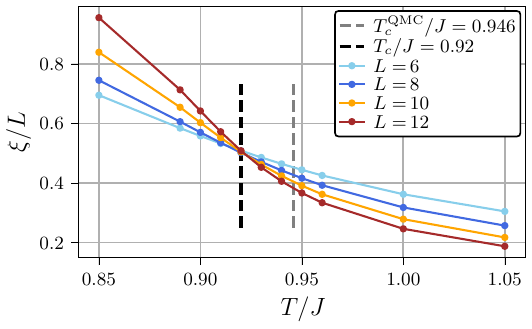}
  \caption{Nearest-neighbor antiferromagnet on simple cubic lattice: PMFRG results for the correlation ratio \cref{eq:correlationRatio} using a one-loop + Katanin flow. The crossing of the data for various system size indicates a critical temperature $T_c=0.92J$ close to the quantum Monte Carlo result $T_c/J=0.946(1)$~\cite{Sandvik1998} (dashed lines). We re-defined $L = [3/(4\pi)N ]^{1/3}$ using the number $N$ of sites correlated to the reference
site instead the maximum of the treated vertex range. Empirically, this yields smoother results especially in the small-$N$ case.}  
  \label{fig:cubicPMFRG}
\end{figure}

In summary, we have used the cubic lattice NNAF to discuss how magnetic ground state order or ordering temperature as well as critical exponents can be detected within PFFRG or PMFRG, respectively. The strength of the methods now rely in the fact that the same studies can be applied to almost arbitrary bilinear spin systems, especially those which are frustrated by competing interactions. Such systems can harbor multiple exotic magnetic ordering patterns in the ground state, see for example the rich phase diagram of the $J_1-J_2-$Heisenberg model on the pyrochlore lattice studied by PFFRG in Ref.~\cite{Iqbal2018}. Of particular interest, however, are paramagnetic regions in the magnetic phase diagram where no magnetic order is detected. We dedicate the following subsection to this topic.


\subsection{Incommensurate order}

\label{sec:incommensurate}
In the most common cases, magnetic orders are invariant under translations by an integer linear combination of lattice vectors. For example, N\'eel order on the square lattice with ${\bm k}_\textrm{N\'eel} = (\pi,\pi)/a$ is invariant under translations by two lattice vectors in either direction.
It is, however, also possible that one or more components of the ordering wavevector are irrational multiples of $\pi/a$. In such cases, the magnetic unit cell becomes infinitely large, and finite-size limitations of methods with periodic boundary conditions pose significant challenges.
As discussed in the previous section, PF- and PMFRG implement infinite systems, limiting only the maximal length of correlations and thus do not share this limitation.
More concretely, the site summation in the magnetic susceptibility in \cref{eq:static_suscep} formally runs over infinitely many sites, although $\chi_{ij} = 0$ for a large enough separation of $i$ and $j$ and hence, wavevectors are continuous and the type of incommensurate order is resolved accurately, apart from a finite width of sharp features.

Previous PFFRG studies have successfully investigated the evolution of incommensurate order under variation of model-specific parameters \cite{Ghosh2019,Buessen2021}. For instance, in Ref.~\cite{Buessen2021}, the ordering behaviour of the spin-1/2 Kitaev-$\Gamma$ model
\begin{equation}
    H = \sum_{\langle i,j \rangle_\gamma} K S^\gamma_i S^\gamma_j + \Gamma \left(S^\alpha_i S^\beta_j + S^\beta_i S^\alpha_j  \right)
\end{equation}
has been investigated on the honeycomb lattice, where $\gamma = x,y,z$ corresponds to one of the three types of honeycomb bonds and $\alpha, \beta$ are the other two spin components.
\Cref{fig:IncommensurateOrder} tracks the continuous evolution of the distance of the peak in the magnetic susceptibility away from the origin ${\bm k} = (0,0)$ as a function of $\alpha$, where $\alpha$ determines $K = - \cos(\alpha)$ and $\Gamma = \sin(\alpha)$. This allows for an accurate resolution of the entire phase diagram which is dominated by incommensurate order.

\begin{figure}
    \centering
    \includegraphics{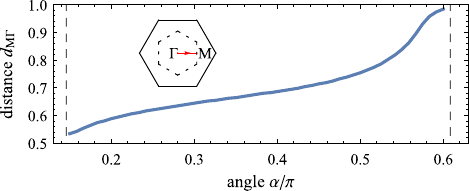}
    \caption{Distance $d_{M\Gamma}$ of peak in the static magnetic susceptibility $\chi(\mathbf{k})$ from the origin of the Brillouin zone for the Kitaev-$\Gamma$ model on the honeycomb lattice. The peak location follows the line between the high-symmetry points $\Gamma$ and $M$ of the first Brillouin zone of the honeycomb lattice. Figure reproduced from \cite{Buessen2021}.}
    \label{fig:IncommensurateOrder}
\end{figure}

\subsection{Paramagnetic phases}
\label{sec:pyro}
A paradigmatic model where frustration prevents the formation of magnetic order is the nearest-neighbor pyrochlore antiferromagnet which, in the classical case, famously realizes a spin liquid \cite{Moessner1998}. The four sites in the unit cell are arranged each at half the fcc lattices vectors $\mathbf{a}_i/2$, $i=1,2,3$, i.e.~$\mathbf{b}_0 = (0,0,0), \mathbf{b}_i = \mathbf{a}_i/2$ such that the lattice realizes an arrangement of corner-sharing tetrahedra.
Establishing an understanding of the nature of its apparently nonmagnetic quantum ground state has motivated a plethora of numerical studies employing various non-FRG methods \cite{Isakov2004,Schafer2020,Derzhko2020,Muller2019,Astrakhantsev2021a,Hagymasi2021,Hagymasi2022}. In \cref{fig:pmfrg_Pyrochlore} we show a comparison of the uniform static susceptibility $\chi(\mathbf{k}=0)$ between PMFRG and error-controlled methods. It can be seen that PMFRG produces \emph{quantitatively} reliable results even at low temperatures, without sharing the limitations of other methods, which are often hamstrung by low momentum-space resolution. We note that due to the rather small values of $\chi(\mathbf{k}=0)$ for antiferromagnets, the effect of two-loop corrections appears more significant. 

\begin{figure}
  \centering
  \includegraphics[width = \linewidth]{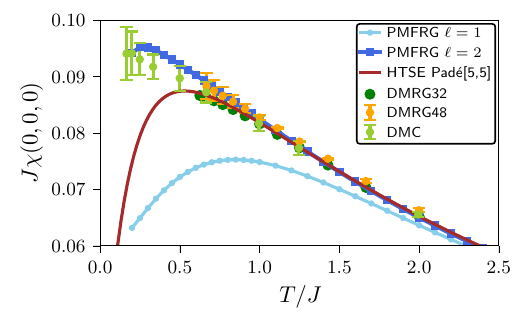}
  \caption{Comparison PMFRG in one-loop ($\ell=1$) and two-loop ($\ell=2$) with high-temperature series expansion (HTSE, red) \cite{Muller2019}, DMC (light green)\cite{Huang2016}, and DMRG for 32 (dark green) and 48 site (orange) clusters \cite{Schafer2020}. Reproduced from Ref.~\cite{Niggemann2021}.} 
  \label{fig:pmfrg_Pyrochlore}
\end{figure}

One of the advantages of PM and PFFRG methods is the implementation of fully translationally invariant lattices. Moreover, only the maximal correlation length is limited such that in disordered systems, where the physical correlation length is finite, size effects are effectively absent.

\begin{figure}
  \centering
  \includegraphics[width = \linewidth]{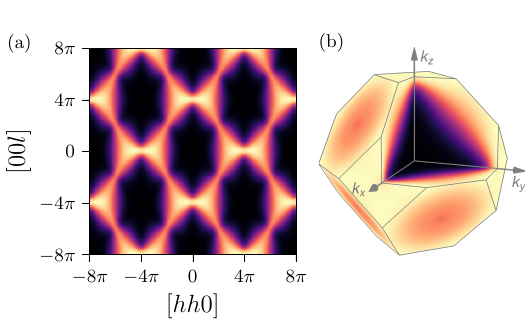}
  \caption{(a) Momentum resolved susceptibility of the NNAF on the Pyrochlore lattice in the $hhl$-plane calculated from PFFRG at the lowest simulated $\Lambda=0.05J$. Broadened pinch points are visible at $(4\pi,0,0)$ and symmetry related points. (b) Susceptibility in the Brillouin zone (truncated octahedron) of the Pyrochlore lattice. Results obtained via PMFRG at low temperatures are observed to be qualitatively equivalent \cite{Niggemann2021a,Niggemann2023}.}  
  \label{fig:pffrg_Pyrochlore}
\end{figure}

The static susceptibility can therefore be obtained in high resolution at the end of the flow. As can be seen in \fig{fig:pyroflow}, the Katanin-corrected PFFRG flow shows no sign of a divergency, as expected for a magnetically disordered ground state. In \fig{fig:pffrg_Pyrochlore}(a) the value of the susceptibility obtained from PFFRG at the lowest simulated RG scale in the $hhl$-plane ($k_x=k_y$). It shows the expected broadening of the pinch points at $(0,0,4\pi)$, resulting from violations of the spin ice rule in the quantum limit (see also~\cite{Niggemann2023} where a similar broadening of higher-fold pinch-points is investigated). \fig{fig:pffrg_Pyrochlore}(b) shows the same susceptibility in the extended Brillouin zone of the Pyrochlore lattice. In accordance with both the featureless flow and the broad susceptibility in the $hhl$-plane, no Bragg peaks are visible in reciprocal space, only broad maxima of the static susceptibility, strengthening the perception of a magnetically disordered ground state.

\subsection{Probes for symmetry breaking in paramagnets}
\label{sec:symmetryBreakingParamag}

The results in the previous section establish an absence of long-range order in the ground state of the NNAF on the Pyrochlore lattice. The FRG equations, however, respect all symmetries of the microscopic Hamiltonian, in particular both spin-rotational invariance and lattice symmetries. Recent studies employing the density matrix renormalization group~\cite{Hagymasi2021} and variational Monte Carlo~\cite{Astrakhantsev2021a}, however, find an inversion symmetry breaking and combined inversion and rotation symmetry breaking ground state, respectively.

To find such tendencies in an unbiased way, a dimer-dimer correlation function of the form $D_{ij,kl} = \ev{(\bm{S}_i\cdot \bm{S}_j)(\bm{S}_k\cdot \bm{S}_l)}-\ev{(\bm{S}_i\cdot \bm{S}_j)}\ev{(\bm{S}_k\cdot \bm{S}_l)}$ would have to be calculated~\cite{Iqbal2019}. The divergence of this quantity then would signal the onset of nematic order. In PFFRG, however, this would necessitate the calculation of a four-particle vertex, which, while formally accessible in the formalism, is numerically too demanding.

To probe for specific symmetry breaking patterns, however, one can resort to the introduction of a small perturbation $\delta$ in accordance with the pattern. For this procedure, the nearest neighbor bonds are grouped into weakened ($W$) and strengthened ($S$) bonds. The perturbation $\delta$ then modifies the Heisenberg couplings $J_{ij} \to J_{ij} \pm \delta$ for $\langle i,j\rangle \in S/W$. 
The flow of the dimer response function 
\begin{equation}
  \eta^\Lambda_\mathrm{dim} = \frac{J}{\delta} \frac{\chi_S^\Lambda-\chi_W^\Lambda}{\chi_S^\Lambda+\chi_W^\Lambda},\label{eq:dimerresponse}
\end{equation}
where $\chi^\Lambda_{S/W}$ is the spin-spin correlator according to \cref{eq:realspacecorrel} on strong/weak bonds, then signals the tendency of the system to accept/reject the symmetry breaking.

If the initial $\eta^{\Lambda\to\infty}_\mathrm{dim} = 1$ grows larger during the RG flow, the symmetry breaking likely is present in the true ground state of the system, whereas a diminishing susceptibility implies a rejection of the pattern. This method has been used to confirm the spatial symmetry breaking on the Pyrochlore lattice from a PFFRG perspective~\cite{Hering2021b}. 

\begin{figure}
    \centering
    \includegraphics{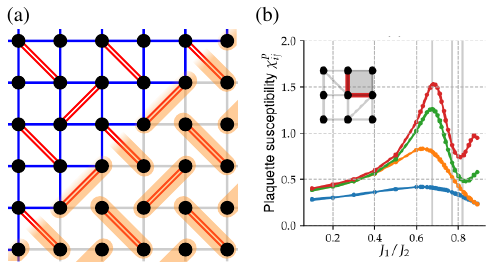}
    \caption{(a) The Shastry-Sutherland lattice (top left) features square-lattice nearest neighbor couplings $J_1$ (blue) and select second nearest neighbor ones $J_2$ (red). For dominant antiferromagnetic $J_{2}$ couplings, the ground state is a product of exact singlets on these bonds (bottom right). (b) Upon increasing $J_1/J_2$, the dimerized state gives way to a plaquette order signaled by a peak in the plaquette susceptibility (inset: pattern of weak and strong bonds in this susceptibility). Reproduced under permission from \mRef{Keles2022}.}
    \label{fig:shastry}
\end{figure}

Another prime example for the application of such a symmetry breaking probe is the determination of the phase boundaries of the two-dimensional Shastry-Sutherland-Model~\cite{Shastry1981} as diagnosed \mRef{Keles2022}. Apart form the analytically dimerized ground state shown in \cref{fig:shastry}(a), which is a product of spin singlets covering the lattice, the model including nearest- ($J_1$) and second-nearest enighbor ($J_2$) interactions is known to host at least two more phases: A plaquette valence bond order and a  N\'eel phase.
In PFFRG the dimerized phase is readily identified by a interaction-independent value of the spin-spin correlation along the bond hosting the singlet. To probe for the plaquette valence bond order, the corresponding susceptibility is introduced, which refers to a pattern of weak and strong bonds in \cref{eq:dimerresponse} as shown in the inset of \cref{fig:shastry}. It shows a pronounced peak at the onset of plaquette order at $J_2/J_1 = 0.67$ as shown in \cref{fig:shastry}(b). 
The ordered N\'eel phase in turn can be directly identified from the spin structure factor. In between the latter phases, a putative spin liquid region was identified in \mRef{Keles2022} by probing the response to another pattern of strong and weak bonds.

Similar prescriptions were used to probe for valence-bond crystal orders on the simple cubic lattice\cite{Iqbal2016b}. Introducing the symmetry breaking term in spin-space, i.e. breaking $\SU(2)$ symmetry, allows accessing spin-nematic tendencies, as has also been investigated on the Pyrochlore lattice~\cite{Iqbal2019} and square lattice~\cite{Iqbal2016c}.

Although this approach can only probe for the specific symmetry breaking patterns assumed to exist \emph{a priori}, it is an important diagnostic for quantum paramagnetic states complementing the pure magnetic long-range order analysis from the flow of the spin-susceptibility.

\subsection{Three-dimensional quantum spin liquid materials}
\label{sec:3DQSL}
Following these more demonstrative applications, we now proceed towards presenting PFFRG applications to real magnetic materials. The field of frustrated quantum magnetism is poised with the arrival of new candidate quantum spin liquid materials based on novel three-dimensional lattices. Most prominently, these include the network of $S=1/2$ Cu$^{2+}$ ions forming a hyper-hyperkagome lattice in PbCuTe$_{2}$O$_{6}$~\cite{Khuntia-2016} and $S=1$ Ni$^{2+}$ ions forming a bi-trillium lattice in K$_{2}$Ni$_{2}$(SO$_{4}$)$_{3}$~\cite{Zivkovic2021}. Indeed, no sign of magnetic long-range ordering has been observed in PbCuTe$_{2}$O$_{6}$ down to 20mK despite a Curie-Weiss temperature of 23K, while in K$_{2}$Ni$_{2}$(SO$_{4}$)$_{2}$, a QSL is stabilized under the application of a magnetic field~\cite{Yao-2023,gonzalez2023}. The magnetic Hamiltonians for these 3D materials feature a complex hierarchy of antiferromagnetic Heisenberg exchange interactions, and are as such not amenable to most numerical quantum many-body approaches. The PFFRG approach has met with remarkable success in explaining the momentum resolved susceptibility~\cite{Chillal2020,gonzalez2023}. As a case in the point, for PbCuTe$_{2}$O$_{6}$, PFFRG predicts a diffuse sphere of scattering at same wave-vectors and similar intensity modulations as those observed in experiment (see Fig.~\ref{fig:pffrg-rydberg}), and is even able to reproduce the subdominant features. It is worth noting that this level of agreement has not been achieved for such a material with many competing interactions on a complicated three-dimensional lattice and $S=1/2$ limit.
\begin{figure}
  \centering
  \includegraphics[width = \linewidth]{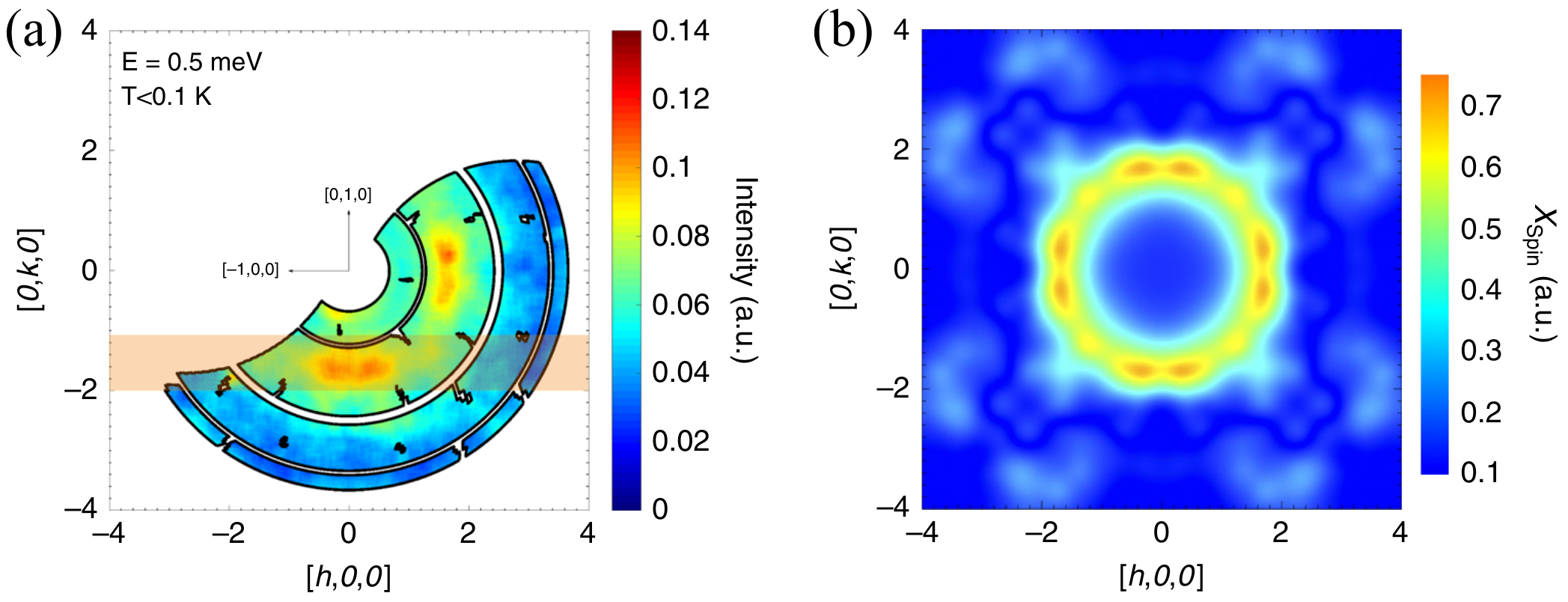}
  \caption{(a) The excitation profile obtained at an energy transfer of $E=0.5$ meV and temperature $T<0.1$ K in the $[hk0]$ plane. (b) The static (real-valued) spin susceptibility calculated using the PFFRG approach for the exchange parameters $J_{1}=1.13$ meV, $J_{2}=1.07$ meV, $J_{3}=0.59$ meV, and $J_{4}=0.12$ meV (corresponding to a Curie-Weiss temperature $\Theta_{\rm CW}=-23$ K). The figure is partially reproduced from Ref.~\cite{Chillal2020}.  }
  \label{fig:pffrg-rydberg}
\end{figure}

\subsection{Thermometry in Rydberg array experiment}
\label{sec:ThermometryAMO}
While the PF/PMFRG has been traditionally applied for solid state quantum magnets, a recent application from the field of cold-atom physics considers a Rydberg atom tweezer array experiment \cite{chenContinuousSymmetryBreaking2023} which realizes a square-lattice XY-model with long-range dipolar interactions $H= J \sum_{ (i,j)} \frac{1}{r_{{ij}}^3}\left(S_i^x S_{j}^x + S_i^y S_{j}^y\right)$. Both the ferromagnetic ($J<0$) as well as the frustrated antiferromagnetic case ($J>0$) was implemented. In contrast to the realm of solid-state applications, Rydberg tweezer array experiments are only conducted on microsecond time scales and the question about thermalization and temperature is non-trivial. In this context, the PMFRG (which assumes thermal equilibrium) was applied to fit a temperature to the measured equal-time spin correlator profile \cite{sbierskiMagnetismTwodimensional2023}. Fig.~\ref{fig:pmfRG_Rydberg} shows the results for the antiferromagnetic case. As the experimental hold-time $t$ increases, more and more decay processes heat up the system which leads to increasing temperatures.
\begin{figure}
  \centering
  \includegraphics[width = \linewidth]{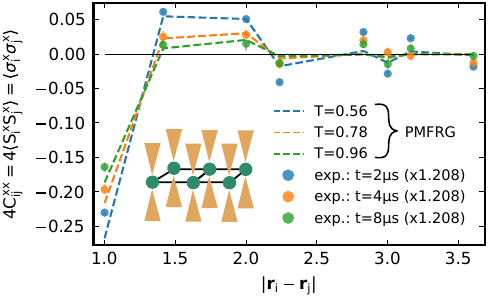}
  \caption{Experimental spin correlation function of an AFM-dipolar XY model realized in a Rydberg atom array shown schematically in the inset. The data was measured after $t =2 \mu s$ (blue dots), $t = 4 \mu s$ (orange dots) and $t = 8 \mu s$ (green dots) and is taken from Ref.~\cite{chenContinuousSymmetryBreaking2023}. The nearest-neighbor coupling was set to one. The measured data was multiplied with a factor of 1.208 to take into account measurement errors \cite{sbierskiMagnetismTwodimensional2023}. The data for small distances are well reproduced by thermal PMFRG simulations (dashed line, infinite system) at temperatures shown in the labels. The figure is partially reproduced from Ref.~\cite{sbierskiMagnetismTwodimensional2023}.  }
  \label{fig:pmfRG_Rydberg}
\end{figure}

\section{Challenges and future directions}
\label{sec:challenges}

In this section, we review the current challenges and limitations of the PFFRG and PMFRG methods. We combine this discussion with an outlook to possible future methodological improvements and propose novel areas of application.

\subsection{Magnetic fields}
\label{subsec:magneticFields}
In all previous sections, the PFFRG and PMFRG have been developed under the assumption that time reversal symmetry is intact, which excludes external magnetic fields. In the most general case a magnetic field would give rise to an extra term in the Hamiltonian of the form
\begin{equation}
\mathcal{H}\rightarrow \mathcal{H} +\sum_i \bm{B}_i\cdot \bm{S}_{i}\;.
\end{equation}
Indeed, all currently published applications of the PFFRG and PMFRG have made the assumption of vanishing external magnetic fields. The only reason for the community's reluctance to include magnetic fields is the increased numerical cost which has various different origins. First, even when ignoring the complications from time-reversal symmetry breaking, the presence of a magnetic field breaks spin-rotation symmetry from SU(2) for a Heisenberg system down to at least U(1). In PFFRG this generates additional contributions to the two-particle vertex besides the spin and density terms in \cref{eq:spindensparam}.

The main complicating effect of magnetic fields, however, comes from time-reversal symmetry breaking which was used extensively in the above calculations to simplify the flow equations and the parametrization of vertex functions. For example, the simple form of the self-energy in PFFRG [see \cref{eq:selfenergy_final}] which is purely imaginary and proportional to the identity in spin space would no longer hold. Instead, magnetic fields give rise to additional contributions proportional to Pauli matrices in spin space, which correspond to the terms $\Sigma^\alpha(\omega) \sigma^\alpha_{\mu'_1\mu_1}$ with $\alpha=1,2,3$ in \cref{eq:paraself}. Since $\Sigma^\alpha$ is found to be real for $\alpha=1,2,3$ this also means that the self-energy and, consequently, the Green's functions are generally composed of a finite real and imaginary part. This doubling of terms becomes a quadrupling when products of two Green's functions are involved such as in the flow equations for the two-particle vertex. Another source of increased numerical costs is that the symmetries in the frequency dependence of the two-particle vertex, listed in \tab{tab:naturalsymm}, are reduced. Altogether, these complications lead to significantly longer numerical runtimes for the solution of the flow equations. This certainly does not generally prohibit the application of the PFFRG or PMFRG in the presence of finite magnetic field but makes it more challenging to obtain numerically stable and converged outcomes. At the time of writing of this review it was not yet entirely clear how severe these complications are in practice, i.e.~how much they come at the expense of accuracy/stability or, more generally, how large the impact of the truncation of flow equations is in the presence of finite magnetic fields.

Apart from these challenges, the opportunities of implementing magnetic fields for future applications should also be explained. One can, generally, pursue two strategies when adding magnetic fields. The first is the inclusion of small fields (e.g.~on the order of a percent of the exchange couplings) which regularizes the breakdown of the flow in magnetically ordered phases. This enables one to continue the RG flow down to the small $\Lambda$ limit such that magnetically ordered phases can be investigated in the physical limit where the $\Lambda$-regulator is absent. Another interesting question for the PMFRG is if the treatment of a finite magnetization allows for the detection of 1st order magnetic phase transitions. The second strategy is to add finite fields (e.g.~on the order of the exchange couplings) to identify phases and phenomena that require magnetic fields. An obvious application would be the investigation of magnetization plateaux in magnetization curves which are known to occur in a variety of frustrated spin systems\cite{Kageyama1999,Honecker2004,Momoi2000}. In the high-field limit the approach would even be strictly error controlled in a perturbative sense since the interacting part of the Hamiltonian is small compared to the field term, where the latter is only quadratic in the pseudo (Majorana) fermions, i.e.~non-interacting. In that case, the method has certain conceptual similarities with an FRG approach for fermionic Hubbard models, where, likewise the case of small ratio of Hubbard interactions and hopping constitutes a controlled limit. 

Besides the advantages of adding magnetic fields for theoretical purposes, such an extension would also lead to further possibilities to connect to experiments since adding a magnetic field is one of the simplest and most straightforward ways to manipulate a quantum magnet.  

\subsection{Accessing real frequency data}
When comparing both PFFRG and PMFRG data to experiments, the most striking limitation of the methods is their formulation on the Matsubara (imaginary frequency) axis. 
As shown in \Sec{sec:susceptibilities}, this only allows to access the static ($\omega=0$) physical quantities directly, although in principle all physical information is already encoded in the vertex. Inelastic neutron scattering experiments, in contrast, directly access the frequency resolved spin-structure factor, which is related to the imaginary part of the dynamical spin susceptibility.

One way to obtain information on the real frequency axis is analytical continuation of susceptibilities in terms of Matsubara frequencies. Performing this procedure numerically, however, is a notoriously difficult problem and so far has only been applied in special situations~\cite{Reuther2014b}.

An alternative approach is to reformulate the whole renormalization procedure in terms of real frequencies directly, utilizing the Keldysh formalism~\cite{Keldysh1965}. Originally formulated for non-equilibrium problems, it is also applicable to systems in thermal equilibrium. The complication in porting FRG to the real axis stems from the fact that under time-evolution of a system the initial and final states at $t=\pm \infty$ do not necessarily have to be the same. This prevents the formulation of a path integral akin to the ones discussed in \Sec{sec:functionalrenormalization} and therefore the whole formalism as discussed there is not applicable.

The Keldysh formalism circumnavigates this complication by performing the time evolution both forward and backward, leading to a doubling of degrees of freedom by introducing a positive and negative evolving time contour branch. The formulation of the flow equations stays unchanged, up to an additional Keldysh index per external leg of a vertex, labelling the respective branch of the time contour~\cite{Gezzi2007,Jakobs2010,Karrasch2010,Ge2023}. This means that the self-energy becomes a $2\times2$ matrix, whereas the two-particle vertex acquires four Keldysh indices and therefore 16 components. Additionally all these quantities are not necessarily real-valued anymore, leading to an additional increase in numerical complexity.

Further complications arise from the regulators discussed in \Sec{sec:regulators}: In the Keldysh context they violate causality, which makes them unsuitable for use in a real-frequency setup. Additionally, the symmetries listed in \tab{tab:naturalsymm} acquire additional Keldysh structure. Both points have only been explored for general fermionic models so far~\cite{Jakobs2010a}, but not for pseudo-fermionic or pseudo-Majorana degrees of freedom in particular, which is, however, vital for a numerically performant implementation of the method.

So far, it is not clear, if the additional run time cost will be outweighed by the potential physical insights from the method, or if PFFRG or PMFRG is suited to be treated in the Keldysh formalism at all.

\subsection{Low temperature problem}
\label{subsec:lowTProblem}

From the discussions in \cref{sec:finiteT} as well as the data presented in \cref{fig:HeisenbergDimerPopov}, we have seen that reaching low temperatures is a major problem for PM and finite-temperature PFFRG: The one-loop + Katanin truncation of the flow equations introduces an error $\mathcal{O}(J^3 / T^2)$ to the vertex. Although this is an asymptotic statement for large $T/J$, it explains why benchmark comparisons and method-inherent consistency checks are likely to fail if $T \ll J$. This is unfortunate since some of the most pressing questions about quantum spin models, such as the formation of exotic low-temperature states remain unanswered from the FRG perspective. Similarly, one could argue that the cutoff $\Lambda$ perturbatively controls the error for the FRG truncation in the $T = 0$ PFFRG formalism and, thus, the association of instabilities in the flow with magnetic order might only be sensible for $\Lambda \gtrsim J$. Despite an early discussion of these concerns in Ref.~\cite{Goettel2012} solid numeric evidence for this statement is currently lacking. 

The straightforward attempt to improve the quantitative accuracy of the FRG are multiloop extensions, see the discussion above in \cref{subsec:truncation}. Indeed, even the inclusion of two-loop $2\ell$ corrections can improve the accuracy of the results as demonstrated in \cref{fig:pmfrg_Pyrochlore} for the Pyrochlore NNAF. However, even loop-converged multiloop FRG can at best reproduce the truncation error of the parquet approximation, which is $\mathcal{O}(J^4 / T^3)$ on the vertex level. Moreover, there is no guarantee that the additional diagrams which improve the error scaling at large $T/J$ also systematically improve the accuracy in the low temperature case. Taking the Heisenberg dimer or the simple-cubic AFM as an example, truncations of the flow equations beyond one-loop + Katanin do not seem to systematically improve the quality of the results \cite{schneiderTamingPseudofermion2022}. At the time of writing, therefore, the role of multiloop extensions of PFFRG and PMFRG remains unclear. 

A promising avenue for future research is the fusion of pseudofermion FRG with non-perturbative strategies, such as the direct enforcement of Ward identities \cite{Kopietz2013} or the use of local vertex quantities in the initial condition of the flow, similar to the DMF$^2$RG \cite{Vilardi2019} method, where the result of a DMFT calculation is used to initialize the FRG flow. To make the latter approach feasible, it is crucial to  find a representation of the flow equations in terms of well-conditioned  objects which (a) do not lead to double-counting of diagrams \cite{Reuther2014b} and (b) are non-divergent \cite{Chalupa_2021}. Recent progress in this direction has been made by rewriting the flow equations in terms of single and multiboson scattering processes \cite{Krien_2021, Gievers_2022, Bonetti_2022}, mitigating the need to compute two-particle irreducible vertices from the (inverse) Bethe-Salpeter equation.

The low-$\Lambda$ or low-$T$ problem of PF- or PMFRG is closely related to the ignorance of (most of the) higher order fermionic vertices with more than four legs corresponding to correlators of more than $n=2$ spins. However, an unbiased study of competing non-magnetic phases (e.g.~dimerized or spin liquid states), might require the calculation of $n=3$ (chiral), $n = 4$ (dimer-dimer) or even higher correlators. One idea to circumvent the daunting numerical cost associated with these objects in the PF- or PMFRG is to abandon a parton spin representation altogether and instead work with the spin operators themselves. This approach, dubbed \textit{spin-FRG} is discussed further in the following \cref{subsec:SpinFRG}.

\subsection{Working without partons: Spin-FRG}
\label{subsec:SpinFRG}

In 2019 Kopietz and coworkers \cite{kriegExactRenormalization2019} suggested a paradigm change in the application of the FRG to spin systems: Instead of computing vertex functions of auxiliary and as such unobservable fermionic partons, their scheme termed spin-FRG is applied directly to correlation (and vertex-) functions of spin operators $\bigl\langle \mathrm{T}_\tau S_{i_1}^{\alpha_{1}}(\tau_{1})S_{i_2}^{\alpha_{2}}(\tau_{2})...S_{i_n}^{\alpha_{n}}(\tau_n)\bigr\rangle$ without the need of any intervening representation. This builds on the earlier insight \cite{pawlowskiAspectsFunctional2007} that FRG flow equations do not necessarily require an unconstrained Grassmann (or real) functional integral representation of the partition function, which does not exist for spin. Instead, flow equations can be derived for a generating functional written in terms of a time ordered exponential of operators. In the Heisenberg case, for example, this functional in terms of source-fields $h_i^\alpha(\tau)$ is given by
\begin{eqnarray}
   \mathcal{G}[\mathbf{h}]&=&\mathrm{ln}\,\mathrm{Tr}\, \mathrm{T}_\tau\exp\! \int_{0}^{\beta}\mathrm{d}\tau \;[-\sum_{(i,j)}J_{ij}\mathbf{S}_{i}(\tau)\cdot\mathbf{S}_{j}(\tau) \\ 
& &+\sum_i\mathbf{h}_i(\tau)\cdot\mathbf{S}_i(\tau)]. \nonumber
\end{eqnarray}
The flow parameter $\Lambda:0\rightarrow 1$ is introduced via $J_{ij}\rightarrow\Lambda J_{ij}$ ($\mathcal{G}[\mathbf{h}]\rightarrow\mathcal{G}_{\Lambda}[\mathbf{h}]$)   
increasing the coupling $J_{ij}$ from zero to its final value. The resulting flow equations for the connected imaginary time-ordered spin correlation functions obtained by functional derivatives with respect to source fields $\delta\mathcal{G}_{\Lambda}[\mathbf{h}]/[\delta h_{i_{1}}^{\alpha_{1}}(\tau_{1})...\delta h_{i_{n}}^{\alpha_{n}}(\tau_{n})]_{\mathbf{h}=0}$ take on a standard bosonic form. However, for the free spin case encountered at the beginning of the flow at $\Lambda=0$ the Legendre transform to the generating functional of vertex functions $\Gamma_{\Lambda}[\mathbf{m}]$ is not defined for source fields with non-trivial time dependence. Kopietz and coworkers bypassed this problem by defining a non-standard hybrid functional \cite{tarasevychDissipativeSpin2021} which treats finite and vanishing frequency cases on unequal footing. Despite this technical complication, the flow equations for the one-line irreducible vertex functions finally take a relatively simple form.
Up to now, performant numerical approaches treating the system of resulting flow equations in its full complexity have not yet been implemented. Instead, several approximations were employed which however lead to promising results \cite{tarasevychRichMan2018,gollSpinFunctional2019,gollZeromagnonSound2020,tarasevychDissipativeSpin2021, tarasevychCriticalSpin2022,ruckriegelSpinFunctional2022,tarasevychSpinFunctional2022} which will be reviewed in the following.

One of the features of spin-FRG is efficiency. In FRG approaches to the Hubbard model, the full parametrization of frequency and momentum dependence for vertices with up to 4 legs is by now standard \cite{schaferTrackingFootprints2021}. Hence, once implemented within spin-FRG, the flows of 3- and 4-point spin correlators should be equally accessible numerically, giving simplified access to dimer-susceptibilities when compared to PF- and PMFRG, see the discussion in \cref{sec:symmetryBreakingParamag}. Moreover, one can hope that the low-temperature problem can be alleviated to a certain extent by the flow of such higher-order spin vertices. In part, this optimism is fueled by the percent-range accuracy for critical temperatures of classical \cite{kriegExactRenormalization2019} and quantum \cite{tarasevychSpinFunctional2022} spin models achieved by the Kopietz group with vertex parametrization approximated even at the 2-point level. Also the inclusion of magnetic fields seems possible at a moderate numerical effort within spin-FRG. Kopietz and coworkers have already demonstrated magnetization calculations and magnon dynamics \cite{gollSpinFunctional2019,gollZeromagnonSound2020} in rather simple non-frustrated settings and with limited numerical ambition. 

Due to the $\SU(2)$ spin algebra, even local free-spin correlation functions $\bigl\langle\mathrm{T}_\tau S_{i}^{\alpha_{1}}(\tau_{1})S_{i}^{\alpha_{2}}(\tau_{2})...S_{i}^{\alpha_{n}}(\tau_n)\bigr\rangle_{c,0}$ are non-trivial at every order $n$ but can still be computed \cite{tarasevychSpinFunctional2022,halbingerSpectralRepresentation2023}. The spin-FRG takes advantage of this non-trivial information by starting the flow at the free-spin limit $\Lambda=0$. Moreover, the correlation functions along the spin-FRG flow are physical at every $J_{\mathrm{eff}}=\Lambda J$ and a single flow produces correlation functions for the full accessible range of $T/J_{\mathrm{eff}}$. As another benefit, the spin-FRG formalism is applicable for general spin length $S$, which merely enters as a parameter in the initial condition. For completeness, we also mention the earlier work by Rançon who used a non-perturbative variant of the FRG to study the XY-spin model \cite{ranconNonPerturbativeRenormalization2014}.

In summary, while the spin-FRG seems attractive for various reasons reviewed above, further work and comparisons to PF- and PMFRG are necessary to gauge its full potential, especially in the framework of three-dimensional frustrated quantum magnets.

\section{Conclusions}
\label{sec:conclusions}

Frustrated quantum magnets which evade any analytical or quasi-exact numerical solution pose a many-body problem with two complementary ways to address them. First, guided by symmetry and topology, one develops a theory to characterize the suggested ground state of the problem. This motif includes, but is not exhausted by, approaches as diverse as mean field theory, variational Monte Carlo, projected entangled pair states encodings, density matrix renormalization group, and tensor network methods in general. Its ultimate goal is to not only learn about the magnetic quantum ground state, but also elementary excitations whose structure is inherited by the given ground state. Second, inspired by reducing the bare model Hamiltonian problem to an effective model where the competing ordering and disordering tendencies are more clearly expressed, one seeks to distill a low-energy model upon exploiting the scale separation between bare exchange couplings and eventual ordering strength or paramagnetic incompressibility. Both ways are intertwined, as the former can readily be applied to the effective model resulting from the latter.
Functional renormalization group serves both purposes in one go. Upon the renormalization flow procedure explicated in the review and as such rendered accessible to everyone, one produces an effective spin exchange model at a lower energy scale dependent on the respective cutoff value. Furthermore, the renormalization group procedure allows to investigate the flow of the susceptibility as function of cutoff, and as such picks up any symmetry breaking propensity along the flow. This means that even though the method is first of all an attempt to retrieve effective Hamiltonians / many-particle vertices at low energies, it also provides a largely unbiased access to explore the landscape of symmetry breaking in a quantum magnet. 
Characterized by an exceptional flexibility, i.e., not limited by either sign problem, dimensionality, lattice geometry, or interaction range, the pseudo-particle functional renormalization group promises to become an indispensable tool in contemporary research on frustrated quantum magnetism, and has already proven so in the past decade. With further improvements on their way to increase their quantitative predictability, there is substantiated hope that PFFRG and PMFRG will eventually help to close the circle of cognition in frustrated magnetism composed of experimental observation, theoretical conceptualization, and mathematical abstraction. 

\ack

We thank 
Maria Laura Baez,
Finn Lasse Buessen,
Kiyu Fukui,
Mat\'{\oldi}as Gonzalez,
Lasse Gresista, 
Pratyay Ghosh,
Max Hering,
Ahmet Keleş,
Peter Kopietz, 
Fabian Kugler,
Vincent Noculak,
Janik Potten,
Stephan Rachel,
Marc Ritter,
Dietrich Roscher,
Marlon Rück,
Yannik Schaden,
Michael Scherer,  
Bendedikt Schneider,
and Jan von Delft 
for helpful discussions.
The work of Y.I.~was performed in part and completed at the Aspen Center for Physics, which is supported by National Science Foundation grant PHY-2210452. The participation of Y.I.~at the Aspen Center for Physics was supported by the Simons Foundation. The research of Y.I.~was supported, in part, by the National Science Foundation under Grant No.~NSF~PHY-1748958 during a visit to the Kavli Institute for Theoretical Physics (KITP), UC Santa Barbara, USA for participating in the programs ``A Quantum Universe in a Crystal: Symmetry and Topology across the Correlation Spectrum'' and ``A New Spin on Quantum Magnets''. Y.I.~acknowledges support from the ICTP through the Associates Programme and from the Simons Foundation through grant number 284558FY19, IIT Madras through the Institute of Eminence (IoE) program for establishing QuCenDiEM (Project No. SP22231244CPETWOQCDHOC), the International Centre for Theoretical Sciences (ICTS), Bengaluru, India during a visit for participating in the program ``Frustrated Metals and Insulators" (Code: ICTS/frumi2022/9). Y.I.~acknowledges the use of the computing resources at HPCE, IIT Madras. 
N. N., J. R., and S.T. acknowledge support from the Deutsche Forschungsgemeinschaft (DFG, German Research Foundation), within Project-ID 277101999 CRC 183 (Project A04).
B.S. is supported by a MCQST-START fellowship and by the Munich Quantum Valley, which is supported by the Bavarian state government with funds from the Hightech Agenda Bayern Plus. 
D.K. and S.T. acknowledge partial funding from the Deutsche Forschungsgemeinschaft (DFG, German Research Foundation) within Project-ID 277146847, SFB 1238 (projects C02 and C03). 
The work in W\"urzburg was supported by DFG Grant No. 258499086-SFB 1170 and the W\"urzburg-Dresden Cluster of Excellence on Complexity and Topology in Quantum Matter, Grant No. 390858490-EXC 2147. 
N.N.~thanks IIT Madras for funding a three-month stay through an International Graduate Student Travel award that facilitated completion of this work. J. R. and R. T. thank IIT Madras for a Visiting Faculty Fellow position under the IoE program which facilitated the completion of this work and writing of manuscript. 
B.S., N.N., D.K., J.R., and S.T.~acknowledge usage of the JUWELS cluster at the Forschungszentrum J\"ulich and the Noctua2 cluster at the
Paderborn Center for Parallel Computing ($PC^{2}$). \\

\bibliography{bibliography}


\clearpage

\appendix

\section{Multi-local PFFRG flow equations in asymptotic frequency parametrization}
\label{app:floweqasymptoticfreq}
For reference, in this appendix we list the multi-local flow PFFRG flow equations as well as the expression for the spin-spin susceptibility in the asymptotic frequency parametrization for the three channels individually. The frequency arguments of the channel derivatives are taken to be in natural parametrization, while the arguments of the vertices on the right-hand side of the equations always list all three parametrizations, as they can be decomposed in a sum of channel contributions. For this, we use the convention
\begin{equation}
  \vertex{i_1i_2}{\mu_{1234}}{\omega_s&\nu_s&\nu_s'}{\omega_t&\nu_t&\nu_t'}{\omega_u&\nu_u&\nu_u'},
\end{equation}
where we also introduce the shorthand $\mu_{1234} = \mu_1\mu_2\mu_3\mu_4$ for spin indices. We refrain from specifying the parametrization in spin space, as the remaining degressof freedom in this space are highly dependent on the Hamiltonian.

Furthermore, the symmetric Katanin substituted bubble derivative
\begin{equation}
P(\omega,\nu) = G^\Lambda(\omega) S_\text{kat}^{\Lambda}(\nu) + S_\text{kat}^{\Lambda}(\omega) G^\Lambda(\nu)
\end{equation}
is used.

\subsection{Self-energy}
The multilocal self-energy flow reads
\begin{equation}
	\begin{split}
		 &\frac{\dd}{\dd \Lambda} \gamma(\omega)=  \int \dd\omega' \frac{\pdv{\Lambda}R(\omega',\Lambda)}{\omega'+\gamma(\omega')} \bigg\{\\
 &\vertex{i_1 i_1}{\mu_{1 1 1 1}}
 {\omega+\omega'&\omega'/2-\omega/2&\omega'/2-\omega/2}{\omega-\omega'&\omega'/2+\omega/2&\omega'/2+\omega/2}{0&\omega'&\omega} \\
 &-\sum_{j,\mu_2} 2\vertex{i_1 j}{\mu_{1 2 1 2}}
 {\omega+\omega'&\omega/2-\omega'/2&\omega'/2-\omega/2}
 {0&\omega&\omega'}
 {\omega-\omega'&\omega/2+\omega'/2&\omega/2+\omega'/2}\bigg\} 
	\end{split}\label{eq:seflownew}
\end{equation}
where $i_1$ and $\mu_1$ are arbitrary site and spin indices, respectively, due to the the self-energy being local and diagonal in spin.

\subsection{S-channel}
\begin{equation}
	\begin{split}
		&\dot{g}_{\mathrm{s}{i_1}{i_2}}^{\mu_{1'2'12}}\left(s, \nu_s, \nu_s'\right) = \frac{1}{2\pi} \int \dd\omega\sum_{\mu_3\mu_4} P(\frac{s}{2}+\omega, \frac{s}{2}-\omega) \\
    &\times \vertex{i_1i_2}{\mu_{3412}}{s&\omega&\nu_s'}{-\nu_s'-\omega&(s+\omega-\nu_s')/2&(s-\omega+\nu_s')/2}{-\nu_s'+\omega&(s+w+\nu_s')/2&(s-\omega-\nu_s')/2}\\
 &\times\vertex{i_1i_2}{\mu_{1'2'34}}{s&\nu_s&-\omega}{\omega-\nu_s&(s+\omega+\nu_s)/2&(s-\nu_s-\omega)/2}{\omega+\nu_s&(s+\nu_s-w)/2&(s-\nu_s+\omega)/2}
	\end{split}
 \end{equation}

\subsection{T-channel}

\begin{equation}
  \begin{split}
  &\dot{g}_{{i_1}{i_2}}^{\mu_{1'2'12}}\left(t, \nu_t, \nu_t'\right) = \frac{1}{2\pi} \int \dd\omega \sum_{\mu_3\mu_4} P(\omega+\frac{t}{2}, \omega-\frac{t}{2}) \bigg\{\\
&- \sum_j  \vertex{i_1 j}{\mu_{1'413}}{\nu_t+\omega&(\nu_t-t-\omega)/2&(\omega-\nu_t-t)/2}{t&\nu_t&\omega}{-\omega+\nu_t&(w+\nu_t-t)/2&(w+t+\nu_t)/2}  \\
&\times \vertex{j i_2}{\mu_{32'42}}{\nu_t'+\omega&(\omega-t-\nu_t')/2&(\nu_t'-t-\omega)/2}{t&\omega&\nu_t'}{-\nu_t'+\omega&(\nu_t'-t+\omega)/2&(t+\omega+\nu_t')/2}\\
&+\vertex{i_1 i_2}{\mu_{1'413}}{\nu_t+\omega&(\nu_t-t-\omega)/2&(\omega-t-\nu_t)/2}{t&\nu_t&\omega}{-\omega+\nu_t&(\omega-t+\nu_t)/2&(t+\nu_t+\omega)/2}  \\
&\times \vertex{i_2 i_2}{\mu_{32'24}}{\nu_t'+\omega&(\nu_t'+t-\omega)/2&(\nu_t'-t-\omega)/2}{w-\nu_t'&(\nu_t'+t+\omega)/2&(\nu_t'-t+\omega)/2}{t&\nu_t'&\omega}\\
&+\vertex{i_1 i_1}{\mu_{1'431}}{\nu_t+\omega&(t+\omega-\nu_t)/2&(\omega-t-\nu_t)/2}{\nu_t-\omega&(t+\nu_t+\omega)/2&(\nu_t-t+\omega)/2}{t&\omega&\nu_t} \\
&\times \vertex{i_1 i_2}{\mu_{32'42}}{\nu_t'+\omega&(-t-\nu_t'+\omega)/2&(\nu_t'-t-\omega)/2}{t&\omega&\nu_t'}{-\nu_t'+\omega&(\nu_t'-t+\omega)/2&(\nu_t'+t+\omega)/2}\\
&\bigg\} 
  \end{split}
\end{equation}
\subsection{U-channel}
\begin{equation}
	\begin{split}
		&\dot{g}_{{i_1}{i_2}}^{\mu_{1'2'12}}\left(u, \nu_u, \nu_u'\right) = \frac{1}{2\pi} \int \dd\omega \sum_{\mu_3\mu_4}P(\omega-\frac{u}{2}u, \omega+\frac{u}{2})\\
 &\times\vertex{i_1i_2}{\mu_{2'413}}{\nu_u+\omega&(\nu_u+u-\omega)/2&(-u+\nu_u-\omega)/2}{\omega-\nu_u&(\omega+u+\nu_u)/2&(-u+\nu_u+\omega)/2}{u&\nu_u&\omega}  \\
 &\times \vertex{i_1i_2}{\mu_{31'42}}{\nu_u'+\omega&(\omega+u-\nu_u')/2&(-u+\omega-\nu_u')/2}{\nu_u'-\omega&(\nu_u'+\omega+u)/2&(-u+\omega+\nu_u')/2}{u&\omega&\nu_u'}
	\end{split}
 \end{equation}
\subsection{Spin-spin correlator}

\begin{equation}
	\begin{split}
    &\chi^{\alpha\beta,\Lambda}_{ij}(\omega) =-\frac{1}{4\pi} \int \dd \omega G(\omega-\nu/2) G(\omega+\nu/2)\delta_{ij}\delta_{\alpha\beta}\\
		   &-\frac{1}{16\pi^2} \int \dd\omega \int \dd\omega' \sigma^\alpha_{\mu_1\mu_1'}\sigma^\beta_{\mu_2\mu_2'}\\
       &\times G(\omega-\nu/2) G(\omega+\nu/2)G(\omega'-\nu/2) G(\omega'+\nu/2)\\
		   &\times\vertex{ij}{\mu_{1'2'12}}{\omega+\omega'&(\omega-\omega'-\nu)/2&(\omega'-\omega-\nu)/2}{\nu&\omega&\omega'}{\omega'-\omega&(\omega+\omega'-\nu)/&(\omega+\omega'+\nu)/2} 
	\end{split}
 \end{equation}

 \section{Numerical implementation}

\subsection{Lattice symmetries}
\label{app:lattice}

Vertices within the PM- and PFFRG take the form $\Gamma_{a,ij}(s,t,u)$ where $a$ indicates the type of vertex (i.e. $a = s,d$ for the PFFRG in Heisenberg systems), $i,j$ refer to sites that are ``effectively interacting'' by means of higher-order interactions and $s,t$ and $u$ are frequency arguments.

Here, we are not interested in frequency arguments and will push them into the definition of the vertex type $a \rightarrow (a,s,t,u)$, $\Gamma^\Lambda_{a,ij}(s,t,u) \rightarrow \Gam{a}{i}{j}$ to save digital ink and highlight the spatial dependence.
In both the PM- and PFFRG each contribution in the site summations can be written in the form
\begin{align}
    \partial_\Lambda \Sigma^\Lambda_{\text{sum}}(\R_i) &= \sum_k \Gam{a}{i}{k} S(\R_k) \nonumber\\    
    \partial_\Lambda \Gam{\text{sum}}{i}{j} &= \sum_k \Gam{a}{i}{k} \Gam{b}{k}{j} P(\R_k, \R_k) \text{,} \label{eq:sitesum}
\end{align}
where $S(\R_k)$ is the single-scale propagator and $ P(\R_k, \R_k) = S(\R_k) G(\R_k)$ the usual bubble propagator.
In particular, the order of indices $kj  \neq jk$, is not generally interchangeable.

For infinitely large systems, we would have infinitely many equations and an infinite site summation for each of them. However, in magnetically disordered phases, the effective interaction between two sites is expected to decay with their distance, so we may neglect vertices with sufficiently large $|\R_i - \R_j|\gg L$, where $L$ is a numerical parameter corresponding to a system size. Effectively, this corresponds to a maximum correlation length $\xi_L \sim L$. We may also make use of lattice symmetries to further reduce the number of vertices we need to consider. These lattice symmetries are \emph{global} transformations $\R_i \rightarrow \operatorname{L}(\R_i)$ which leave the lattice (and hence vertices) invariant
\begin{equation}
    \Gam{a}{i}{j} = \Gamma^\Lambda_a(\operatorname{L}(\R_i), \operatorname{L}(\R_j)).
\end{equation} 
As an example, all lattices are by definition invariant under translations along any of the lattice vectors $\bm{a}_l$. This is most important, as it allows us to only consider vertices in which the first site argument, $\R_i$, lies in the first unit cell. Together with the assumption of finite length vertices, we may restrict ourselves to a finite number of $N_\text{sites}$ sites surrounding this reference unit cell. Here, it is convenient refer to a site by an integer combination of lattice vectors together with one of the unit cell's basis vectors $\bmR_i = [n_1, n_2, n_3, b] \equiv n_1 \bm{a}_1 + n_2 \bm{a}_2 + n_3 \bm{a}_3 + \bm{b}_b$. Beside translation, there can also be other symmetries, such as rotations or mirror symmetries, which may in particular transform a site from one sublattice to another. This further reduces the number of reference sites $\R_i$ that we need to consider, see for example \cref{fig:Pyrochlore}. Usually, many (or even \textit{all}) sites in the unit cell are \textit{equivalent}. Therefore, it suffices to identify a number $\NU \leq \text{max}(b)$ of \textit{inequivalent} sites in our reference unit cell that are distinct by all symmetries. Where neccessary, we will then use an index $x_i = 1,2,\dots ,\NU$ which shall label the \textit{type} of site $i$.  

The implementation of these symmetries is as follows: First, given a reference site $\R_i$, we identify all pairs of sites within the maximum correlation distance $\bmR_{kj} \equiv (\R_i, \R_j)$ and subsequently reduce this set to a minimum set of \emph{inquivalent pairs} $\Rin_{ij} $ which can be used to reconstruct all other pairs via lattice symmetries. 

Subsequently, we pre-compute all the terms that appear in the site summation for each inequivalent vertex $\Gam{a}{i}{j}$ on the lhs. of \cref{eq:sitesum} and perform lattice symmetries such that each term contains only the symmetry inequivalent vertices. In short, the steps that are needed are summarized below, with more detailed descriptions following in the remainder of this section.

\begin{itemize}
    \item \textbf{Generate inquivalent pairs:} We need to first generate a minimal list of sites which are in inequivalent pairs to each reference site $\{\Rin_{ij}\}= {\bmR_{1},\bmR_{2},\dots}$. In the FRG, we then solve differential equations for vertices represented as four-dimensional arrays, where each index corresponds to a particular pair $\Rin_{ij}$ and the other three are for frequency arguments. 
    At this step, we also generate a list $\{(x_i,x_j)\}$ of the types of sites $i$ and $j$ which will be needed to index the propagators in the FRG. It is recommended to save both lists in order to re-identify the inequivalent pairs with the real-space structure after FRG results are obtained.
    \item \textbf{Compute the site summation: } For each of these pairs $\Rin_{ij}$ we need to evaluate the site-summation and apply lattice symmetries to express the pairs $\bmR_{ki}$ and $\bmR_{kj}$ through inequivalent pairs computed via FRG. Here, we will save the corresponding indices of each inequivalent pair in an abstract matrix $M_{kl}$ with dimensions $\Nsum \times \NPairs $, i.e.  the number of terms appearing in the summation $\Nsum$ and the number of inequivalent pairs $\NPairs$. Each matrix element will contain a tuple of four positive integers: $M_{kj} = (i_{ki},i_{kj},x,m)_{kj}$. The first two, $i_{ki},i_{kj} \leq \NPairs$, are the indices of inequivalent pairs corresponding to $\bmR_{ki}$ and $\bmR_{kj}$, or in other words the indices of the vertices appearing in a term in the sum when computed in the FRG. $x$ refers to the type of site $k$ which is needed for $P(\R_k,\R_k) \equiv P_{x_k,x_k}$ and $S(\R_k) \equiv S_{x_k}$ in \cref{eq:sitesum}. The integer $m$ is the multiplicity of the term, i.e. the number of times the term appears in the sum. Initially these multiplicities are set to one. After the matrix is constructed, we may identify duplicate entries for each column index $j$, and reduce them by adding the multiplicities.
    \item \textbf{Construct mapping arrays:} Some positive and negative frequency arguments of vertices are related by exchanges of sites $\Gam{a}{i}{j} \leftrightarrow \Gam{x'}{j}{i}$. As we only compute positive values of the frequencies in the spin FRG, we need to provide a list of indices that maps an inequivalent pair $\Rin_{ij}$ to its corresponding site-swapped pair $\text{invpairs}[\Rin_{ij}] = \bmR_{ji}$. Likewise, we need to determine an array containing our couplings $J_{ij} \rightarrow J_{\Rin_{ij}}$ for each inequivalent pair, as our main program will not have any information about the actual geometry.
    Finally, there are a few terms in the flow equations which contain local, or \emph{onsite} vertices $\Gam{a}{i}{i}$, so the FRG needs to know the positions of onsite pairs in our pair list. This can be either fixed by a suitable sorting convention, or by providing a short list with the appropriate indices.
\end{itemize}

\subsection*{Step 1: generation of inequivalent pairs}
At the heart of any efficient implementation lies an identification of a list of inequivalent pairs $\{\Rin_{ij}\}$ which characterizes all the vertices $\Gam{a}{i}{j}$ that are needed for a complete treatment of the flow equations.

Starting from \emph{each} reference site $i$, we may generate a list of \textit{paired sites}, which are within a given distance to site $i$. One unbiased approach is to progressively add nearest neighbors to the list, starting from site $i$, such that in the end distances up to $N_\text{Len}$ nearest neighbor pairs are included, but other choices are also possible such as including all sites within a sphere of radius $L$. We simply append all those for the other reference sites to the \textit{same list}, such that it is of the form $[\bmR_{11}, \bmR_{12}, \dots , \bmR_{21}, \bmR_{22},\dots ]$)! In order to keep track of the types of sites corresponding to the pairs, we simultaneously generate a list ``PairTypes'' with equal length which contains the corresponding types of sites $(x_i,x_j)$. As a result, the arrays representing vertices within FRG need four dimensions (one for real space pairs and three for frequencies) and whenever we need to evaluate propagators we look up the types of sites $i$ and $j$.
Afterwards we reduce all parts of this list that are redundant by one of the symmetries. We have already made use of several symmetries by fixing $\R_i$, so not all of them will be of further use. To be more precise, we are now only interested in transformations which leave our \textit{reference site} invariant, as this allows for symmetries $\Gamma^\Lambda_a(\Rin_{ij}) = \Gam{a}{1}{j} = \Gamma^\Lambda_a(\R_1, \operatorname{L}(\R_{j'})) =  \Gamma^\Lambda_a(\Rin_{ij'})$. A simple example of this step is given in \cref{fig:pairs_Tetramer}.

To treat the case of more than one site per unit cell, let us 
consider the pyrochlore lattice: As shown in \cref{fig:Pyrochlore}, we can fix our reference site $i$ to be the white site, located at the origin. For $N_\text{Len} = 1$, there are seven pairs in our initial list, one for each corner of the tetrahedra and the onsite pair. Inversion leaves the white site invariant, but it maps the coordinates of all other sites to negative ones, reducing the number of inequivalent pairs to four. In the same way, the white site does not transform under $C_3$ rotations and mirror reflections around the $x=y$-plane. We may thus systematically find our inequivalent pairs by iterating over the list of all pairs, and deleting all pairs that are obtained by applying a symmetry transformation to the current element. For the pyrochlore lattice, this means we may divide our paired sites into $2 \times 3 \times 2 = 12$ equivalent sectors and only consider one of them which by itself reduces numerical effort by a factor of $12$. 

It is advisable to sort this list, for instance after the separation distance of each pair so that the onsite pair will always be the first element of this list. 
In the FRG, we may compute the susceptibility $\chi_{ij} \rightarrow \chi_{\Rin_{ij}}$ for each inequivalent pair. When evaluating the results, it is thus necessary to save which number corresponds to which pair of sites, so it is advisable to save the list of inequivalent pairs in lattice or real-space coordinates that was generated in step 1.
It is necessary to generate a mapping between a given arbitrary pair of sites and the corresponding inequivalent one. This mapping is easily represented by a dictionary, which can be generated similar to the previous step, i.e. by applying the full list of point group symmetries to each inequivalent pair, such that the inequivalent pairs give the values and the generated pairs are the keys, as shown in \cref{alg:pairToInequiv}.

\begin{figure} 
    \centering
    \includegraphics[width = 0.5\linewidth]{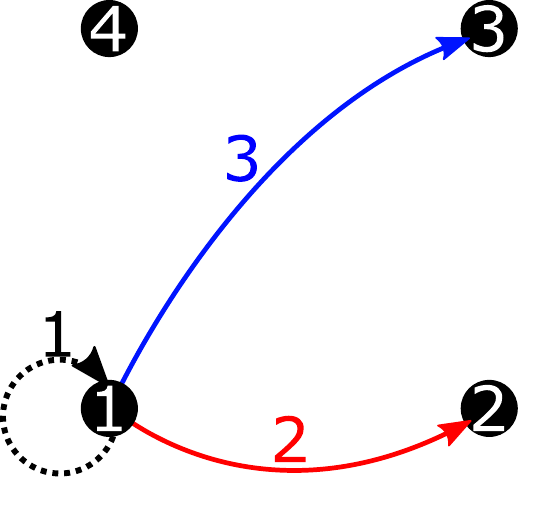}
    \caption{Displayed in different colors are all inequivalent pairs. All other possible pairings are equivalent, i.e., the pair $(4,2)$ is equivalent to $(1,3)$ by a $C_4$ rotation around the center.}
    \label{fig:pairs_Tetramer}
\end{figure}
\begin{figure}
    \centering
    \includegraphics[width = 0.95\linewidth]{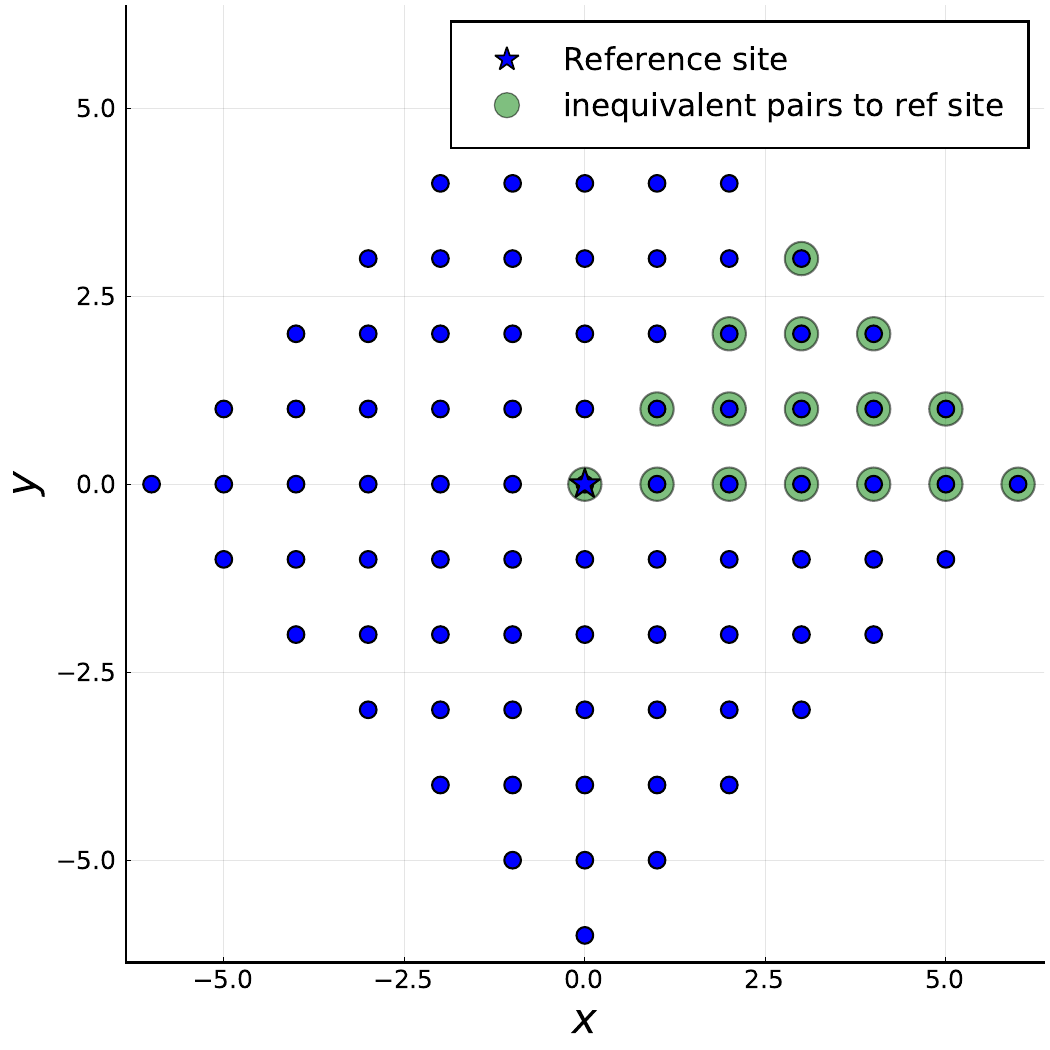}
    \caption{Blue circles: Sites correlated to the reference site at the origin (star), and inequivalent sector of pair sites  (green circles) that are found relative to the reference site.}
    \label{fig:Square_inequivalent}
\end{figure}

\begin{figure}
    \centering
    \includegraphics[width = 0.95\linewidth]{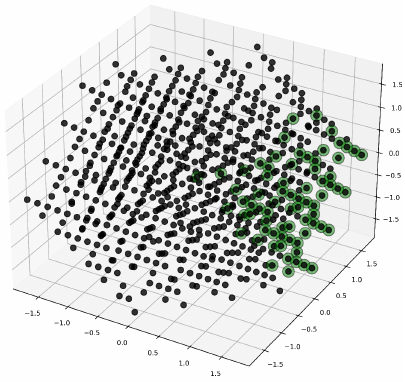}
    \caption{ Sites of the pyrochlore lattice within a finite number of nearest neighbor bonds to the reference site at $(0,0,0)$. Highlighted in green are the sites that correspond to the symmetry inequivalent pairs $\Rin_{0,j}$ with respect to the reference site. Using translation and point group symmetries, each other pair occuring in the flow equations can be mapped onto this selection.} \label{fig:Pyrochlore}
\end{figure}
\begin{figure}
    \centering
    \includegraphics[width = 0.95\linewidth]{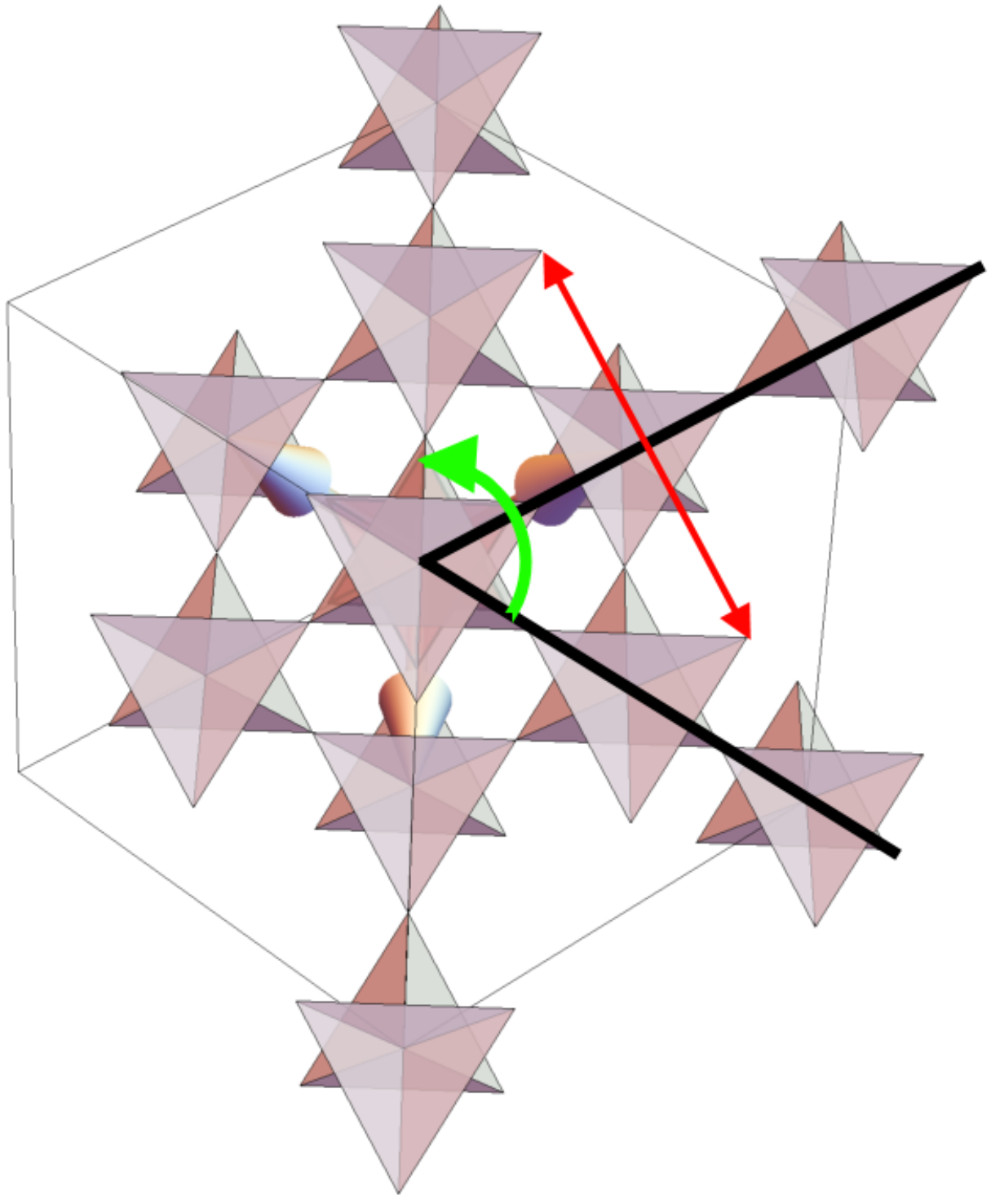}
    \caption{Selection of inequivalent pairs in the pyrochlore lattice. Using the $C_3$ rotation symmetry (green) and the mirror symmetry (red) together with inversion at the origin, only $1/12$ of all sites are needed to reconstruct arbitary pairs $\bmR_{ij}$.} \label{fig:PyrochloreProjections}
\end{figure}

\begin{algorithm}
    \caption{Find inequivalent pair from pair of sites}
    \begin{algorithmic} 
    \REQUIRE sites $\bmR_k ,\bmR_j$  \hfill//in lattice coords with basis $\bmR_k = [k_1,k_2,k_3,b_k]$
    \REQUIRE Symmetries[]  \hfill// Array of symmetries
    \REQUIRE RefSites[]  \hfill// Array of sublattice indices for reference sites
    
    \FOR{Sym $\in$ Symmetries \&\& $R_k.b \notin $ Refsites }   
        \STATE $\bmR_k = \text{Sym}(\bmR_k )$ \hfill// there might also be several distinct symmetries to use here
        \STATE $\bmR_j = \text{Sym}(\bmR_j )$
    \ENDFOR
    \STATE $\bmR_k =  [0,0,0,b_i]$ \hfill// global translation of $\bmR_k$ to first unit cell 
    \STATE $\bmR_j =  [j_1-k_1,j_2-k_2,j_3-k_3,b_j]$ 
    \RETURN $(\bmR_k ,\bmR_j)$
    \end{algorithmic}
    \label{alg:pairToInequiv}
\end{algorithm}
\subsection*{Step 2: computing the site summation}
After the inequivalent pairs $\Rin_{ij}$ are identified, we may perform the site summation for each corresponding vertex $\Gamma_a(\Rin_{ij})$. For instance, selecting the onsite pair $\Rin_{ii}$ for the simple system of four sites in \cref{fig:pairs_Tetramer} we obtain from \cref{eq:sitesum}
\begin{align}
    \partial_\Lambda \Gamma^\Lambda_{\text{sum}}(\Rin_{ii}) &= \sum_k \Gam{a}{i}{k} \Gam{b}{k}{j} P(\R_k, \R_k) \nonumber \\
    &= \Gam{a}{1}{1} \Gam{b}{1}{1} P(\R_1,\R_1) \nonumber \\
    &+ \Gam{a}{1}{2} \Gam{b}{2}{1} P(\R_2,\R_2) \nonumber \\
    &+ \Gam{a}{1}{3} \Gam{b}{3}{1} P(\R_3,\R_3) \nonumber \\
    &+\Gam{a}{1}{4} \Gam{b}{4}{1} P(\R_4,\R_4) \nonumber \\
    \partial_\Lambda \Gamma^\Lambda_{\text{sum}}(1)&= \Gamma^\Lambda_a(1) \Gamma^\Lambda_b(1) P(1) +\Gamma^\Lambda_a(2) \Gamma^\Lambda_b(2) P(1) \nonumber \\
     &+ \Gamma^\Lambda_a(3) \Gamma^\Lambda_b(3) P(1) +\Gamma^\Lambda_a(2) \Gamma^\Lambda_b(2) P(1) \text{,}
    \label{eq:onsiteSum}
\end{align}
where in the last step symmetries were used so that each pair of sites could be replaced by its corresponding inequivalent pair. Due to the equivalence of sites, all propagators are equal and, hence, so are the second and fourth term in this sum. Doing this step for all other pairs allows us to implement the site summation in the FRG as

\begin{align}
    \partial_\Lambda \Gamma^\Lambda_{\text{sum}}(l) &=  \sum_{k}^{N_\text{pairs}} m_{k}(l) \Gamma^\Lambda_a(i_k) \Gamma^\Lambda_b(j_k) P_{x_k}.
    \label{eq:splitsum_gen}
\end{align}
Here, $i$ and $j$ and $l$ do not label the distinct sites but rather inequivalent pairs. These indices are found by mapping each pair of sites to the corresponding inequivalent pair using the dictionary generated in the previous step.
During this step it is thus also necessary to save $x_k$, ideally in a matrix $M_{kl}$ together with the site pairs and the multiplicity. 
As demonstrated in \cref{eq:onsiteSum}, due to point-group symmetries certain terms $\Gamma^\Lambda_a(i_k) \Gamma^\Lambda_b(j_k)$ will appear several times, meaning they may be added up to a multiplicity $m_k(l)$. This is not strictly neccessary but it is easy to implement and will reduce the time that is spent in the $k$ summation of the FRG. 

With this at hand, the site summation may be easily computed: We first prepare a ``matrix'' of size $\NPairs \times N_\text{sites}$. Then, for each $j = 1,2,\dots \NPairs$ we sum over all sites in the system, map the pairs $(i,k)$ and $(k,j)$ inequivalent pairs with at indices $i_k$ and $j_k$ in our list of inequivalent pairs and finally write these indices in our matrix. One can then search for multiplicities of these pairs in each row of the matrix and further reduce the number of columns by saving the multiplicity for each pair of vertices. This matrix can then simply be passed as an argument to the FRG code, which can then evaluate the sum by inserting its elements as indices to vertex functions without any further information about the particular lattice geometry.
Note that the site summation in the self energy is also contained within this matrix, as the summation of the onsite pair will contain a simple summation of all sites in the system, which can be seen from \cref{eq:onsiteSum}.
\subsection*{Constructing mapping arrays}
The final step is simple: In general we have $\Gamma_{ij} \neq \Gamma_{ji}$. Since we want to swap signs of frequency arguments when evaluating vertices, we need to give our program a list which maps an inequivalent pair $\Rin_{i,j}$ to its inverted pair $\bmR_{j,i} = \Rin_{i,j'}$ for some $j'$. This is actually just a special case of the operations we have done within the site summation and thus we may make use of the dictionary in \cref{alg:pairToInequiv} to generate an array ``$\text{invpairs}[j] = j'$''. This array is also passed to the FRG and consequently used whenever we need to change the sign in either the $s$ or $u$ frequency.
Similarly, the initial couplings $J_{ij}$ should be passed to the FRG. If we have sorted our pairs according to the distance to the reference site, we may easily set nearest, next- nearest and further couplings. More complicated couplings can also be set as long as we know which physical pair of sites a particular inequivalent pair index corresponds to.

\subsection{Frequency content of vertex functions}
\label{app:frequencydiscr}

Not only the numerical treatment of the spatial, but also the frequency dependence of both the self-energy and the vertex function need some consideration. As all frequency arguments can take on countably (in case of finite-T Matsubara frequencies) or over--countably (in the $T=0$ case) many values, any numerical implementation has to approximate the Matsubara frequency axis.

In the finite $T$ case, considering only the first $N_\omega$ discrete Matsubara frequencies, i.e. all $\i \omega_n$ with $|n|<N_\omega$ has proven to be most effective~\cite{Niggemann2021a}. If the flow equations require a quantity at a larger frequency than considered in the grid, constant extrapolation is used as a zero-order approximation to the asymptotic behavior of the vertex discussed in \cref{subsec:frequencyparam}. Additionally, energy conservation excludes some combinations of Matsubara indices for the three frequency arguments of the vertex for being unphysical. Using the parametrization in terms of transfer frequencies $s,t,u$, as used in \cref{subsec:PMFRG}, e.g., requires $n_s+n_t+n_u$ to be odd. As unphysical combinations will never occur in the FRG flow, these can be excluded~\cite{Niggemann2021}.

At $T=0$, Matsubara frequencies become continuous. Therefore a discrete frequency meshing has to be imposed on all frequency-dependent functions. The structure of the vertex function, as illustrated in \cref{subsec:frequencyparam} is such that there is more structure around the origin in frequency space, while larger frequencies are dominated by asymptotic with less features. To efficiently capture all information in the vertex, denser mesh points are needed for low frequencies compared to higher ones~\cite{Ritter2022}.

To achieve this, purely logarithmically spaced meshes~\cite{Reuther2010,Iqbal2016b,Buessen2021b} or combinations of linear spacing around zero frequency and logarithmic tails~\cite{PFFRGSolver,Kiese2020,Thoenniss2020} have been used. As the location of the features in frequency space shifts with $\Lambda$, adapting the meshes to the current form of the vertex leads to improved numerical accuracy. Sophisticated scanning routines have been put forward to achieve low numerical errors, which is especially needed in multiloop implementations, as due to the iterative nature of the multiloop corrections errors will proliferate~\cite{Kiese2020,Thoenniss2020,Ritter2022}.

The frequency integration on the right hand-side of the FRG flow equations for $T=0$ combined with the non-equal spacing of the frequency meshes necessitates a means to extract vertex values  at an arbitrary point $(\omega,\nu,\nu')$ in three-dimensional frequency space.
To this end, a multi-linear interpolation scheme
\begin{equation}
    \begin{split}
        \Gamma(&\omega, \nu, \nu') = \big[\\ &\Gamma(\omega_{i_<}, \nu_{i_<}, \nu'_{i_<}) (\omega_{i_>} - \omega) (\nu_{i_>} - \nu) (\nu'_{i_>} - \nu')  \\
        +&\Gamma(\omega_{i_<}, \nu_{i_<}, \nu'_{i_>}) (\omega_{i_>} - \omega) (\nu_{i_>} - \nu) (\nu' - \nu'_{i_<})  \\
        +&\Gamma(\omega_{i_<}, \nu_{i_>}, \nu'_{i_<}) (\omega_{i_>} - \omega) (\nu - \nu_{i_<}) (\nu'_{i_>} - \nu')  \\
        +&\Gamma(\omega_{i_<}, \nu_{i_>}, \nu'_{i_>}) (\omega_{i_>} - \omega) (\nu - \nu_{i_<}) (\nu' - \nu'_{i_<})  \\
        +&\Gamma(\omega_{i_>}, \nu_{i_<}, \nu'_{i_<}) (\omega - \omega_{i_<}) (\nu_{i_>} - \nu) (\nu'_{i_>} - \nu')  \\
        +&\Gamma(\omega_{i_>}, \nu_{i_<}, \nu'_{i_>}) (\omega - \omega_{i_<}) (\nu_{i_>} - \nu) (\nu' - \nu'_{i_<}) \\        
        +&\Gamma(\omega_{i_>}, \nu_{i_>}, \nu'_{i_<}) (\omega - \omega_{i_<}) (\nu - \nu_{i_<}) (\nu'_{i_>} - \nu')  \\        
        +&\Gamma(\omega_{i_>}, \nu_{i_>}, \nu'_{i_>}) (\omega - \omega_{i_<}) (\nu - \nu_{i_<}) (\nu' - \nu'_{i_<})  \\
        &\big] \frac{1}{(\omega_{i_>} - \omega_{i_<}) (\nu_{i_>} - \nu_{i_<}) (\nu'_{i_>} - \nu'_{i_<})},
    \end{split}
\end{equation}

is used, where the indices $i_>(i_<)$ indicate the nearest larger (smaller) frequency in the grid on the respective frequency axis. This scheme can be used for the full vertex in the transfer frequency parametrization or, turning to the asymptotic frequency paramatrization, in the three different diagrammatic channels separately. The vertex function asymptotics as well as the self-energy are interpolated using the two- and one-dimensional version of this scheme.

 For an efficient implementation of the asymptotic parametrization, the kernel functions $K$ defined in \eq{eq:kernelfuncs} are not the most suitable choice, as in this formulation calculating the value of a single channel at a specific frequency point amounts to the interpolation and subsequent summation of all three functions. Therefore, one can define computationally more favorable functions $Q$ according to~\cite{Wentzell2020,Kiese2020}
\begin{align}
    Q_3^c(\omega_c,\nu_c,\nu_c') &= \dot{g}_c(c,\nu_c,\nu_c')\\
    Q_2^c(\omega_c,\nu_c) &=  \lim_{|\nu_c'|\to\infty}\dot{g}^c(\omega_c,\nu_c,\nu_c')\\
    \bar{Q}_2^c(\omega_c,\nu_c') &= \lim_{|\nu_c|\to\infty}\dot{g}^c(\omega_c,\nu_c,\nu_c')\\
    Q_1^c(\omega_c) &= \lim_{|\nu_c|,|\nu_c'|\to\infty}\dot{g}^c(\omega_c,\nu_c,\nu_c').
\end{align}

Depending on the exact value of the frequencies, only a single evaluation of $Q_3^c$ or, if one or two freuqencies are outside of the mesh, $Q_2^c$ or $Q_1^c$ is required, respectively.

These new functions can be related to the original kernels according to
\begin{align}
    \begin{split}Q_3^c(\omega_c,\nu_c,\nu_c') &=  K_1^c(\omega_c) + K_2^c(\omega_c,\nu_c) \\&+ \bar{K}_2^c(\omega_c, \nu_c') + K_3^c(\omega_c,\nu_c,\nu_c')\end{split}\\
    Q_2^c(\omega_c,\nu_c) &= K_1^c(\omega_c) + K_2^c(\omega_c,\nu_c)\\
    \bar{Q}_2^c(\omega_c,\nu_c') &=  K_1^c(\omega_c) + \bar{K}_2^c(\omega_c, \nu_c')\\
    Q_1^c(\omega_c) &= K_1^c(\omega_c).
\end{align}

Numerically, the asymptotic parts can be calculated setting the according frequencies to a large value. The numerical advantage of the $Q$ functions defined here, however, comes at a price: in the parametrization using the kernels $K$, the frequency discretization on all axes for all asymptotic classes, i.e. $K_1$, $K_2$, $\bar{K}_2$ and $K_3$ can be chosen independently, such that the numerically cheaper to calculate $K_1$ class can be augmented by a higher resolution mesh. 
Using the sum of these kernels in form of the $Q$s defined above, we do not have this choice anymore. Although one would naively expect that the same split can be done for the limiting functions, i.e. $Q_1$, $Q_2$, $\bar{Q}_2$ and $Q_3$, in a real implementation, this will lead to interpolation artifacts at the boundaries of the frequency mesh, which lead to unphysical errors in the flow. The reduced accuracy in the kernel functions with lower frequency degrees of freedom, however, turned out to not alter the accuracy of the calculation as a whole, cf. \mRef{Ritter2022}.

Therefore, the only split we introduce in the frequency meshes is to allow for different meshes for the bosonic and fermionic axes, with the latter being the same for both $\nu$ and $\nu'$. The reason for this is the symmetry under exchange of $\nu$ and $\nu'$, as discussed in \cref{subsec:frequencyparam}. We also allow the $t$-channel contributions to be defined on a different set of meshes than the $s$- and $u$-channel, with the latter utilizing the same discretization, again due to the symmetry between both under exchange of the two fermionic arguments. This split in the diagrammatic channels turns out to be crucial for tracking the interplay between magnetic ordering tendencies in the $t$ channel and paramagnetic behavior stemming from the $s$- and $u$-channels at low RG scale $\Lambda$. 
Additionally, spin- and density-part of the vertex are allowed to be defined on possibly different meshes, as their frequency content qualitatively differs.

\subsection{Frequency integration}
The evaluation of the loops on the right-hand side of the FRG equations in the $T=0$ case calls for a quadrature rule to be used. In early implemenations, simple trapezoidal quadrature using the mesh points as integration points was used~\cite{Buessen2021b}. As discussed in \cref{app:frequencydiscr}, this leads to a good resolution only around the origin of the integration domain. As the integrand at $1\ell$ level always includes a propagator $G^\Lambda$ and a single-scale propagator $S^\lambda$ and the latter is more sharply peaked, this leasds to remarkably accurate results, when the integration domain is shifted, such that the single-scale propagator is peaked at the origin.

Using a more symmetric frequency parametrization, as in \cref{app:floweqasymptoticfreq}, this is no longer the case and adaptive routines have to be implemented, see e.g. \mRef{Ritter2022}.

\subsection{Differential equation solver}
As for numerical quadrature, the integration of the FRG flow has to be done numerically. As multiple orders of magnitude in $\Lambda$ have to be covered, the step-size during the flow should be adapted, with initially larger steps becoming smaller while approaching either $\Lambda=0$ or a flow breakdown. While an adaptive step-size Euler method performs quite well~\cite{Reuther2010,Iqbal2016b,Buessen2021b}, adaptive Runge-Kutta methods have proven themselves to decrease numerical cost while maintaining numerical control over the integration error~\cite{Ritter2022,Niggemann2021a}. Lately, in a model study on itinerant FRG, adaptive step-size multi-step methods have been found to outperform even these methods by requiring a lower number of evaluations of the right-hand side~\cite{Beyer2022}. Similar results should hold true for PF/PMFRG applications as long as the tolerances are high enough. For solutions with low tolerance, higher-order methods are favourable in which case the stability region of multi-step methods shrinks while standard higher order Runge-Kutta solvers benefit from their growing stability region, allowing for larger stepsizes~\cite{Rackauckas2021}.

\end{document}